\documentclass[referee]{aa}  
\usepackage{graphicx}
\usepackage{txfonts}
\usepackage{booktabs}
\usepackage[colorlinks]{hyperref}
\usepackage{tablefootnote}

\hypersetup{%
  citecolor=black,
  linkcolor=black
}

\begin{document} 

\title{From exo-Earths to exo-Venuses}
\subtitle{Flux and Polarization Signatures of Reflected Light}

\author{G. Mahapatra,
          \inst{1}\fnmsep\thanks{Now at SRON Netherlands Institute 
                                 for Space Research, Leiden, 
                                 The Netherlands}
          \and
          F. Abiad \inst{1}
          \and
          L. Rossi\inst{2}
          \and
          D.M.Stam \inst{1}
          }

\authorrunning{Mahapatra et al.}
\titlerunning{From exo-Earths to exo-Venuses}

\institute{Faculty of Aerospace Engineering, Delft University of 
           Technology, Delft, The Netherlands \\
           \email{g.mahapatra@tudelft.nl}
%           \thanks{xx}
           \and
           CNRS/INSU, LATMOS-IPSL, Guyancourt, France\\
           }

\date{Accepted 11 January, 2023}

%%%%%%%%%%%%%%%%%%%%%%%%%%%%%%%%%%%%%%%%%%%%%%%%%%%%%%%%%%%%%%%%%%%%%%%
\abstract
{Terrestrial-type exoplanets in or near stellar habitable zones appear 
to be ubiquitous. It is, however, unknown which of these planets have
temperate, Earth-like climates or e.g.\ extreme, Venus-like climates.}
{Technical tools to distinguish different types of terrestrial-type 
planets are crucial for determining whether a planet could be habitable 
or incompatible with life as we know it. We investigate the potential 
of spectropolarimetry for distinguishing exo-Earths from exo-Venuses.}
{We present numerically computed fluxes and degrees of linear 
polarization of starlight that is reflected by exoplanets with atmospheres 
in evolutionary states ranging from similar to the current Earth to similar 
to the current Venus, with cloud compositions ranging from pure water to 
75\%~sulfuric acid solution, for wavelengths between 0.3~and 2.5~$\mu$m. 
We also present flux and polarization signals of such planets in 
stable but spatially unresolved orbits around the star Alpha Centauri~A.}
{The degree of polarization of the reflected starlight shows larger 
variations with the planetary phase angle and wavelength than the total 
flux. Across the visible, the largest degree of polarization is reached
for an Earth-like atmosphere with water clouds, due to Rayleigh scattering 
above the clouds and the rainbow feature at phase angles near 40$^\circ$. 
At near-infrared wavelengths, the planet with a Venus-like CO$_2$ atmosphere 
and thin water clouds shows the most prominent polarization features 
due to Rayleigh-like scattering by the small cloud droplets. 
A planet in a stable orbit around Alpha Centauri A would leave temporal 
variations on the order of 10$^{-13}$ W/m$^3$ in the total reflected
flux and 10$^{-11}$ in the total degree of polarization as the planet orbits
the star and assuming a spatially unresolved star-planet system. 
Star-planet contrasts are in the order of 10$^{-10}$ and vary proportionally with planetary flux.}
% Conclusions
{Current polarimeters appear to be incapable to distinguish between 
the possible evolutionary phases of spatially unresolved terrestrial exo-planets, 
as a sensitivity close to 10$^{-10}$ would be required to discern 
the planetary signal given the background of unpolarized starlight. 
A telescope/instrument capable of achieving planet-star contrasts 
lower than 10$^{-9}$ should be able to observe the large variation of the 
planet's resolved degree of polarization as a function of its phase angle 
and thus be able to discern an exo-Earth from an exo-Venus based on 
its clouds' unique polarization signatures.}

\keywords{Exo-planets --
                Venus --
                Radiative transfer --
                Polarimetry
               }
\maketitle

%-------------------------------------------------------------------
\section{Introduction}
\label{sect_introduction}

Despite having similar sizes, being formed around the same time
and from similar materials, it is clear that the Earth and Venus have 
evolved into dramatically different worlds. 
While it is generally acknowledged that Venus once had much larger 
amounts of water than today, it is still debated whether Venus was 
once more Earth-like with oceans of water 
before the runaway-greenhouse-effect took off \citep{donahue1982venus}, 
or whether the atmospheric water vapour never actually condensed on 
the surface \citep{turbet2021day}.
\cite{bullock2001recent} conducted a detailed study of the 
possible evolution of Venus's climate over long time periods 
starting with a water vapour enriched atmosphere.
Terrestrial-type exoplanets are also expected to harbour a 
wide variety of atmospheric compositions with maybe only a few planets
hospitable to life as we know it. 
Various climate models suggest that the likelihood of a planetary atmosphere
exhibiting a Venus-like runaway-greenhouse-effect is higher than that 
of an atmosphere in an Earth-like, N$_2$-dominated state 
\citep[and references therein]{lincowski2018evolved}.
A study by \cite{kane2020could} even shows that Jupiter's migration 
might have stimulated the runaway-greenhouse-effect on 
Venus, suggesting that there could be more Venus-analogs than Earth-analogs 
in planetary systems with Jupiter-like planets.

As planned powerful telescopes and dedicated, sensitive 
detection techniques will allow us to characterize  
smaller exoplanets in the near-future, it will become 
possible to probe terrestrial-type planets in and near the 
habitable zones of solar-type stars and to find out whether they resemble
Earth or Venus, or something else all together.
The high-altitude cloud deck on an exo-Venus would make it difficult to
use a technique like transit spectroscopy for the characterization of
the planet as the clouds themselves would block the transmission of the
starlight and apart from a spectral dependence of the cloud optical  
thickness which could leave a wavelength dependent transmission 
through the cloud tops, the microphysical properties of the cloud
particles, such as their composition, shape and size distribution would 
remain a mystery. 
Also, the clouds would inhibit measuring trace gas column densities
as they would block the planet's lower atmosphere and only allow transit 
spectroscopy of the highest regions of the atmosphere 
\citep[see, e.g.][and references therein]{lustig2019detectability}. 
Indeed, a Venus-like ubiquitous cloud deck could possibly be mistaken 
for the planet's surface, as one would only measure the transmittance 
through the gaseous atmosphere above the clouds, possibly inferring the 
atmosphere to be thin and eroding \citep{lustig2019mirage}. 

\cite{2021ApJ...922...44J} modelled the photochemistry of 
some of the primary sulphuric chemical species that should be
responsible for the formation and sustenance of Venus's sulfuric acid 
solution clouds, such as SO$_2$, OCS and H$_2$S, and found that the 
abundances of such species above the cloud deck would depend heavily 
on the effective temperature and distance to the parent star, 
with their abundances decreasing with increasing temperature and 
being depleted, as we see on Venus, in the presence of a star 
like the Sun. Thus it would be challenging to rule out the possibility of an
exoplanet being a Venus analog solely on the basis of the detection 
of such chemical species in transmission spectra 
\citep{2021ApJ...922...44J}.
Indeed, the full characterization of rocky exoplanets and their 
classification appears to require the direct imaging of starlight that
is reflected by such planets and/or the thermal radiation that is 
emitted by them.
While telescopes able to perform such measurements are not yet 
available, plans are underway for their development and deployment
\citep{keller2010epol,luvoir2019luvoir}.

While most of telescopes and instruments are designed for only 
measuring total fluxes of exoplanets, including 
(spectro)polarimetry is also being considered. The main
reason to include (spectro)polarimetry 
\citep[see e.g.][and references therein]{rossi2021spectropolarimetry}
is that it increases the
contrast between star and planet, as the stellar flux will be
mostly unpolarized when integrated over the disk \citep{kemp1987optical},
while the flux of the reflected starlight will usually be (linearly) 
polarized. And in addition, (spectro)polarimetry can be used for the
characterization of planetary atmospheres and surfaces.
As a classic example of the latter, \citet{hansen1974interpretation} used
Earth-based measurements of the disk-integrated degree of
polarization of sunlight that was reflected by Venus 
in three spectral bands and across a broad phase angle range,
to deduce that the particles forming Venus's main cloud deck 
consist of 75$\%$ sulfuric acid solution, that the effective radius 
of their size distribution is 1.05~$\mu$m, and that the effective
width of the distribution is 0.07.
They also derived the cloud top altitude (at 50~mbars)
by determining the amount of Rayleigh scattering in the gas 
above the cloud tops at a wavelength of 0.365~$\mu$m.
This was later confirmed by the Pioneer Venus mission which performed 
in-situ measurements using a nephelometer on a probe that descended 
through the clouds \citep{knollenberg1980microphysics}.

Polarimetry proved to be an effective technique for disentangling 
Venus's cloud properties because the scattering particles leave a unique
angular polarization pattern in the reflected sunlight depending
on the particles' micro- and macro-physical properties 
\citep[for an extensive explanation of the application of polarimetry for the characterization of planetary atmospheres, see][]{hansen1974light}. 
While multiple scattered light usually has a low degree of polarization,
and thus dilutes the angular polarization patterns of the
singly scattered light, the angles where the absolute degree of polarization 
reaches a local maximum and/or where it is zero (the so-called 'neutral points') 
are preserved and thus still allow for the characterization of the 
particles.

Another factor in the successful application of (spectro)polarimetry for 
the characterization of Venus's clouds and hazes is that with Earth-based
telescopes, 
inner planet Venus can be observed at a wide range of phase angles,
thus allowing observations of the angular variation of the degree of 
polarization due to the light that has been singly scattered by the
atmospheric constituents.
In our solar system, only Venus, Mercury, and the Moon can be observed
at a large phase angle range with Earth-based telescopes (ignoring 
the proximity of Mercury to the Sun). 
To effectively apply polarimetry to the outer planets in the solar system, 
a polarimeter onboard a space mission would be needed.
An example of such an instrument was OCCP onboard NASA's Galileo mission
\citep[][]{1992SSRv...60..531R} that orbited Jupiter.
Regarding exoplanets, however, the range of observable phase angles depends 
on the inclination angle of the planetary orbits: for a face-on orbit, the
planet's phase angle will be 90$^\circ$ everywhere along the orbit,
while for an edge-on orbit, the phase angle will range from close to 
0$^\circ$ (when the planetary disk is fully illuminated) 
to 180$^\circ$ (when the night-side of the planet is in view).
The precise range of accessible phase angles would of course depend
on the observational technique and e.g.\ the use of a coronagraph 
or star-shade.

Here we investigate the 
total flux and degree of polarization of starlight 
that is reflected by terrestrial-type exoplanets, focusing on the
possible evolutionary stages of Venus as described by 
\citet{bullock2001recent}. Our goal is to identify characteristic 
signatures that could help to identify the properties of 
exo-Venuses, thus to guide the design of future telescope instruments.
We compute the disk-integrated total and polarized fluxes 
of light reflected between wavelengths of 0.3 to 2.5~$\mu$m.
First, we study the single scattering properties of spherical 
cloud droplets of pure water (H$_2$O) or 75$\%$ sulphuric acid 
(H$_2$SO$_4$) 
in order to identify potentially distinct signatures for each 
particle type as a function of wavelength and planetary phase angle. 
Second, we compute the multiple scattered flux and polarization 
signals that 
are integrated over the planet's illuminated disk as functions 
of the planet's phase angle. Third, we compute the signals of 
the planets in the four evolutionary phases 
in stable orbits around the nearby solar-type Alpha Centauri A, 
simulating the observations of such planets if they are spatially 
unresolved from their parent star.

The outline of this paper is as follows. 
In Sect.~\ref{sect:methods}, we define the fluxes and polarization of 
planets, and we describe our numerical algorithm 
and the four model planets in the evolutionary phases as described by
\citet{bullock2001recent}.
In Sect.~\ref{sect:planetphaseresults}, we present the 
total and polarized fluxes as computed for planets that are 
spatially resolved from their star and for planets that are
spatially unresolved. In the latter case, the planet's signal is
thus combined with the stellar light. We specifically assume that 
our model planet orbits the solar-type star Alpha Centauri~A. 
In Sect.~\ref{sect:discussion&summary}, we summarize our results
and present our conclusions.

%-------------------------------------------------------------------
\section{Numerical method}
\label{sect:methods}

%--------------------------------------------------------------------
\subsection{Flux and polarization definitions}
\label{sect:definitions}

In this paper, we 
present the flux and polarization signals of starlight that 
is reflected by potentially habitable exoplanets that orbit 
solar-type stars, and in particular, Alpha Centauri A.
Because these planets will be very close in angular distance
to their parent star, they will usually be spatially unresolved,
i.e.\ it will not be possible to spatially separate the 
planet's signal from that of its parent star. 
The flux vector ${\bf F}_{\rm u}$ (u= 'unresolved') 
that describes the light of the star and its spatially unresolved
planet, and that arrives at a distant observer is then written as
\begin{equation}
    {\bf F}_{\rm u}(\lambda,\alpha)= 
    {\bf F}_{\rm s}(\lambda) + 
    {\bf F}_{\rm p}(\lambda,\alpha),
\label{eq_1}                             
\end{equation}
with ${\bf F}_{\rm s}$ the star's flux vector and 
${\bf F}_{\rm p}$ that of the planet. 
Furthermore, $\lambda$ is the wavelength
(or wavelength band), and $\alpha$ is the planetary phase angle, 
i.e.\ the angle between the star and the observer as measured
from the center of the planet. 
We assume that the light of the star is captured
together with the starlight that is reflected by the planet. 
A telescope with a coronagraph or star-shade would of course 
limit the amount of captured direct starlight, depending on 
its design and the angular distance between the star and the 
planet.

A flux (column) vector is given by \citep[][]{hansen1974light}
\begin{equation}
    {\bf F} = [ F, Q, U, V ],
\label{eq_2}    
\end{equation}  
with $F$ the total flux, $Q$ and $U$ the linearly polarized
fluxes, and $V$ the circularly polarized flux. The dimensions
of $F$, $Q$, $U$, and $V$ are W~m$^{-2}$, or W~m$^{-3}$ when 
defined per wavelength. 

Measurements of FGK-stars, such as the Sun and Alpha Centauri A, 
indicate that their (disk-integrated) 
polarized fluxes are virtually negligible 
\citep{kemp1987optical,cotton2017intrinsic}, 
thus we describe the star's flux (column) vector that arrives 
at the observer located at a distance $D$ as
\begin{equation}
    {\bf F}_{\rm s}(\lambda)
         = F_{\rm s}(\lambda) \hspace*{0.1cm} {\bf 1}
         = \frac{R_{\rm s}^2}{D^2} \hspace*{0.1cm} 
           \pi B(\lambda,T_{\rm s})
           \hspace{0.1cm} {\bf 1}, 
\label{eq_3a}
\end{equation}
with $\pi B$ the stellar surface flux,
$T_{\rm s}$ the star's effective temperature,
$R_{\rm s}$ the stellar radius, 
and ${\bf 1}$ the unit (column) vector.
The parameter values that we adopt for the Alpha Centauri A 
system are listed in Table~\ref{table1}.

Because of the huge distances to stars and their planets,
flux vector ${\bf F}_{\rm p}$ of the starlight that 
is reflected by an exoplanet pertains to the planet as a whole, 
thus integrated across the illuminated and visible part of 
the planetary disk. It is given by
\citep[see e.g.][]{rossi2018pymiedap}
\begin{eqnarray}
   {\bf F}_{\rm p}(\lambda,\alpha) & = & 
          A_{\rm G}(\lambda) \hspace*{0.1cm} 
          {\bf R}_{\rm p}(\lambda,\alpha) 
          \hspace*{0.1cm} \frac{r_{\rm p}^2}{D^2} 
          \frac{R_{\rm s}^2}{d^2} \hspace{0.1cm}
           \pi B(\lambda,T_{\rm s})
                        \hspace*{0.1cm} {\bf 1} 
                        \label{eq_4} \\
         & = &  A_{\rm G}(\lambda) \hspace*{0.1cm} 
          {\bf R}_{\rm 1p}(\lambda,\alpha) 
          \hspace*{0.1cm} \frac{r_{\rm p}^2}{D^2}
           \frac{R_{\rm s}^2}{d^2} 
           \hspace*{0.1cm} \pi B(\lambda,T_{\rm s}).
\label{eq_5}
\end{eqnarray}
Here, $A_{\rm G}$ is the planet's geometric albedo, 
${\bf R}_{\rm p}$ the matrix describing the reflection by the
planet and ${\bf R}_{\rm 1p}$ its first column, 
$r_{\rm p}$ is the planet's radius, 
$d$ the distance between the star and
the planet, and $D$ the distance to the observer. 
The planet's reflection is normalized such that planetary
phase function $R_{\rm 1p}$,
which is the first element of ${\bf R}_{\rm 1p}$,
equals 1.0 at $\alpha= 0^\circ$.

The contrast $C$ between the total flux of the 
planet and the total flux of the star is then given by
\begin{equation}
    C(\lambda,\alpha)= \frac{F_{\rm p}(\lambda,\alpha)}{F_{\rm s}(\lambda)} 
                     = A_{\rm G}(\lambda) \hspace*{0.1cm} 
                       R_{\rm 1p}(\lambda,\alpha)
                       \frac{r_{\rm p}^2}{d^2},
\label{eq_C}
\end{equation} 
with $F_{\rm p}$ the first element of the planetary flux vector 
${\bf F}_{\rm p}$.
Using the parameters from Table~\ref{table1}, the contrast $C$ 
between a planet with the radius of Venus at a 
Venus-like distance from Alpha Centauri~A 
equals about $2 \cdot 10^{-9} A_{\rm G}$ at $\alpha=0^\circ$ 
(at this phase angle, the planet would actually be precisely 
behind the star with respect to the observer 
and thus out of sight).

The degree of polarization of the spatially resolved planet
(without including any direct light of the star) is defined as
\begin{equation}
    P_{\rm p}= 
         \frac{\sqrt{Q_{\rm p}^2 + U_{\rm p}^2}}
         {F_{\rm p}},
\label{eq_P}
\end{equation}
where we ignore the planet's 
circularly polarized flux $V_{\rm p}$ as it is expected
to be very small compared to the linearly polarized fluxes
\citep{rossi2018circular}. 
We also ignore the circularly polarized fluxes in our radiative
transfer computations, as this saves significant amounts of computing
time without introducing 
significant errors in the computed total and linearly polarized
fluxes \citep[see][]{stam2005errors}. 

Fluxes $Q_{\rm p}$ and $U_{\rm p}$ are defined with respect to 
the planetary scattering plane, 
which is the plane through the planet, the star and the observer.
In case the planet is mirror-symmetric with respect to the 
planetary scattering plane, 
linearly polarized flux $U_{\rm p}$ equals zero and we can use an 
alternative definition of the degree of polarization that includes
the polarization direction as follows
\begin{equation}
\label{eqn:dolp}
    P_{\rm p}= -\frac{Q_{\rm p}}
                     {F_{\rm p}}. 
\end{equation}
If $P_{\rm p} > 0$ ($P_{\rm p} < 0$), the light is polarized 
perpendicular (parallel) to the reference plane.

In case a planet is not completely spatially resolved from its parent 
star, and the background of the planet on the sky is thus filled with
(unpolarized) starlight,
the observable degree of polarization $P_{\rm u}$ can be written 
as (cf.\ Eqs.~\ref{eq_C}-\ref{eq_P})
\begin{equation}
    P_{\rm u} = 
    \frac{\sqrt{Q_{\rm p}^2+U_{\rm p}^2}}{F_{\rm p} + x F_{\rm s}} = 
    \frac{F_{\rm p}}{F_{\rm p} + x F_{\rm s}} \hspace*{0.1cm} P_{\rm p}
    = \frac{C}{C+x} P_{\rm p},
\label{eq_pux}
\end{equation}
with $x$ the fraction of the stellar flux that is in the background,
which will
depend on the angular distance between the star and the planet,
on the starlight suppressing techniques that are employed,
such as a coronagraph or star-shade, and on the spatial 
resolution of the telescope at the wavelength under consideration. 
This equation also holds for the signed degree
of polarization as given by Eq.~\ref{eqn:dolp}.
If $x=1$, the planetary and the stellar flux are 
measured together.
In that case,
\begin{equation}
     P_{\rm u} = \frac{C}{C+1} P_{\rm p} \approx C P_{\rm p}.
\label{eq_pu}
\end{equation}
Here we used the fact that the contrast $C$ will usually be 
very small (on the order of 10$^{-9}$ as shown earlier).

The planet's degree of polarization $P_{\rm p}$ and the contrast 
$C$ both depend on $\lambda$ and $\alpha$, but generally in a 
different way. The dependence of $P_{\rm u}$ on 
$\lambda$ and $\alpha$ will thus generally differ from that of 
either $P_{\rm p}$ or $C$.

%%%%%%%%%%%%%%%%%%%%%%%%%%%%%%%%%%%%%%%%%%%%%%%%%%%%%%%%%%%%%%%%%%%%%
\begin{table}[]
    \centering
    \caption{The values of the parameters describing the planetary
             system of Alpha Centauri A used in our numerical 
             modelling\tablefootnote{The orbital distance $d$ of the planet has been chosen
             such that it receives the same stellar flux as Venus
             receives from the Sun, and in accordance with the
             orbit stability requirements for a planet
             around Alpha Centauri A 
             predicted by \cite{2016AJ....151..111Q}.
             For the radius of the Sun, $R_{\rm Sun}$, we
             use 695,700~km.
             }.}
    \begin{tabular}{l|c|c}
        Parameter (unit) & Symbol & Value \\ \hline
        Stellar radius ($R_{\rm Sun}$) & $R_{\rm s}$ & 1.2234 \\
        Stellar effective temperature (K) & $T_{\rm s}$ & 5790 \\
        Planet radius (km) & $r_{\rm p}$ & 6052 \\
        Planet orbital distance (AU) & $d$ & 0.86 \\
        Planet orbital period (yr) & $P$ & 0.76 \\
        Distance to the system (ly) & $D$ &  4.2 \\
        Angular separation (arcsecs) & $S$ &  0.67 \\
    \end{tabular}
    \vspace{0.3cm}

    \label{table1}
\end{table}
%%%%%%%%%%%%%%%%%%%%%%%%%%%%%%%%%%%%%%%%%%%%%%%%%%%%%%%%%%%%%%%%%%%%%

%--------------------------------------------------------------------
\subsection{Our radiative transfer algorithm}
\label{sect_algorithm}

Our procedure to compute the flux vector 
${\bf F}_{\rm p}$ (Eq.~\ref{eq_5}) 
of the starlight that is reflected by the planet,
is described in \citet[][]{rossi2018pymiedap}.
The radiative transfer algorithm is based on an efficient 
adding-doubling algorithm \citep{de1987adding} and fully 
includes polarization
for all orders of scattering. With this algorithm, through the 
use of a Fourier-series expansion of the planetary reflection 
matrix ${\bf R}_{\rm p}$, the reflected flux vector
can be computed for any planetary phase angle $\alpha$.

Our model planetary atmospheres consist of horizontally homogeneous
layers. For each layer, we prescribe the total optical thickness
$b$, the single-scattering albedo $a$, and the single-scattering
matrix ${\bf P}$. Our layered model atmosphere is bounded below 
by a Lambertian reflecting
surface (i.e.\ the light is reflected isotropically and unpolarized)
with an albedo $a_{\rm surf}$.

A layer's optical thickness $b$ at a wavelength $\lambda$ is the sum
of the optical thicknesses of the gas molecules, $b^{\rm m}$, and, 
if present, the cloud particles, $b^{\rm c}$. We ignore other 
atmospheric particles, such as haze particles.
The single-scattering matrix {\bf P} of a mixture of
gas molecules and cloud particles in a layer is given by 
\begin{equation}
    {\bf P}(\Theta,\lambda) = 
    \frac{b_{\rm sca}^{\rm m}(\lambda)~{\bf P}^{\rm m}(\Theta,\lambda) 
    + b_{\rm sca}^{\rm c}(\lambda)~{\bf P}^{\rm c}(\Theta,\lambda)}
    {b_{\rm sca}^{\rm m}(\lambda) + b_{\rm sca}^{\rm c}(\lambda)},
\end{equation}
with subscript `${\rm sca}$' referring to `scattering',
thus $b_{\rm sca}= a b$, with $a$ the single scattering albedo. 
Furthermore, ${P^{\rm m}}$ is the single-scattering matrix of the
gas molecules, and ${P^{\rm c}}$ that of the cloud particles.
$\Theta$ is the single scattering angle: 
$\Theta= 180^\circ - \alpha$. 
 
We use two types of model atmospheres to study the influence of an 
exoplanet's atmospheric evolution on the reflected light signals: 
an Earth-like and a Venus-like atmosphere. 
For our Earth-like atmosphere, we define the pressure and temperature 
across 17~layers, representing a mid-latitude summer profile 
\citep[following][]{stam2008spectropolarimetric}.
For our Venus-like atmosphere, we use 71~layers with pressure and
temperature profiles
from the Venus International Reference Atmosphere (VIRA) 
\citep{kliore1985venus}, representing a mid-latitude afternoon profile. 
With these vertical profiles, and assuming anisotropic Rayleigh
scattering \citep{hansen1974light}, we compute each layer's 
single scattering matrix ${\bf P}^{\rm m}$ and 
the scattering optical thickness $b_{\rm sca}^{\rm m}$. 
We neglect absorption, thus $b^{\rm m} = b^{\rm m}_{\rm sca}$.
The depolarization factor for computing ${\bf P}^{\rm m}$ and 
$b_{\rm sca}^{\rm m}$ for anisotropic 
Rayleigh scattering depends on the atmospheric composition. 
For the Earth-like atmosphere, we use a (wavelength independent)
depolarization factor of 0.03, which is representative for dry air, 
and for the Venus-like atmosphere, we use 0.09, which is 
representative for a pure CO$_2$ atmosphere
\citep{hansen1974light}. 
We use wavelength-independent refractive indices of 1.00044 
and 1.00027 for the Venus-like and the Earth-like model 
atmospheres, respectively; note that this assumption 
has a negligible effect on the reflected total and polarized fluxes.

The cloud particles in our model atmospheres are spherical and 
distributed in size according to a two-parameter gamma size distribution 
\citep[see][]{hansen1974light} that is described by an effective radius
$r_{\rm eff}$ 
and an effective variance $v_{\rm eff}$. The terrestrial clouds
are located between 1 and 3~km altitude, and the  
Venusian clouds, depending on their evolutionary phase, 
between 47 and 80~km. The cloud optical thickness has a uniform 
vertical distribution through the altitude range
(see Fig.~\ref{fig:cloudevolution}).

The single-scattering properties of the cloud particles are computed 
using Mie-theory \citep{de1984expansion}, as these particles 
are expected to be spherical.
For these computations we specify the wavelength $\lambda$ 
and $n_{\rm r}$, the refractive index of the cloud particles. 
The cloud particles are composed of either pure water or a sulphuric 
acid solution with varying concentration. We use the refractive 
index of water from \cite{hale1973optical} and that of sulphuric 
acid with 75~$\%$ acid concentration from \cite{palmer1975optical}. 
We use a negligible value for the imaginary part of the particles' 
refractive indices, $n_{\rm i}$ = 10$^{-8}$. 

%---------------------------------------------------------------------
\subsection{Cloud properties through the planet's evolution}
\label{sect:cloudproperties}

%%%%%%%%%%%%%%%%%%%%%%%%%%%%%%%%%%%%%%%%%%%%%%%%%%%%%%%%%%%%%%%%%%%%%%
\begin{figure*}[!t]
\centering
\includegraphics[width=0.8\textwidth]{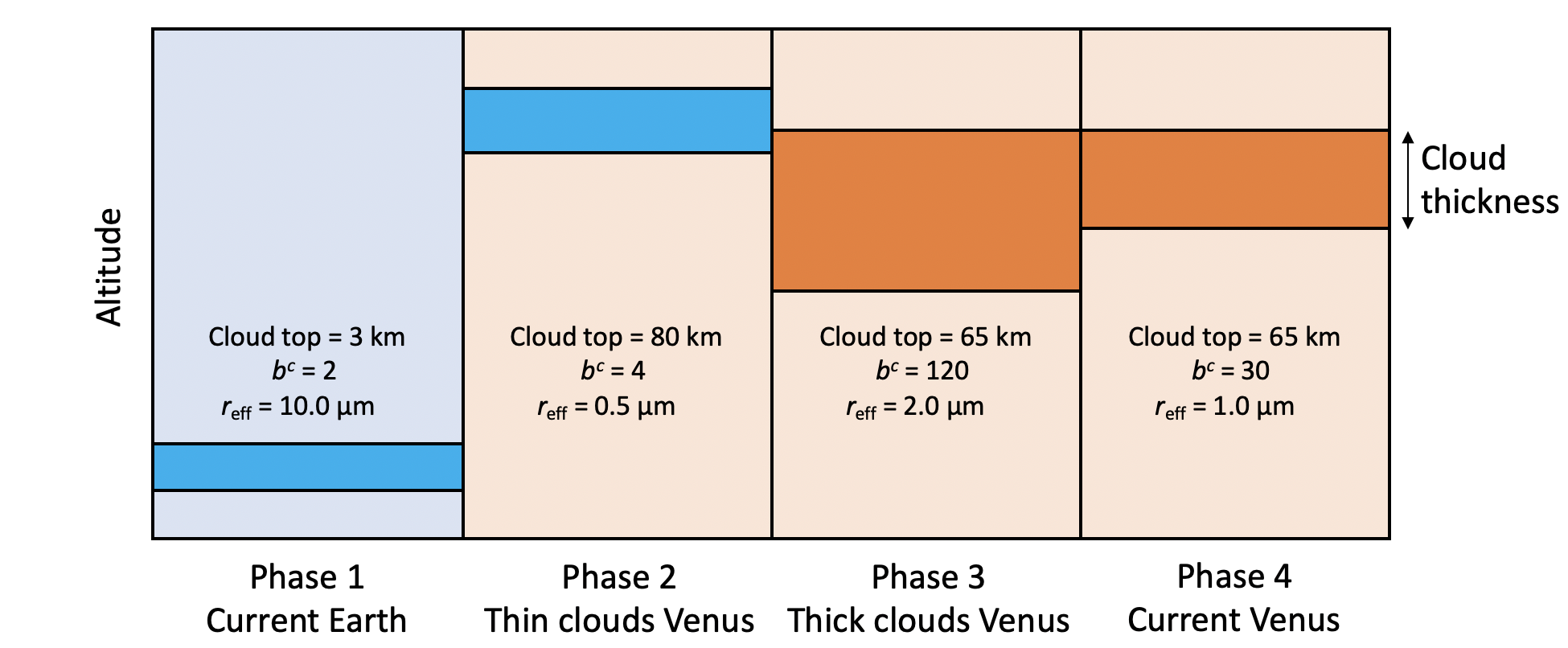}
\caption{The four evolutionary phases of the model planets
         \citep{bullock2001recent}.
         In Phases 1 and 2, the clouds consist of liquid water droplets,
         and in Phases 3 and 4, of liquid sulfuric acid solution droplets.
         The cloud optical thickness is indicated by $b^{\rm c}$
         and the cloud particle effective radius by $r_{\rm eff}$.
         For the effective variance $v_{\rm eff}$ of the size distributions
         in Phases 1-3, we use 0.1, and for Phase~4,
         $v_{\rm eff}=0.07$.}
\label{fig:cloudevolution}
\end{figure*}
%%%%%%%%%%%%%%%%%%%%%%%%%%%%%%%%%%%%%%%%%%%%%%%%%%%%%%%%%%%%%%%%%%%%%%

It is suspected that early Venus had a thin, Earth-like atmosphere 
and (possibly) an Earth-like ocean that was later lost due to the runaway
greenhouse effect \citep{donahue1982venus,kasting1988runaway,way2020venusian}. 
As the planet's surface heated up, the water would have evaporated 
and enriched the atmosphere with water vapor.
The macroscopic cloud properties for the 4 evolutionary phases that 
we will use are illustrated in Fig.~\ref{fig:cloudevolution}.

We start our evolutionary model of Venus assuming 
Earth-like conditions (Phase 1), i.e. an atmosphere 
consisting of 78$\%$ N$_2$ and 22$\%$ O$_2$.
Model simulations showed that the actual dependence of 
the total and polarized flux
signals on the percentage of oxygen appeared to be negligible.
Hence we used the present day Earth atmosphere as the 
Earth-like atmosphere model while the actual percentage of 
oxygen on an exo-planet could be different.
The cloud particles have an effective radius 
$r_{\rm eff}$ of 10~$\mu$m in agreement with ISCCP \citep{isccp_particle} 
and an effective variance $v_{\rm eff}$ of 0.1.
The total cloud optical thickness $b^{\rm c}$ is 10.0 at 
$\lambda= 0.55~\mu$m and the cloud layer extends from 2~to 4~km.

The next evolutionary phases are also 
inspired by the Venus climate model of \cite{bullock2001recent}. 
In Phase 2, the atmosphere is Venus-like as it consists of 
pure CO$_2$ gas, and has relatively thin liquid water clouds with 
$b^{\rm c}=4$, and with the cloud tops at 80~km.
For this phase, we use $r_{\rm eff}$ of 0.5~$\mu$m, which is 
smaller than the present day 
value, because the atmosphere is expected to be too hot for strong
condensation to take place thus preventing the particles to grow larger.
In Phase 3, the clouds are thick sulphuric-acid solution clouds, with 
$b^{\rm c}=120$ and the cloud tops at 65~km, because the atmosphere 
is cool enough to allow condensation and/or coalescence of saturated
vapour over a large altitude range. Since the region of condensation 
covers a large altitude range, the particles can grow 
large until they evaporate. In this phase, 
$r_{\rm eff}=2~\mu$m, which is twice the effective radius 
of the present day Venus cloud particles. 
For both phases 2 and~3, we use $v_{\rm eff}=0.1$. 
In Phase 4, the clouds have the present-day properties of 
Venus's clouds with $b^{\rm c}= 30$ and the cloud tops at 65~km 
\citep{rossi2015preliminary,ragent1985particulate}. 
For the cloud particle sizes in this phase, we use 
$r_{\rm eff}=1.05$~$\mu$m and $v_{\rm eff}=0.07$ following the values 
derived by  \cite{hansen1974interpretation}. 
We ignore the absorption by cloud particles in the UV in all of 
our Venus-like clouds to avoid adding complexity and because 
the exact nature and location of the UV-absorption is still 
under debate \citep[][]{titov2018clouds}.

%%%%%%%%%%%%%%%%%%%%%%%%%%%%%%%%%%%%%%%%%%%%%%%%%%%%%%%%%%%%%%%%%%%%%
\begin{figure*}[h]
\centering
\includegraphics[width=0.94\textwidth]{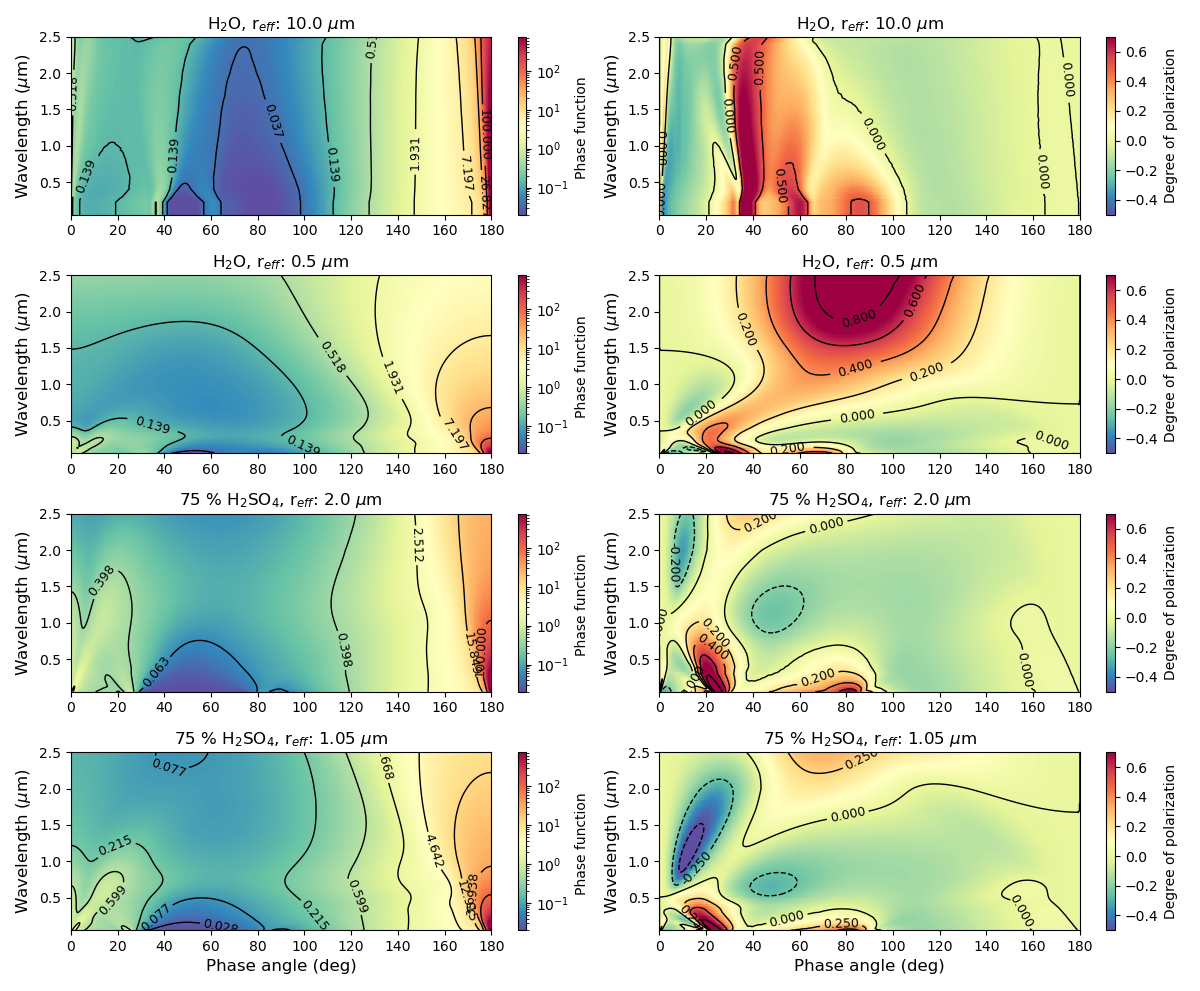}
\caption{The total flux and degree of polarization of 
         incident unpolarized light that has been singly 
         scattered by four different types of cloud particles,
         as functions of the phase angle $\alpha$ and the wavelength
         $\lambda$.
         Left column: the total flux or phase function
         (single scattering matrix element $P_{\rm 11}$). 
         Right column: the degree of polarization 
         ($-P_{\rm 21}/P_{\rm 11}$).
         First row: H$_2$O particles with  
         $r_{\rm eff}= 10~\mu$m and $v_{\rm eff}=0.1$
         (belonging to the Phase 1 model planet: 'Current Earth'); 
         Second row: 
         H$_2$O particles with $r_{\rm eff}= 0.5~\mu$m and 
         $v_{\rm eff}=0.1$
         (Phase 2: 'Thin clouds Venus');
         Third row: 
         75\% H$_2$SO$_4$ particles with $r_{\rm eff}= 2.0~\mu$m 
         and $v_{\rm eff}= 0.1$ (Phase 3: 'Thick clouds Venus'); 
         Fourth row: 75\% H$_2$SO$_4$ particles with 
         $r_{\rm eff}= 1.05~\mu$m and $v_{\rm eff}= 0.07$
         (Phase 4: 'Current Venus').}
\label{fig:singlescattering}
\end{figure*}
%%%%%%%%%%%%%%%%%%%%%%%%%%%%%%%%%%%%%%%%%%%%%%%%%%%%%%%%%%%%%%%%%%%%%

Figure~\ref{fig:singlescattering} shows the phase function 
(i.e.\ single scattering matrix element $P_{\rm 11}$) and 
the degree of linear 
polarization for unpolarized incident light (the ratio of single
scattering matrix elements $-P_{\rm 21}$/$P_{\rm 11}$) that has been 
singly scattered by the four different types of cloud particles
as functions of $\alpha$ (i.e.\ 
180$^\circ$ - $\Theta$), for a range of wavelengths $\lambda$.

As can be seen in Fig.~\ref{fig:singlescattering},
the phase functions show strong 
forward scattering peaks (near $\alpha=180^{\circ}$, thus when the 
night-side of the planet would be turned towards the observer)
that decrease with increasing $\lambda$, 
thus with decreasing effective particle size parameter 
$x_{\rm eff}= 2\pi r_{\rm eff}/\lambda$ (for the large
H$_2$O cloud particles, with $r_{\rm eff}=10$~$\mu$m, this 
decrease is not readily apparent from the figure).
The H$_2$O particles with $r_{\rm eff}=10~\mu$m show
a moderate local maximum in the phase function around 
$\alpha=40^\circ$, which is usually referred to
as the primary rainbow \citep[see e.g.][]{hansen1974light}.
The large H$_2$O and the H$_2$SO$_4$ particles also produce 
higher fluxes towards $\alpha=0^\circ$ that are referred to
as the glory 
\citep{laven2008effects,munoz2014glory,markiewicz2014glory,rossi2015preliminary,markiewicz2018aerosol}. For the small H$_2$O
particles and the H$_2$SO$_4$ particles at larger wavelengths, 
the phase functions become more isotropic and the glory 
and other angular features disappear.

Figure~\ref{fig:singlescattering} also shows the degree of 
linear polarization of the singly scattered light.
This degree of polarization appears to be more 
sensitive to the particle composition than the scattered flux, 
especially for $\lambda$ between 0.5 and 2~$\mu$m, 
where H$_2$O particles yield relatively high positive degrees 
of polarization (perpendicular to the scattering plane)
between phase angles of about 20$^\circ$ and 100$^\circ$, 
whereas the H$_2$SO$_4$ particles impart a mostly negative 
degree of polarization through a broad range of 
phase angles, except for narrow regions around 
$\alpha= 20^\circ$ and $80^\circ$. 
The tiny, $r_{\rm eff}=0.5$~$\mu$m, water droplets 
have a strong, broad positive polarization region 
for $\lambda \geq 1~\mu$m, where they are so small 
with respect to the wavelength that they scatter like 
Rayleigh scatterers.

As mentioned before
\citep[see e.g.][]{hansen1974light,hansen1974interpretation},
patterns in the single scattering degree of polarization are 
generally preserved when multiple scattered light is added, 
as the latter usually has a low degree of polarization,
and thus adds mostly total flux, which subdues angular features,
but does not change the angular pattern (local maxima, minima,
neutral points) itself. 
The single scattering angular features in the polarization
will thus also show up in the polarization
signature of a planet as a whole, and can be used for 
characterisation of the cloud particle properties and thus
possibly of various phases in the evolution of a Venus-like
exoplanet. This will be investigated in the next section.

%--------------------------------------------------------------------
\section{Results}
\label{sect:planetphaseresults}

Here we present the disk-integrated total flux and 
degree of polarization of incident unpolarized starlight that is
reflected by the model planets at different wavelengths $\lambda$ 
and for phase angles $\alpha$ ranging from 0$^\circ$ to 180$^\circ$. 
The actual range of phase angles at which an exoplanet can
be observed depends on the inclination angle $i$ of the 
planet's orbit 
(the angle between the normal on the orbital plane and the direction
towards the observer): 
$\alpha$ ranges between $90^\circ - i$ to $90^\circ + i$.
Obviously, at $\alpha=0^\circ$, the planet would be precisely 
behind its star, and at 180$^\circ$ 
it would be precisely in front of its star (in transit). 
Other phase angles might be inaccessible due to 
restrictions of inner working angles of telescopes and/or instruments.
For completeness, we include all phase angles in our computations.

Section~\ref{sect_resolved} shows results for
spatially resolved planets and Sect.~\ref{sect_unresolved} for 
planets that are spatially unresolved from their star. 
In particular, we show these latter results for a model
planet orbiting the star Alpha Centauri~A at a distance where
the incident stellar flux is similar to the solar flux that 
reaches Venus.
Because our model planets are all
mirror-symmetric with respect to the reference plane, 
their linearly polarized flux $U_{\rm p}$ equals zero 
and will not be discussed further.

%%%%%%%%%%%%%%%%%%%%%%%%%%%%%%%%%%%%%%%%%%%%%%%%%%%%%%%%%%%%%%%%%%%%%
\subsection{Flux and polarization of spatially resolved planets}
\label{sect_resolved}

%--------------------------------------------------------------------
\begin{figure*}[!th]
\centering
\includegraphics[width=0.94\textwidth]{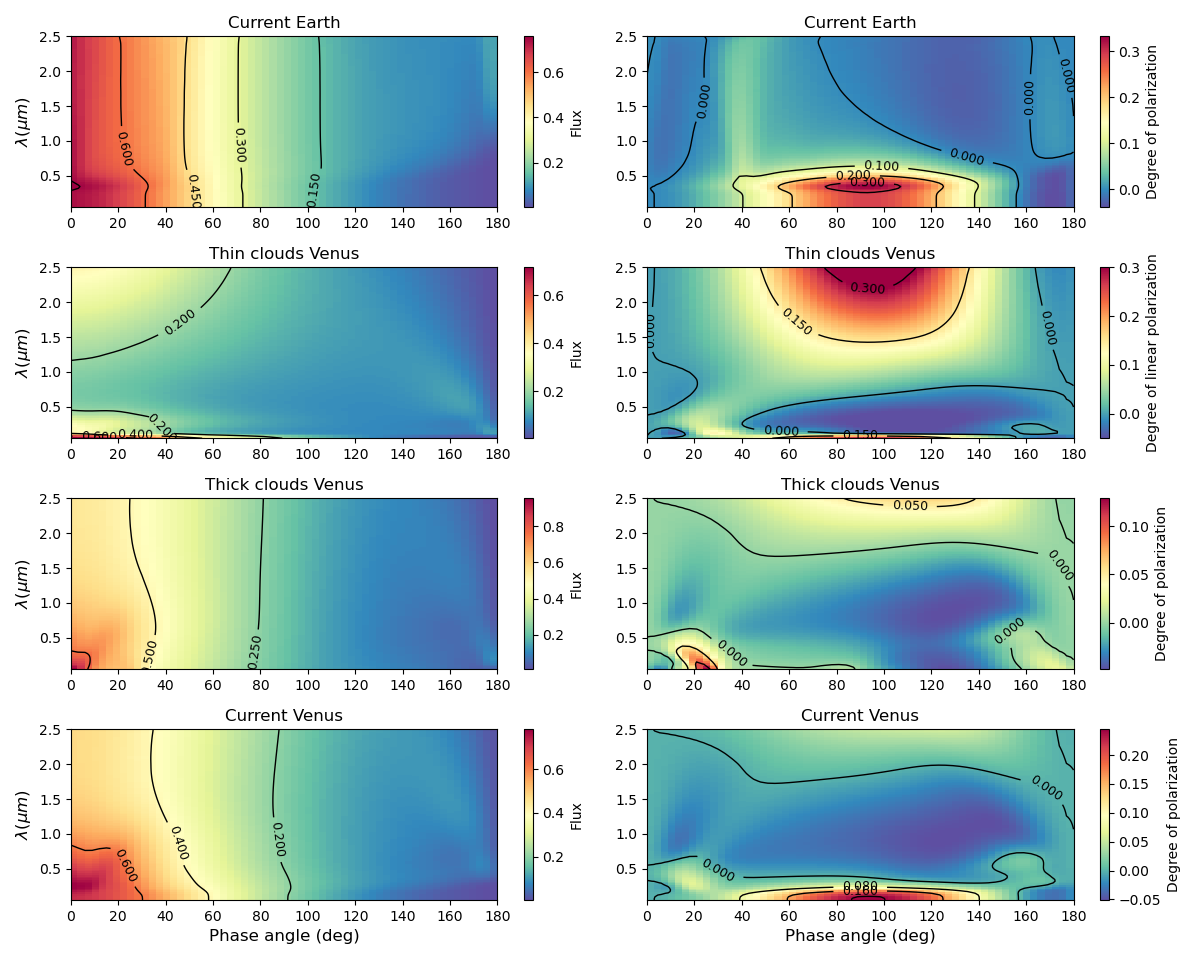}
\caption{Left column: The total flux (or phase function); 
         and right column: The degree of polarization of incident
         unpolarized
         starlight that is reflected by the model planets in the 
         4 evolutionary phases
         as functions of $\alpha$ and $\lambda$.
         First row: Phase 1 ('Current Earth'); Second row: Phase 2 
         ('Thin clouds Venus'); Third row: Phase 3 
         ('Thick clouds Venus'); 
         Fourth row: Phase 4 ('Current Venus'). 
         The phase functions are normalised such that at $\alpha=0^\circ$,
         they equal the planet's geometric albedo $A_{\rm G}$.}
\label{fig:phasecontour}
\end{figure*}
%--------------------------------------------------------------------

\begin{figure}[b!]
\includegraphics[width=0.48\textwidth]{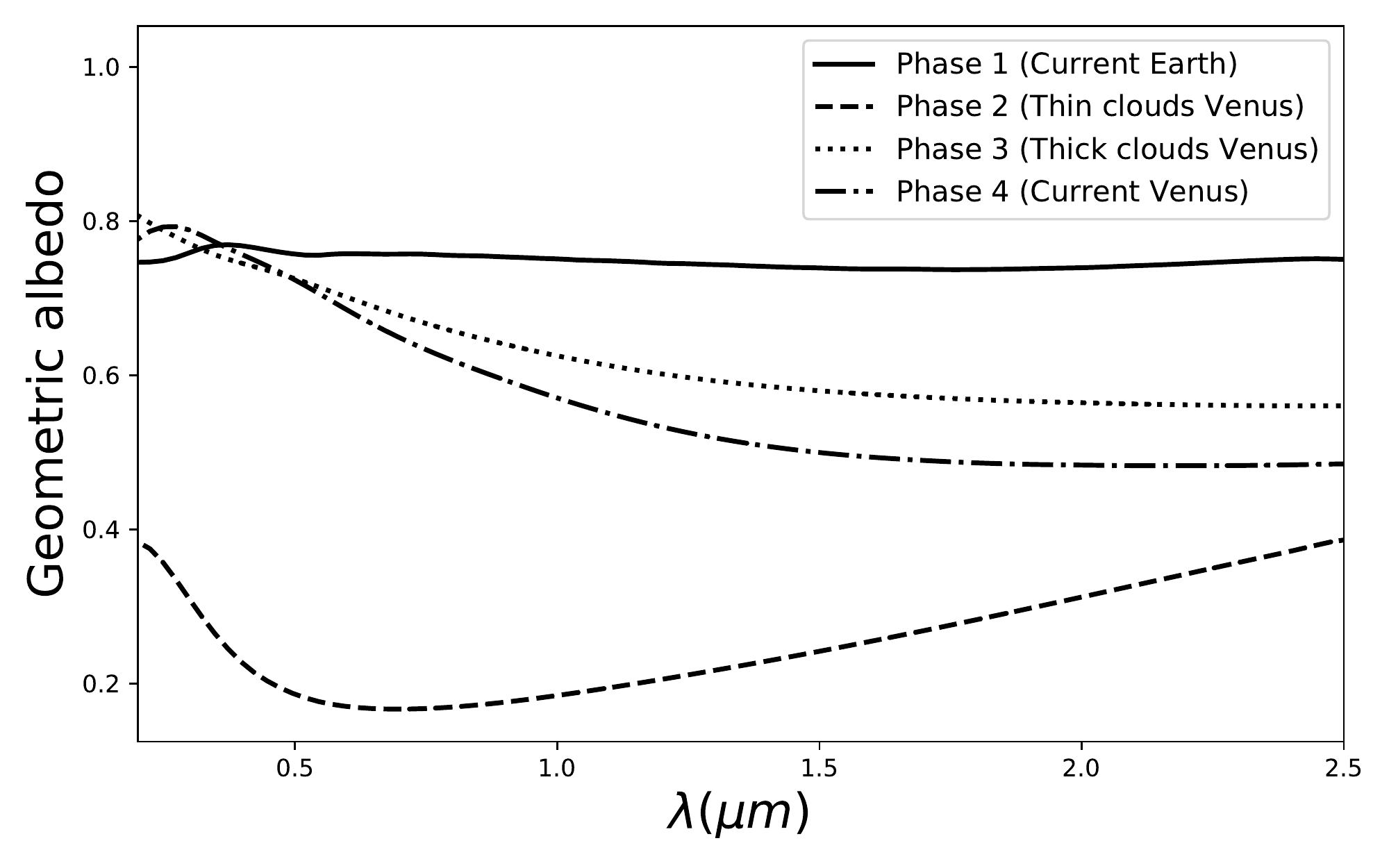}
\caption{The planets' geometric albedo's $A_{\rm G}$ as 
         functions of the wavelength $\lambda$ for the four evolutionary 
         phases.}
\label{fig:geometricalbedo}
\end{figure}

Figure~\ref{fig:phasecontour} shows the total flux $F_{\rm p}$ 
(the planetary phase function) and degree 
of polarization $P_{\rm p}$ as functions of $\alpha$ and $\lambda$
for the four evolutionary phases illustrated in
Fig.~\ref{fig:cloudevolution}. The total fluxes are normalized such that at $\alpha=0^\circ$, 
they equal the planet's geometric albedo $A_{\rm G}$ (see Eq.~\ref{eq_5}). 
Figure~\ref{fig:geometricalbedo} shows $A_{\rm G}$ of the planets in the
four evolutionary phases as functions of the wavelength $\lambda$. 
Table~\ref{table:geometricalbedo} lists the geometric albedo's at 
0.5, 1.0, 1.5 and 2.0~$\mu$m.
The 'Current Earth' (Phase~1) shows very little variation in $A_{\rm G}$,
and the 'Thin clouds Venus' (Phase~2) has the lowest albedo because of 
the small cloud particles and the small cloud optical thickness.  
The geometric albedo's of the 'Thick clouds Venus' (Phase~3) and the 
'Current Venus' (Phase~4) are very similar.
Thus across the wavelength region investigated in this paper, the 
'Current Earth' (Phase~1) has the highest geometric albedo.

\begin{table}[!h]
    \centering
    \caption{The model planets' geometric albedo's $A_{\rm G}$
         for the four evolutionary phases at four wavelengths.}
    \begin{tabular}{l|c|c|c|c} 
         $\lambda$ ($\mu$m) & 0.5 & 1.0 & 1.5 & 2.0 \\ \hline
         Current Earth      & 0.757 & 0.752 & 0.740 & 0.739 \\ 
         Thin clouds Venus  & 0.186 & 0.184 & 0.240  & 0.310 \\
         Thick clouds Venus & 0.727 & 0.627 & 0.580  & 0.564 \\ 
         Current Venus      & 0.726 & 0.573 & 0.500 & 0.484 \\
    \end{tabular}
    \label{table:geometricalbedo}
\end{table}

%--------------------------------------------------------------------
%--------------------------------------------------------------------

For each model planet, the total flux $F_{\rm p}$ decreases with 
increasing $\alpha$
mostly because less of the planet's observable disc is illuminated. 
The planet with thin H$_2$O clouds (Phase 2) is 
very dark over all $\alpha$'s because the cloud optical thickness 
$b^{\rm c}$ is small and the surface is black. 
The total fluxes show vague 
similarities with the single scattering phase functions of the
cloud particles (Fig.~\ref{fig:singlescattering}).
In particular, for the 'Current Earth' (Phase 1) with large H$_2$O 
particles, $F_{\rm p}$ increases slightly around $\alpha=40^\circ$,
the rainbow angle.
Also, the decrease of $F_{\rm p}$ with $\lambda$ is stronger
for the Venus-type planets with H$_2$SO$_4$ clouds (Phases 3 and 4) 
than for the planet with the large H$_2$O particles 
('Current Earth', Phase 1),
because the single scattering phase function of the sulfuric 
acid particles decreases stronger with $\lambda$ than that of 
the water droplets (see Fig.~\ref{fig:singlescattering}). 

Unlike $F_{\rm p}$, the degree of polarization $P_{\rm p}$ of 
each of the model planets, shows angular and spectral features that 
depend strongly on the cloud properties and should 
thus allow distinguishing between the different evolutionary
phases. In Phase~1 ('Current Earth'), $P_{\rm p}$ is high and
positive up till $\lambda=0.5~\mu$m and around $\alpha=90^{\circ}$, 
which is due to Rayleigh scattering by the gas above the 
clouds. Starting at the shortest $\lambda$, $P_{\rm p}$ 
increases slightly with $\lambda$ before decreasing. 
This is due to the slightly larger
contribution of multiple scattered light, with a lower degree of
polarization, at the shortest wavelengths.
A Rayleigh-scattering peak is also seen for Phase~4 
('Current Venus'), except there, the peak decreases more rapidly 
with $\lambda$ because the clouds are higher in the 
atmosphere and there is thus less gas above them.
In Phase 3 ('Thick clouds Venus'), the Rayleigh-scattering
peak is suppressed by the contribution of low polarized 
light that is reflected by the thicker clouds below the gas. 
In Phase 2 ('Thin clouds Venus'), the relatively thin clouds are 
higher in the atmosphere
than in Phase 1 ('Current Earth'), which is why the 
Rayleigh-scattering peak only occurs at the very shortest wavelengths 
(the peak is hardly visible in Fig.~\ref{fig:phasecontour}).
Because the Phase~2 cloud particles are small 
($r_{\rm eff}=0.5~\mu$m), they themselves give rise to a 
Rayleigh-scattering peak at $\lambda \geq 1.0~\mu$m.

The two model planets with the H$_2$O cloud particles 
(Phases~1 and~2) show a narrow
region of positive polarization between 30$^\circ$ and 40$^\circ$,
which is the rainbow peak 
(see Fig.~\ref{fig:singlescattering}).
On exoplanets, this local maximum in $P_{\rm p}$ 
could be used to detect liquid water clouds on exoplanets 
\citep{karalidi2011flux,karalidi2012looking,2007AsBio...7..320B}.
In Phase~1 ('Current Earth'), the rainbow
region starts near the Rayleigh scattering peak of the gas 
and extends towards the largest wavelengths.
In Phase~2 ('Thin clouds Venus'),
with the small water droplets, the rainbow only occurs at the
shortest wavelengths. With increasing wavelength, it
broadens and disappears into the cloud particles' 
Rayleigh scattering peak. 

The H$_2$SO$_4$ cloud particles (Phases 3 and 4) have their own 
specific polarization patterns, such as the broad negative polarization 
region at $\alpha \gtrapprox 80^\circ$,
which can be traced back to their single scattering patterns
(Fig.~\ref{fig:singlescattering}).
In Phase~3 ('Thick clouds Venus'), the cloud particles
give rise to a sharp positive polarization peak at the shortest
wavelengths and for 20$^\circ \leq \alpha \leq 30^\circ$.
In Phase~4 ('Current Venus'), there is a broader, lower, 
positive polarization branch across this phase angle range, which 
resembles the positive polarization branch of the tiny H$_2$O droplets 
in Phase~2 ('Thin clouds Venus'). However, at the longer wavelengths,
the phase angle dependence
of the polarization of the latter planet is very different
which should help to distinguish between such planets.
This emphasizes the need for measurements at a wide range 
of wavelengths and especially phase angles (if the planet's orbital
inclination angle allows this).

%--------------------------------------------------------------------
\subsection{Flux and polarization of spatially unresolved planets}
\label{sect_unresolved}

%--------------------------------------------------------------------
% Figure system sketch
%--------------------------------------------------------------------
\begin{figure*}[!t]
\centering
\includegraphics[width=0.8\textwidth]{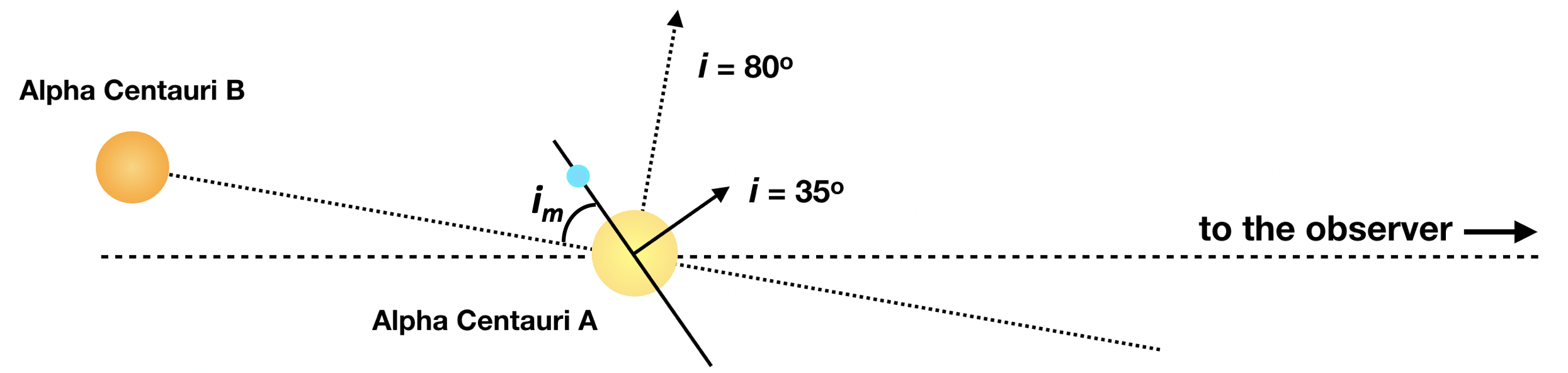}
\caption{A sketch of the geometries within the Alpha Centauri system:
         the orbital plane of the stars Alpha Centauri A and B is
         inclined by about 80$^\circ$ with respect to the 
         observer on Earth. 
         Our model planet (the blue dot) orbits Alpha Centauri~A. 
         In this sketch, the line of nodes of the planet's orbit
         was chosen to coincide with that of the stellar orbits. 
         The inclination angle $i_{\rm m}$ of the planet's 
         orbit with respect to the stellar orbital plane is 
         45$^\circ$, 
         and the inclination angle of the planet's orbit 
         with respect to the observer
         is 80$^\circ - 45^\circ = 35^\circ$.
         The phase angles of the planet in this sketch
         would range from
         $90^\circ - 35^\circ= 55^\circ$ to 
         $90^\circ + 35^\circ= 125^\circ$.
         }
\label{fig:system_sketch}
\end{figure*}
%--------------------------------------------------------------------

In the previous section, we showed the signals of 
spatially resolved planets, thus without background starlight. 
When observing an exoplanet in the habitable zone of a solar-type
star, it will be difficult to avoid the starlight. 
Here, we show the total flux of the planet $F_{\rm p}$,
the star-planet contrast $C$ (see Eq.~\ref{eq_C}), 
the spatially resolved degree of polarization of the planet 
$P_{\rm p}$ (thus without the starlight) and the 
spatially unresolved degree of polarization of the combined 
star-planet signal $P_{\rm u}$ (thus including the starlight).
While the total planet fluxes shown in Fig.~\ref{fig:phasecontour}
were normalized at $\alpha=0^\circ$ 
to the planets' geometric albedo's $A_{\rm G}$,
here they are computed according to Eq.~\ref{eq_5}, and thus
depend on the parameters of the planet-star system.
We assume our model planets orbit Alpha Centauri A.

The solar-type star Alpha Centauri~A is part of a double star system,
and the orbital parameters of ourplanets are chosen based on 
the stable planet orbital distances and orbital inclination 
angles around this star as predicted by \cite{2016AJ....151..111Q}. 
Figure~\ref{fig:system_sketch} shows a sketch of the system.
We use a planetary orbital distance $d$ of 0.86~AU, such that 
each model planet receives a stellar flux similar to the solar 
flux received by Venus.
Additional system parameter values are listed in Table~\ref{table1}.
According to \cite{2016AJ....151..111Q}, stable orbits around 
Alpha Centauri~A can be found for a range of angles between
the planetary orbital plane and the plane in which the two stars 
orbit, and thus for a range of inclination angles~$i$ 
of the planetary orbit.

%--------------------------------------------------------------------
\begin{figure}[b!]
\centering
\includegraphics[width=0.45\textwidth]{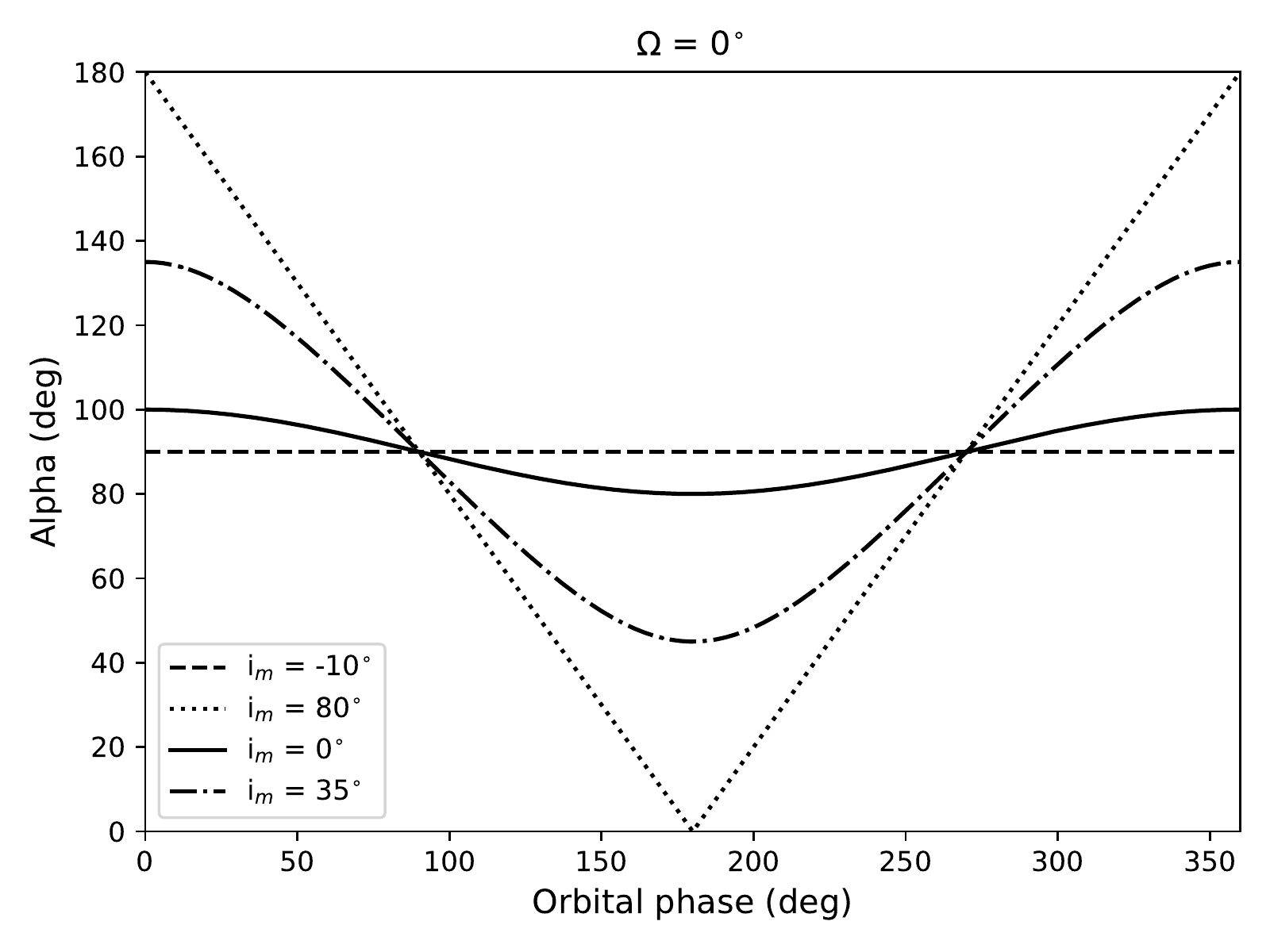}
\includegraphics[width=0.45\textwidth]{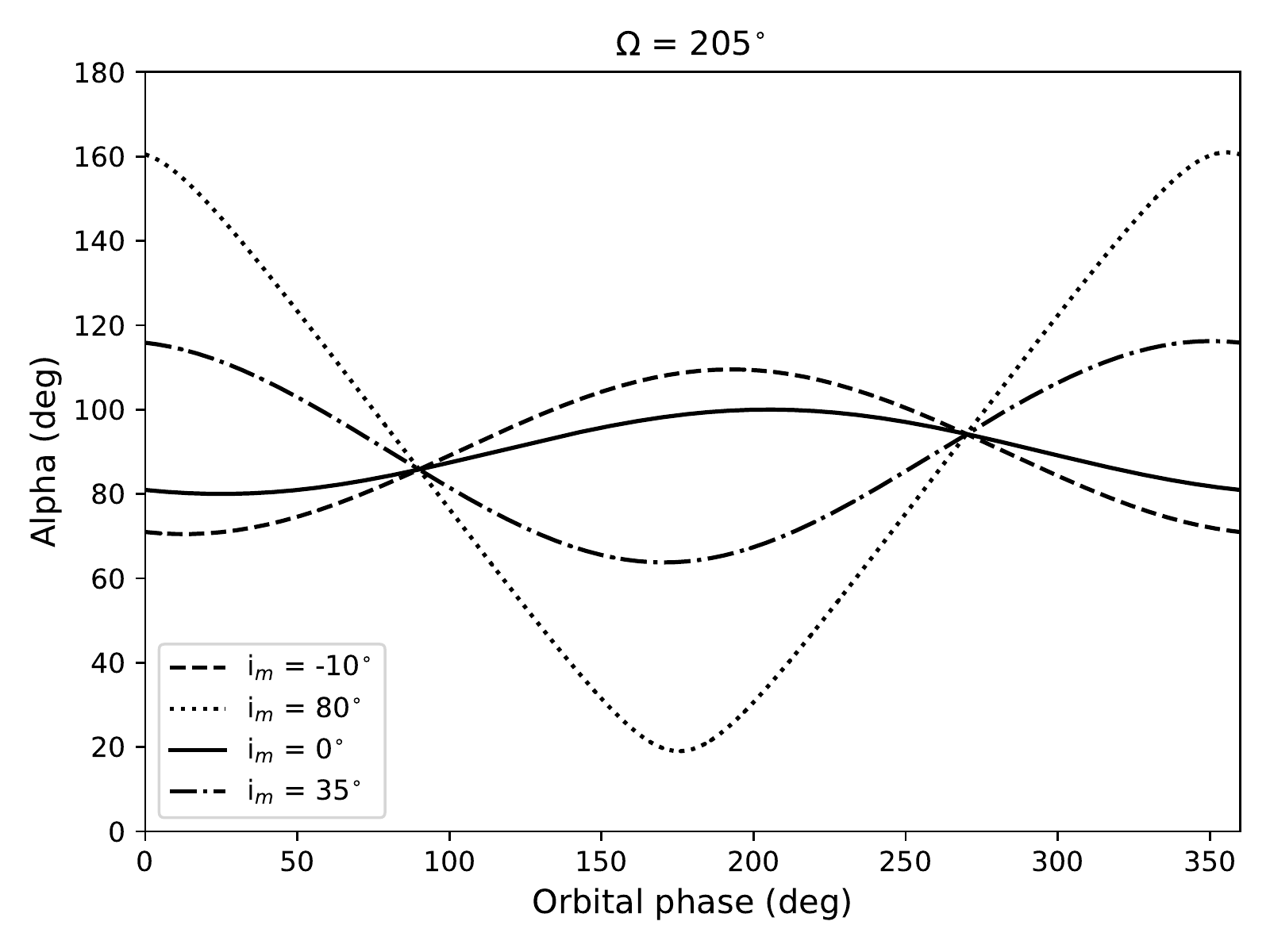}
\caption{Variation of the planet's phase angle $\alpha$ along the
         planet's orbit around Alpha Centauri A for different 
         mutual inclination angles $i_{\rm m}$ of the planetary 
         orbit with respect to the orbital plane of the two stars.
         Top: $\Omega$, the longitude of the ascending node of the
         planet's orbit, is 0$^{\circ}$. 
         For $i_{\rm m}= -10^{\circ}$, the planet 
         is then in a face-on orbit ($i=0^\circ$), while for 
         $i_{\rm m}= 80^{\circ}$, 
         it is in an edge-on orbit ($i=90^\circ$). 
         Bottom: $\Omega=205^\circ$, and the planetary orbit is 
         aligned with the node of the stellar orbital plane.}
\label{fig:AlphaVsOrbitalPhase}
\end{figure}
%--------------------------------------------------------------------

Figure~\ref{fig:AlphaVsOrbitalPhase} shows the variation of 
the planetary phase angle $\alpha$ along a planetary orbit for 
two values of the longitude of the orbit's ascending node $\Omega$:
for $\Omega= 0^{\circ}$ (the line connecting the 
planet's ascending and descending nodes is perpendicular to the
line to the observer) and for $\Omega= 205^{\circ}$ which 
represents the configuration of Earth with Alpha Centauri A.
The orbital phase of the planet is defined such that at an orbital 
phase angle of 0$^\circ$, $\alpha=180^{\circ}$. 
The inclination angle $i_{\rm m}$ is 
the angle between the plane in which the stars move
and the planetary orbital plane. 
For $\Omega= 0^{\circ}$, $i_{\rm m}=-10^{\circ}$ 
would yield a face-on planetary orbit ($i=0^\circ$)
with $\alpha=90^\circ$ everywhere along the orbit. 
For $i_{\rm m}= 80^{\circ}$, the orbit is edge-on 
($i=90^\circ$) and $\alpha$ varies between 0$^{\circ}$ and
180$^{\circ}$. 
Figure~\ref{fig:AlphaVsOrbitalPhase} also shows the range of 
$\alpha$ for $i_{\rm m}=0^\circ$ and 35$^\circ$. 
According to \cite{2016AJ....151..111Q}, the latter is the 
most probable orientation of a stable planetary orbit.
For these two cases, the accessible 
phase angles range from 80$^{\circ}$ to 100$^{\circ}$, and 
from 45$^{\circ}$ to 135$^{\circ}$, respectively. 
For $\Omega=205^\circ$, the maximum range of $\alpha$ would 
be from 20$^\circ$ to 160$^\circ$, depending on $i_{\rm m}$.

%--------------------------------------------------------------------
\begin{figure*}[!]
\centering
\includegraphics[width=0.32\textwidth]{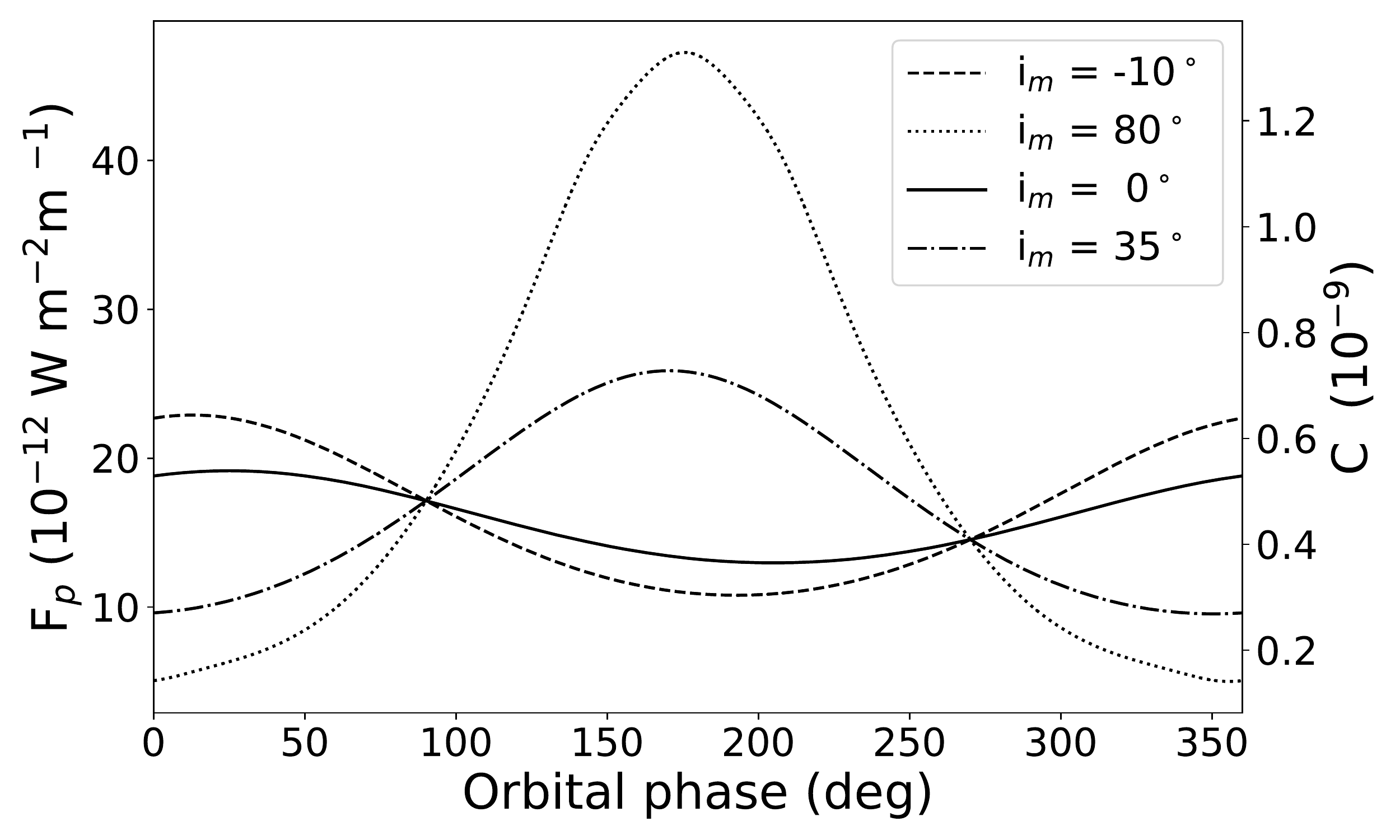}
\includegraphics[width=0.32\textwidth]{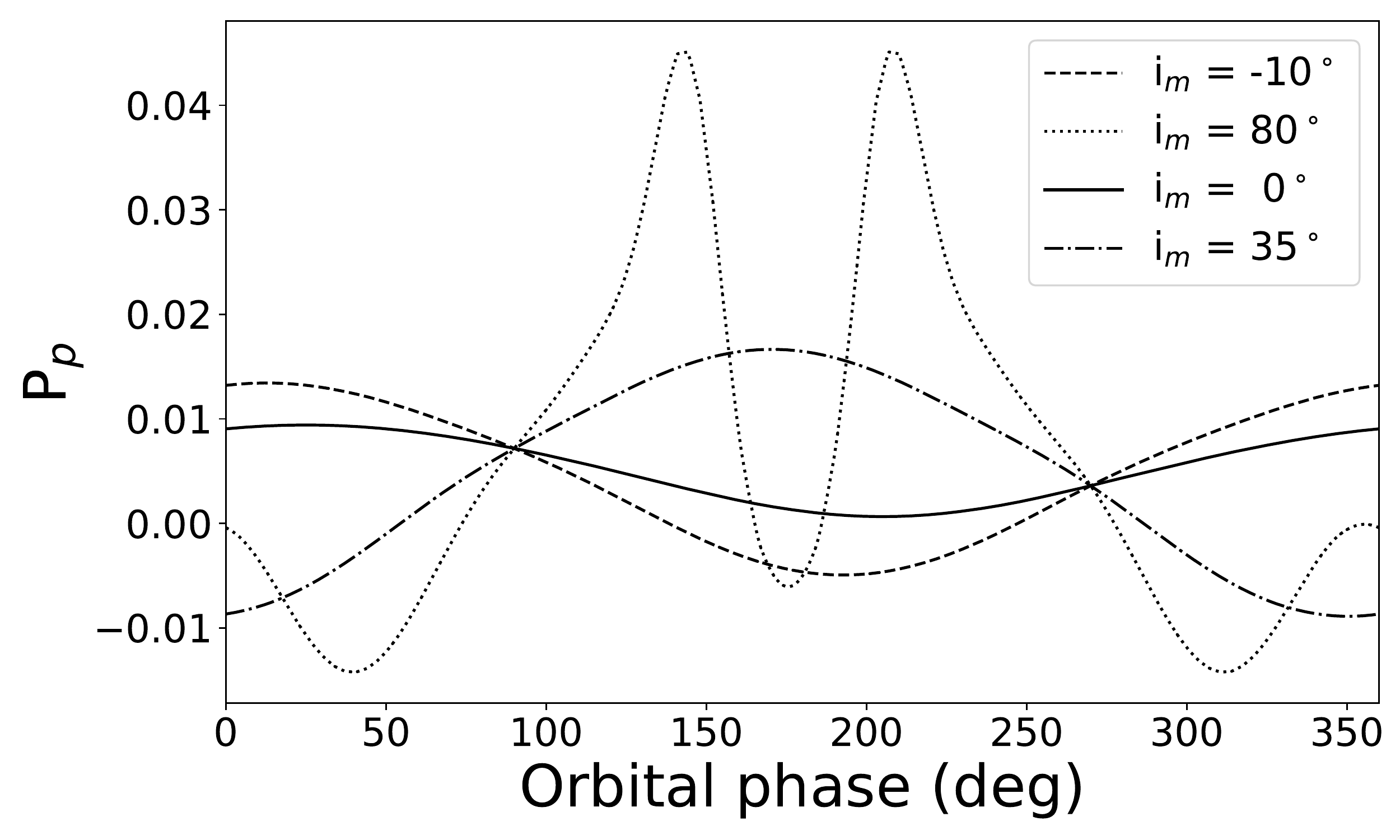}
\includegraphics[width=0.32\textwidth]{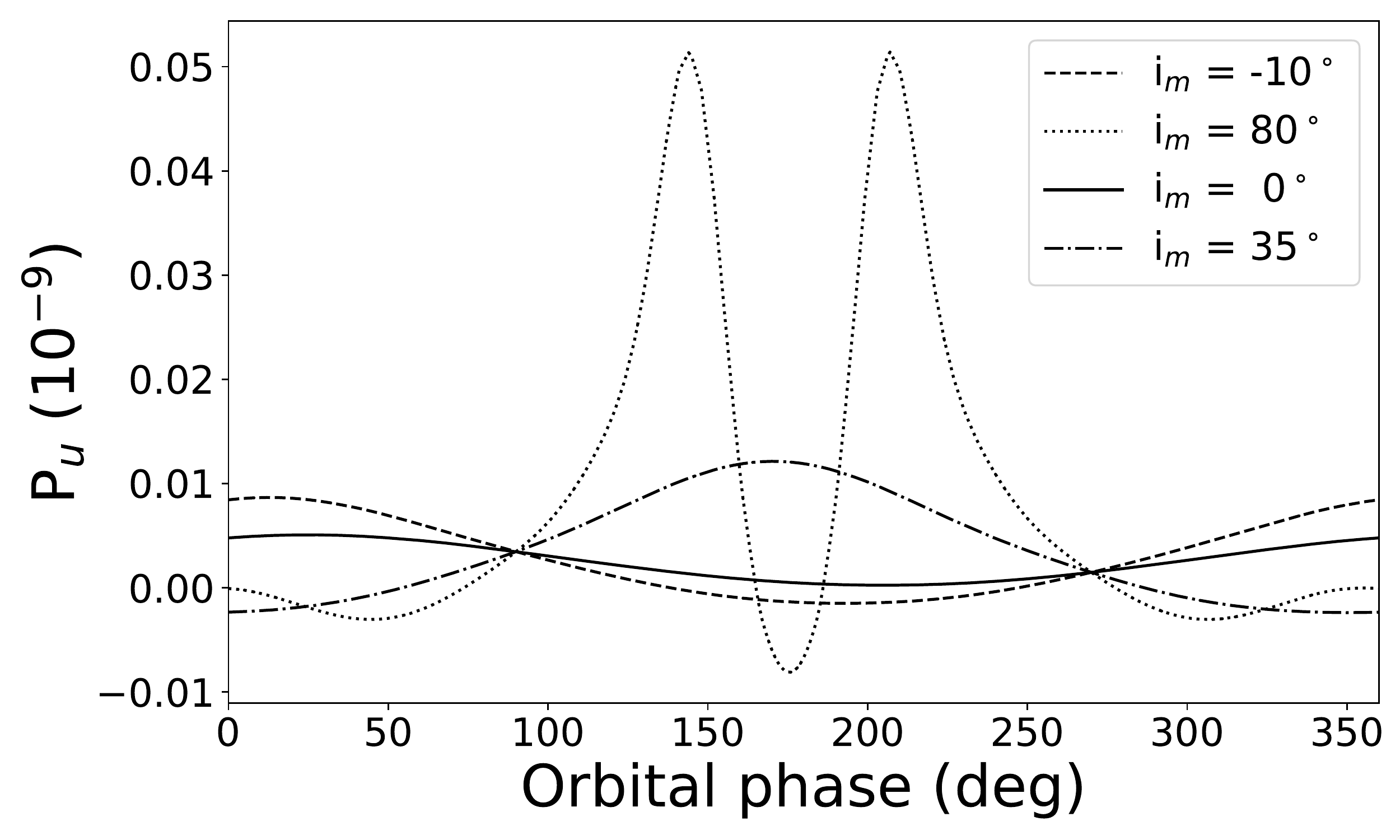}\\

\includegraphics[width=0.32\textwidth]{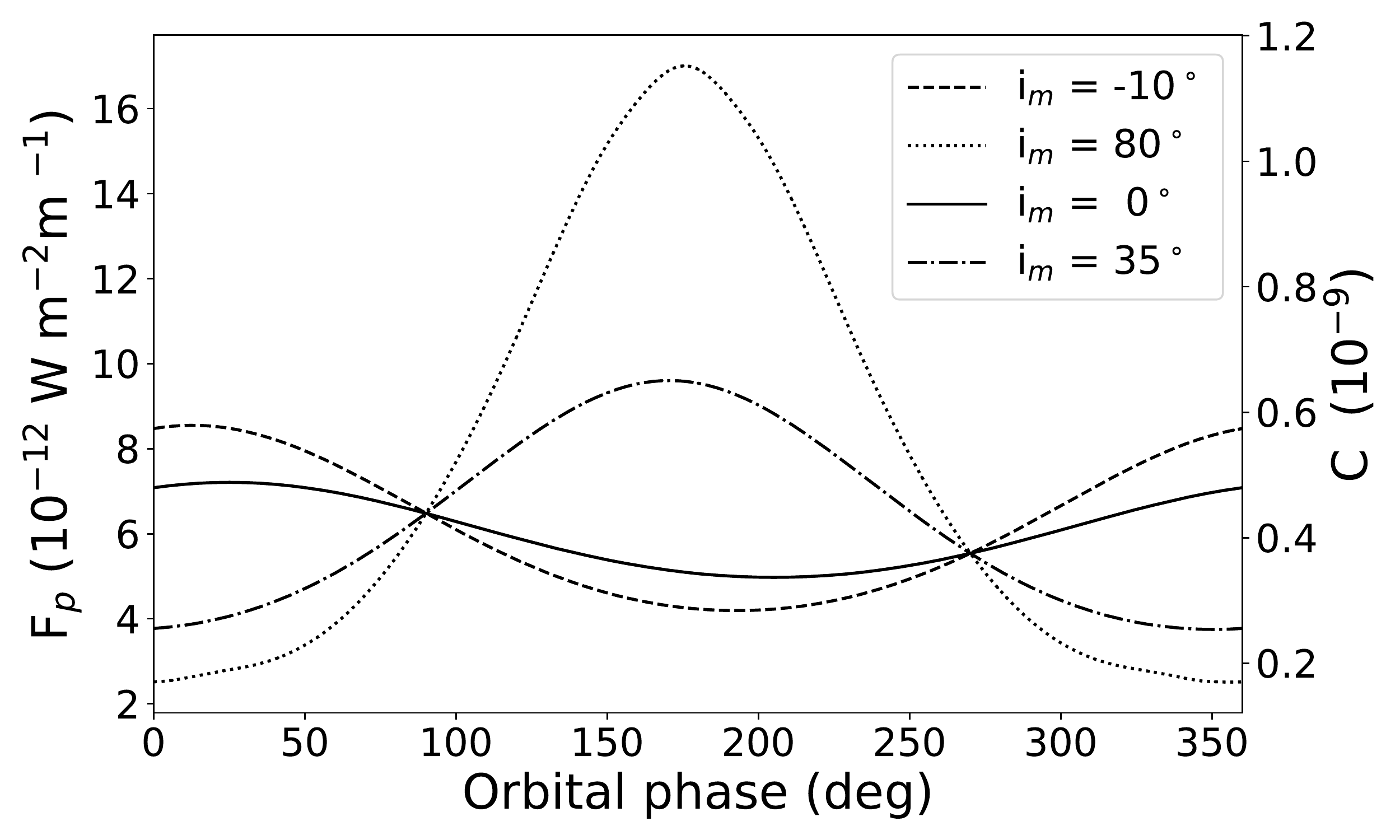}
\includegraphics[width=0.32\textwidth]{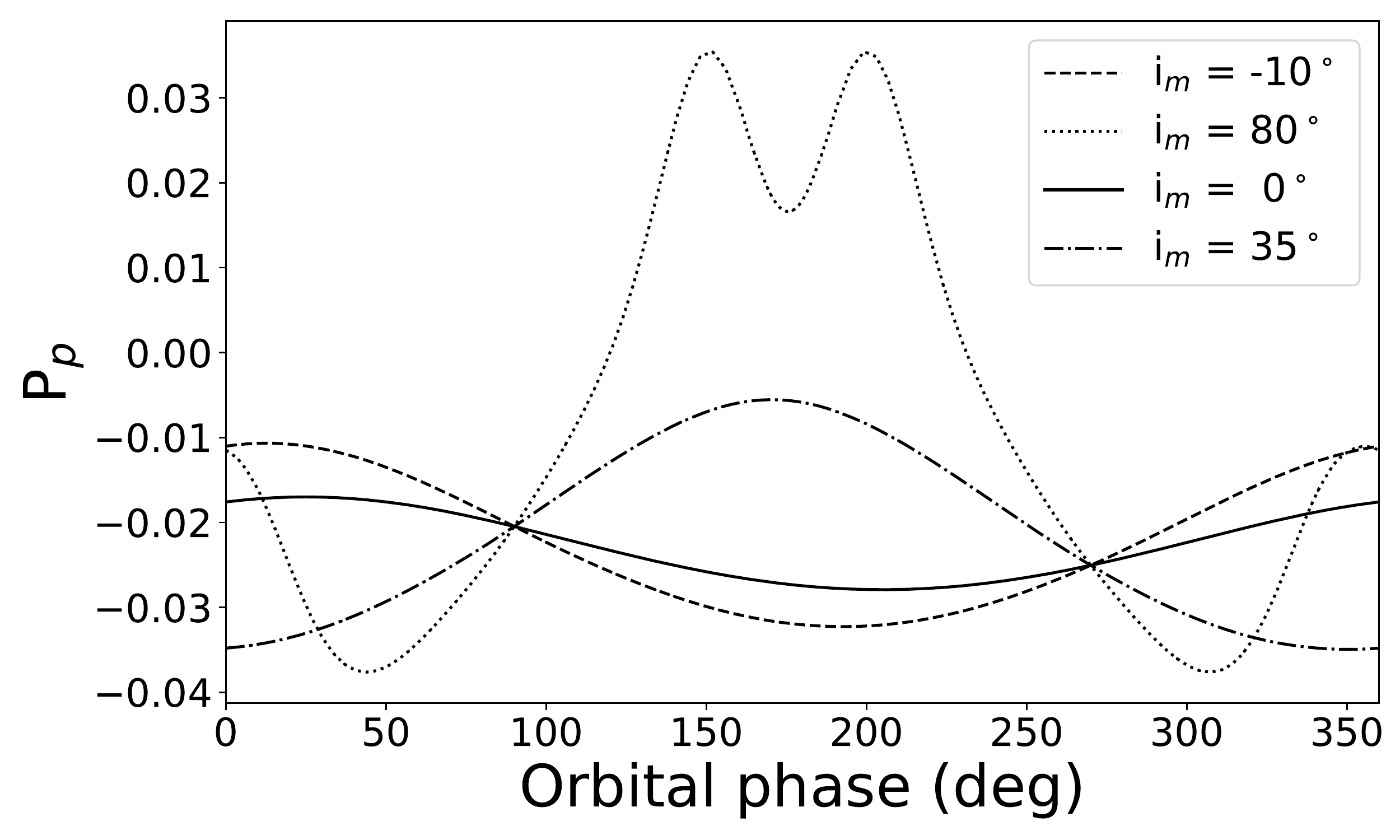}
\includegraphics[width=0.32\textwidth]{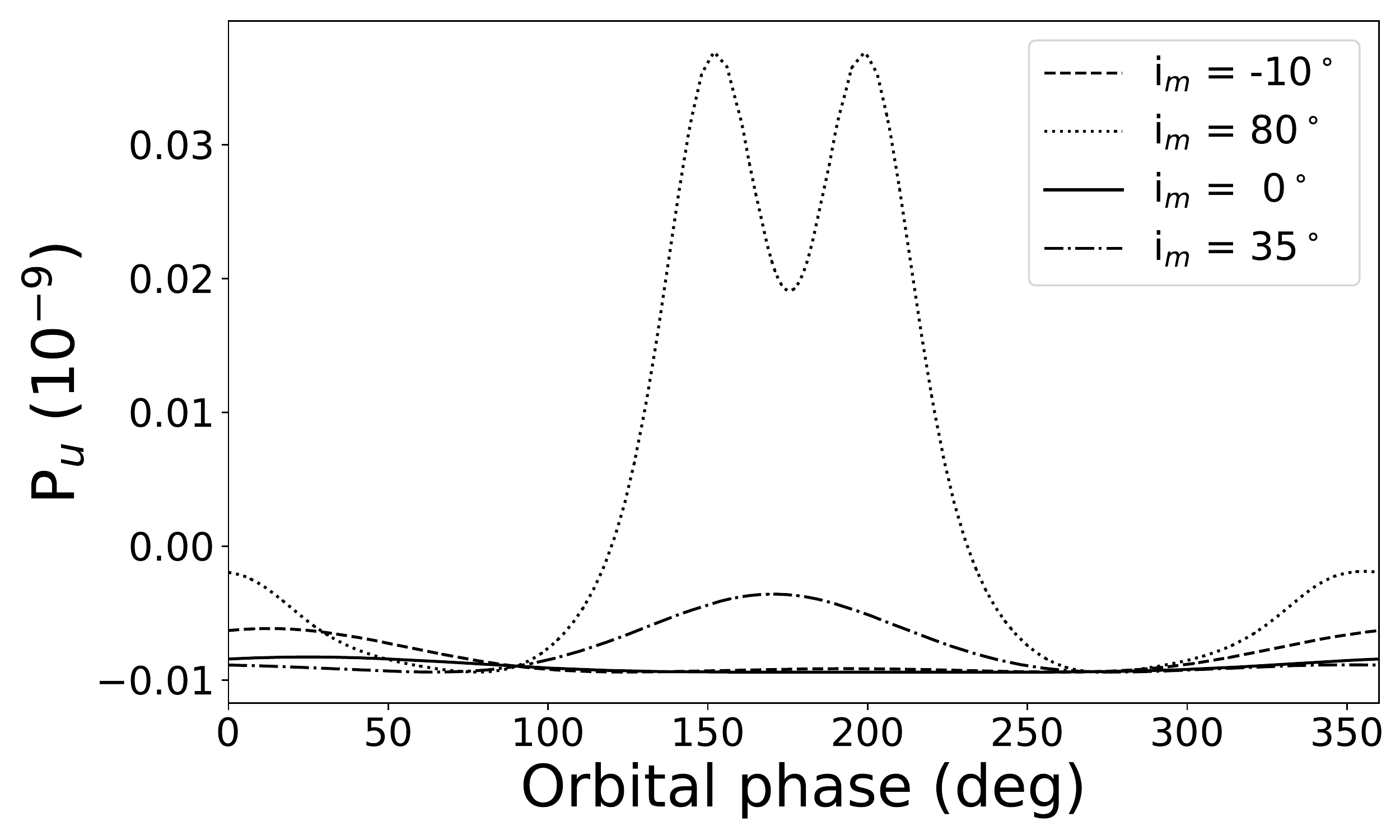}\\

\includegraphics[width=0.32\textwidth]{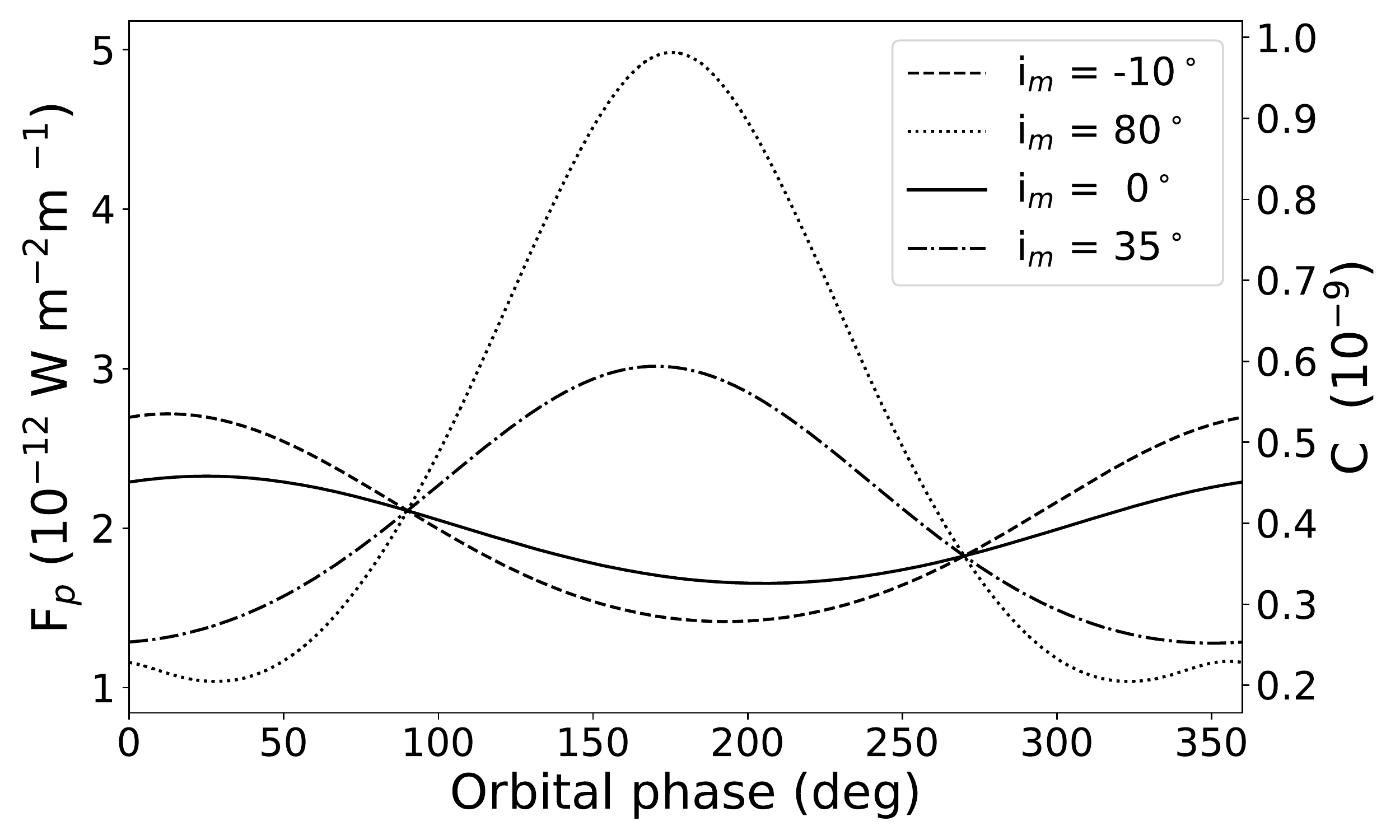}
\includegraphics[width=0.32\textwidth]{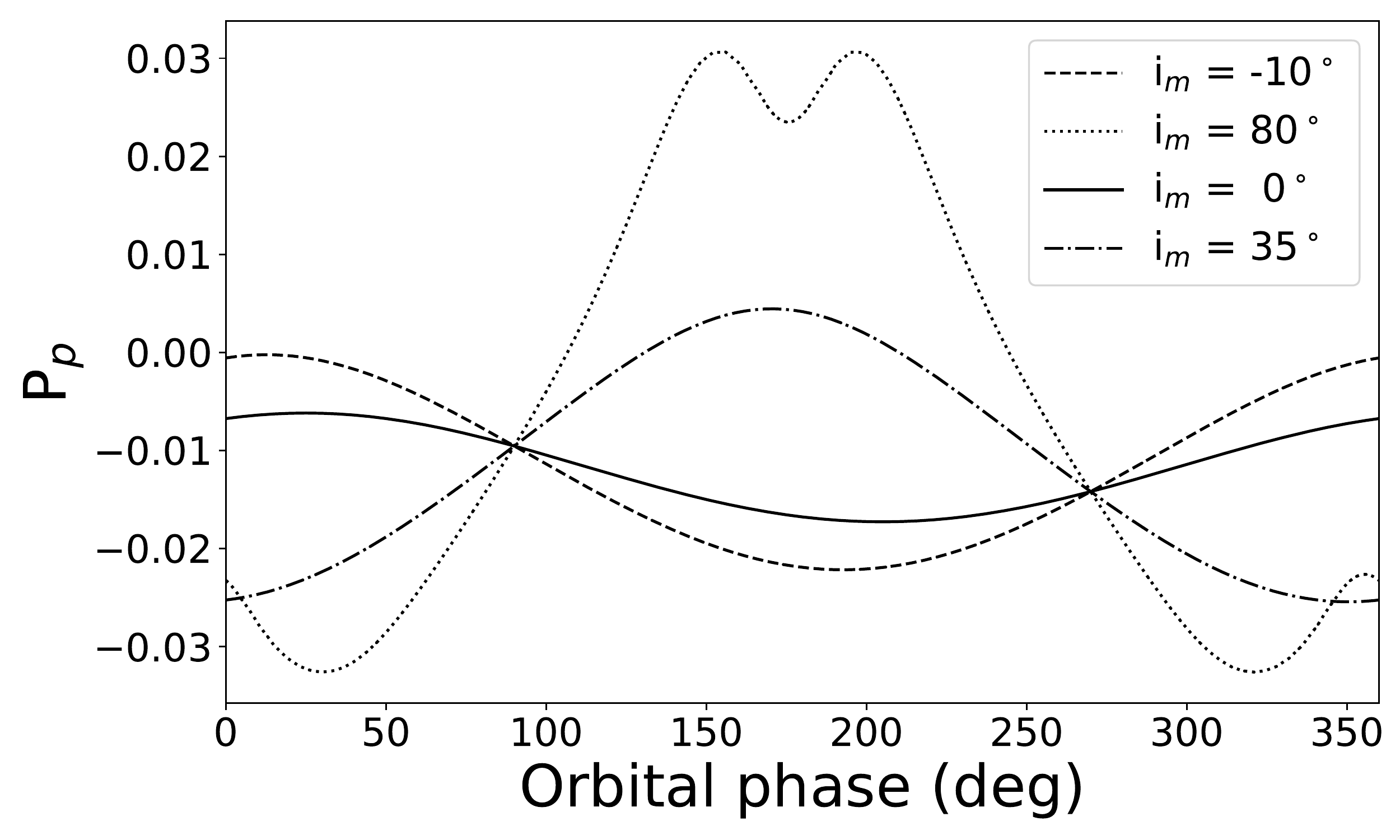}
\includegraphics[width=0.32\textwidth]{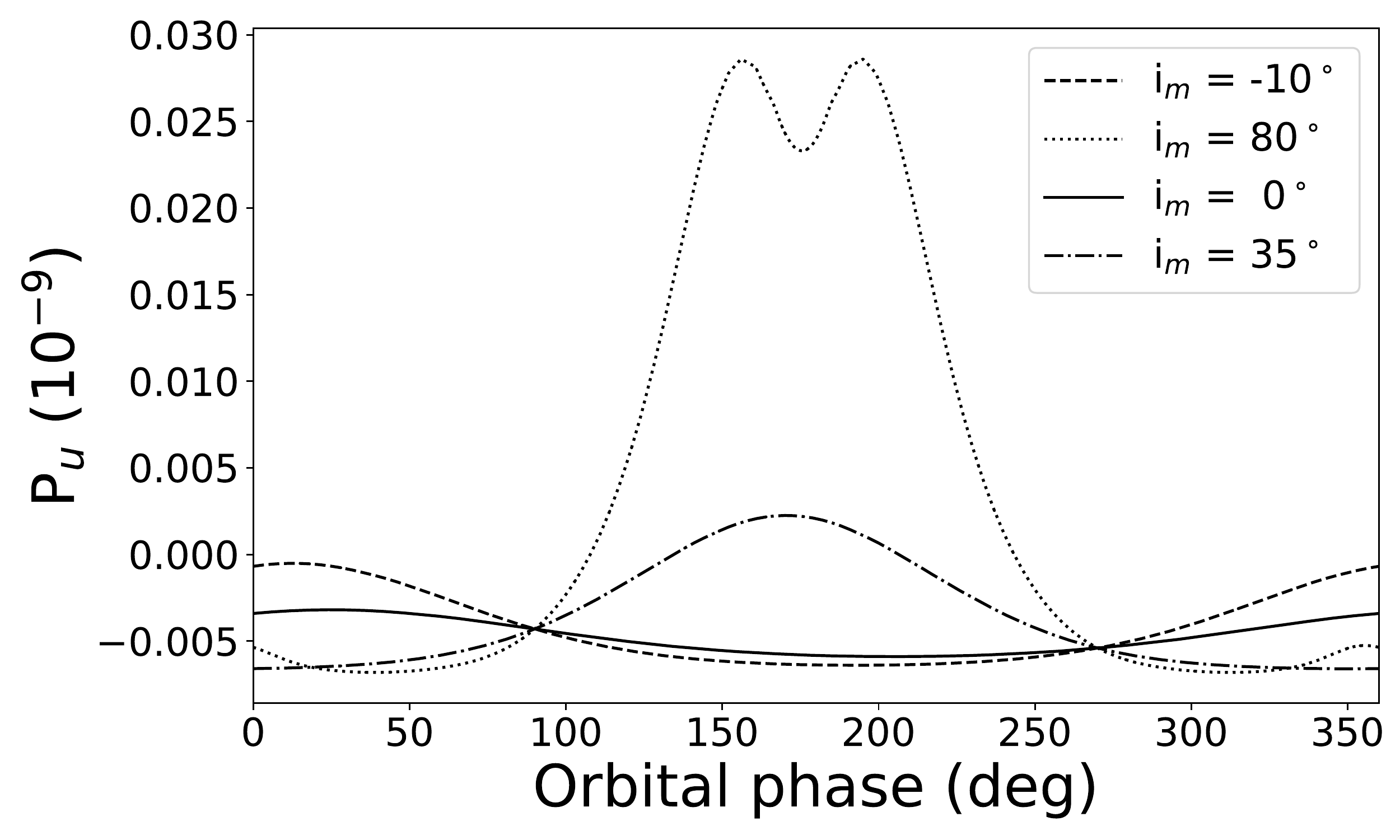}\\

\includegraphics[width=0.32\textwidth]{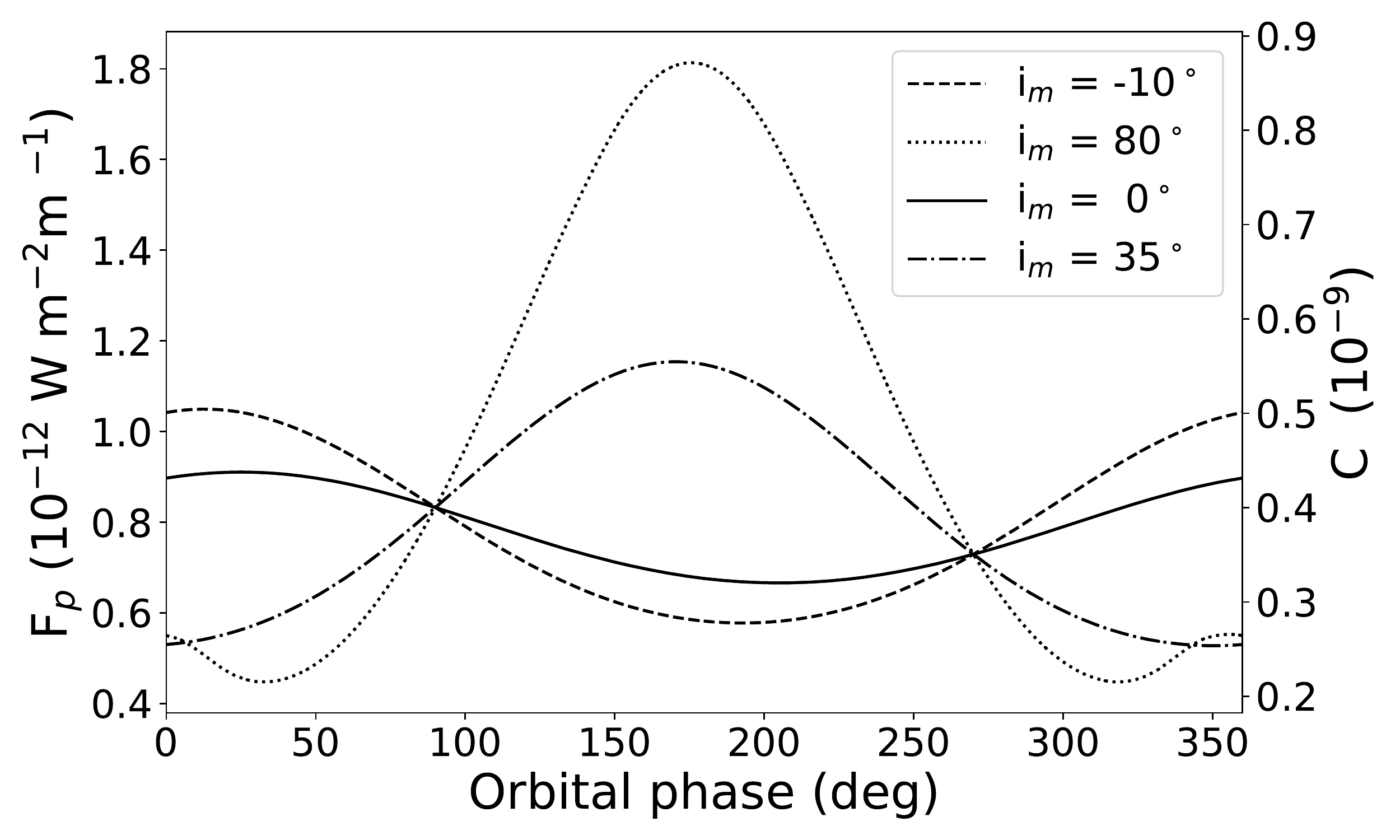}
\includegraphics[width=0.32\textwidth]{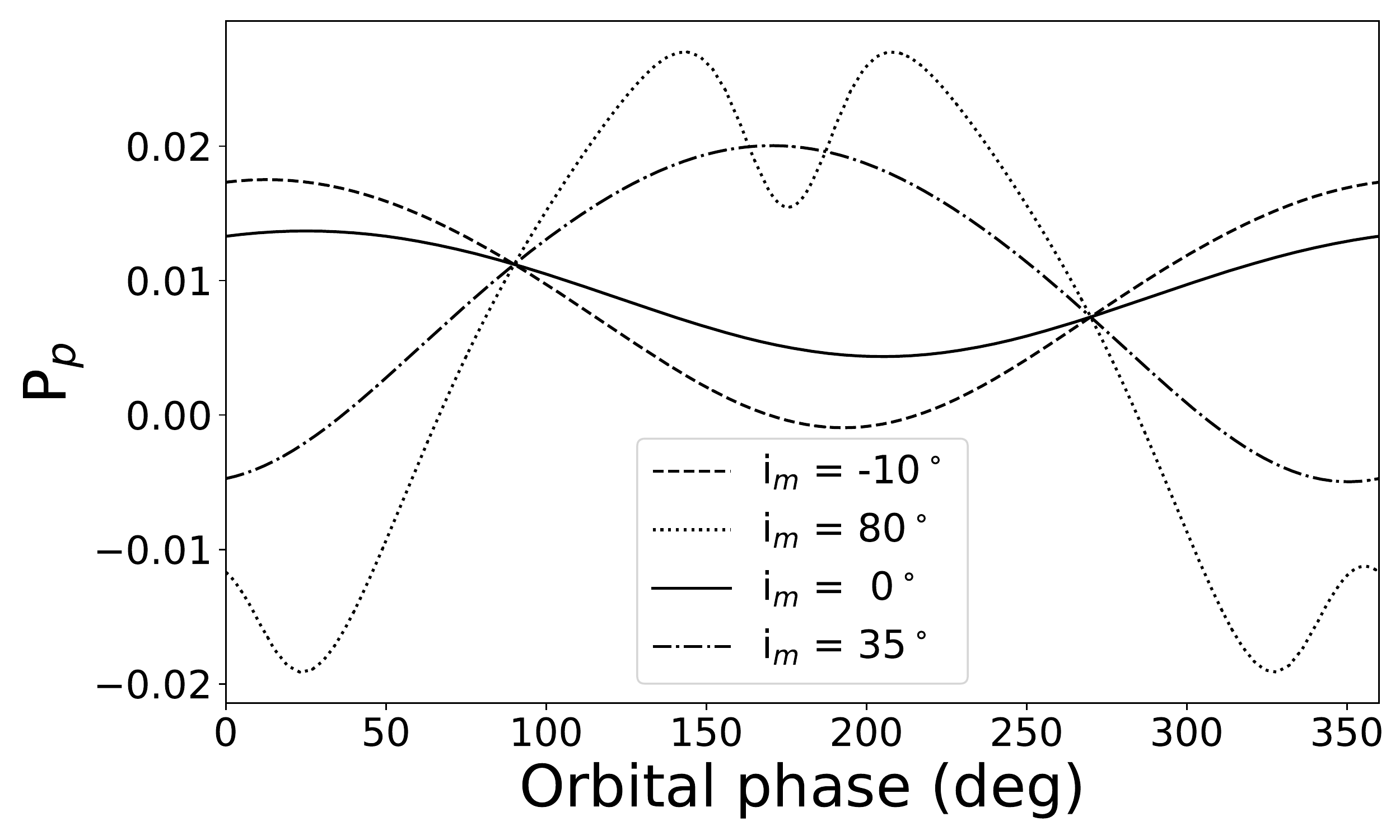}
\includegraphics[width=0.32\textwidth]{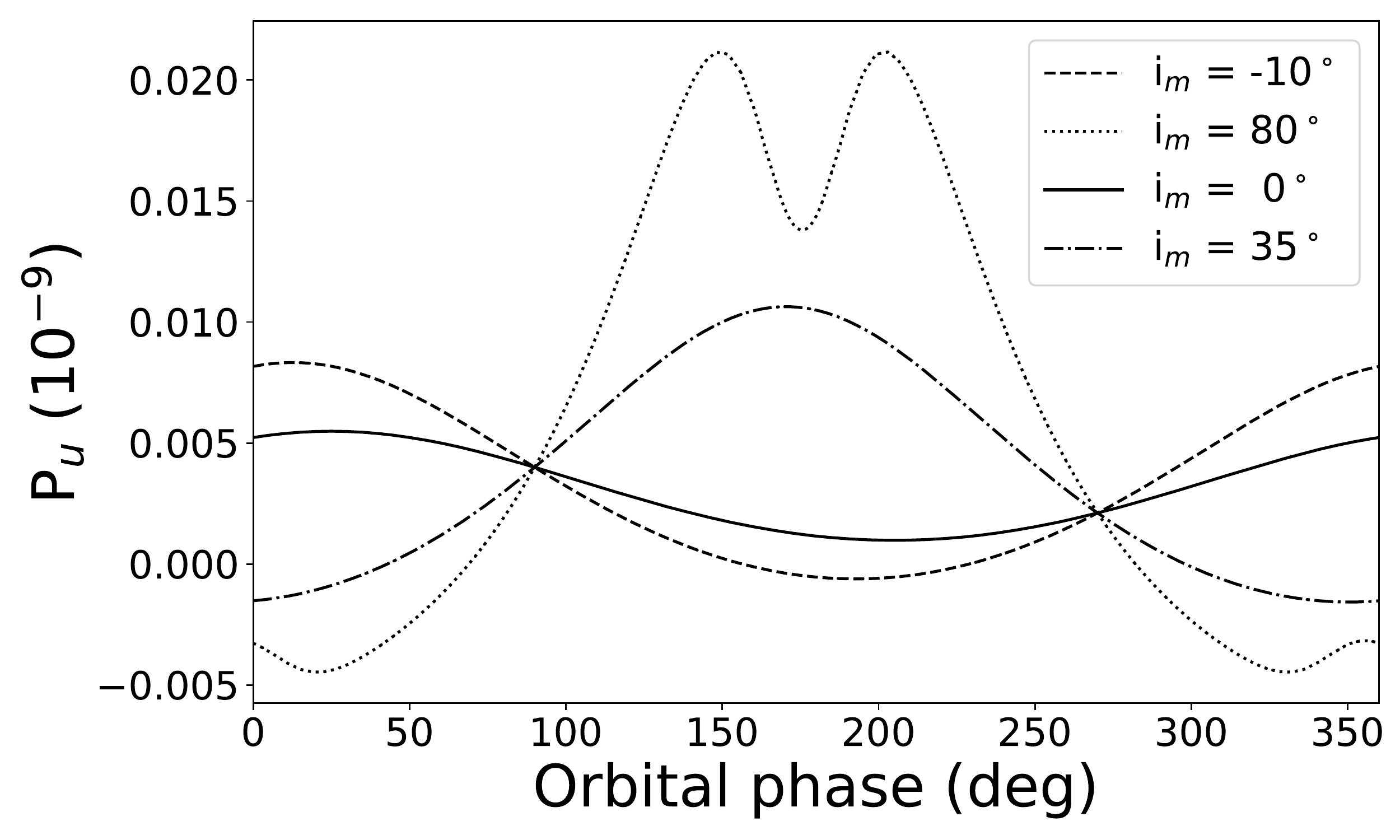}\\
\caption{The total planetary flux $F_{\rm p}$ (in W m$^{-3}$) 
         and the star-planet contrast $C$, the degree of 
         polarization $P_{\rm p}$ of the spatially resolved planet, 
         and $P_{\rm u}$, the degree of polarization of the star and 
         the spatially unresolved planet, all for a 'Current Venus' 
         model planet (Phase 4)
         as functions of the planet's orbital phase for four 
         mutual inclination angles $i_{\rm m}$ and four 
         wavelengths $\lambda$:
         $0.5~\mu$m (row 1); $1.0~\mu$m (row 2);
         $1.5~\mu$m (row 3); and $2.0~\mu$m (row 4).
         The longitude of the ascending node of the planetary orbit,
         $\Omega$, is 205$^{\circ}$.}
\label{fig:orbitfluxes1}
\end{figure*}
%--------------------------------------------------------------------

Figure~\ref{fig:orbitfluxes1} shows $F_{\rm p}$, $P_{\rm p}$ 
and $P_{\rm u}$ (the spatially unresolved planet, thus with 
starlight included) 
for Phase~4 ('Current Venus') as functions of the planet's orbital 
phase for $\Omega=205^\circ$, four values of $i_{\rm m}$ 
(-10$^\circ$, 80$^\circ$, 0$^\circ$, and 35$^\circ$),
and four wavelengths (0.5, 1.0, 1.5, and 2.0~$\mu$m). 
The plots for $F_{\rm p}$ also show the contrast $C$.
Because $C$ is the ratio of the planetary flux $F_{\rm p}$ 
to the stellar flux $F_{\rm s}$ (Eq.~\ref{eq_C}), 
its variation with the orbital phase is proportional to 
that of $F_{\rm p}$.

At the two orbital phases in each plot
where all the lines cross, the planetary phase angles $\alpha$ 
are the same (see Fig.~\ref{fig:AlphaVsOrbitalPhase})
and thus all $F_{\rm p}$ and $P_{\rm p}$ are the same. 
The plots for $F_{\rm p}$ appear to be very similar for the 
different wavelengths, apart from a difference in magnitude
which is mainly due to the decrease of the stellar flux that is 
incident on the planet with increasing wavelength, 
although the planetary albedo $A_{\rm G}$ 
and phase function $R_{\rm 1p}$ also decrease with increasing
$\lambda$ as can be seen in Fig.~\ref{fig:phasecontour}.
This wavelength dependence of $F_{\rm p}$ also
causes the decrease of the contrast $C$ with increasing wavelength
(i.e.\ the planet darkens with increasing $\lambda$), as
$C$ is independent of the wavelength dependence
of the stellar flux (see Eq.~\ref{eq_C}).
The shape differences between the $F_{\rm p}$ (and $C$) 
curves are due to the wavelength dependence of the planetary 
flux that can also be seen in Fig.~\ref{fig:phasecontour}.

At each wavelength $\lambda$, the largest variation 
in $F_{\rm p}$ with the orbital phase is
seen for $i_{\rm m}= 80^{\circ}$, because for that configuration 
the variation of $\alpha$ along the orbit is largest
(see Fig.~\ref{fig:AlphaVsOrbitalPhase}). 
The degree of polarization $P_{\rm p}$ of the planet shows 
significant variation with the orbital 
phase at all wavelengths. A particularly striking feature 
for the geometry with $i_{\rm m}=80^\circ$ is the double peak 
close to the orbital phases of 150$^\circ$ and 200$^\circ$.
As can be seen in Fig.~\ref{fig:AlphaVsOrbitalPhase} 
for $\Omega=205^\circ$ and $i_{\rm m}=80^\circ$, 
$\alpha$ decreases from about 160$^\circ$ at an orbital phase of 
0$^\circ$, to 20$^\circ$ at an orbital phase around 175$^\circ$, 
to then increase again with increasing orbital phase.
Tracing this path of $\alpha$ through the $P_{\rm p}$ panel in the 
bottom row of Fig.~\ref{fig:phasecontour} explains 
the double peaked behaviour of $P_{\rm p}$ and its wavelength 
dependence as shown in Fig.~\ref{fig:orbitfluxes1}.
For the other values of $i_{\rm m}$, the phase angle range that is
covered along the orbit is smaller, and therefore the variation
in $P_{\rm p}$ is also smaller. 

The degree of polarization of the spatially unresolved planet,
$P_{\rm u}$, shows similar variations along the orbital phase
as $P_{\rm p}$, except that most features are flattened out
because of the addition of the unpolarized stellar flux, 
which is independent of the orbital phase angle.
The double peaked feature for $i_{\rm m}=80^\circ$ remains
strong, however, as at those orbital phase angles, the 
contrast $C$ is relatively large and thus the influence of 
the added stellar flux relatively small.
The variation in the polarisation of the unresolved system
due to the orbiting planet is on the order of 10$^{-11}$.

%-----------------------------------------------------------------------
\begin{figure*}[!]
    \centering
    \includegraphics[width=0.32\textwidth]{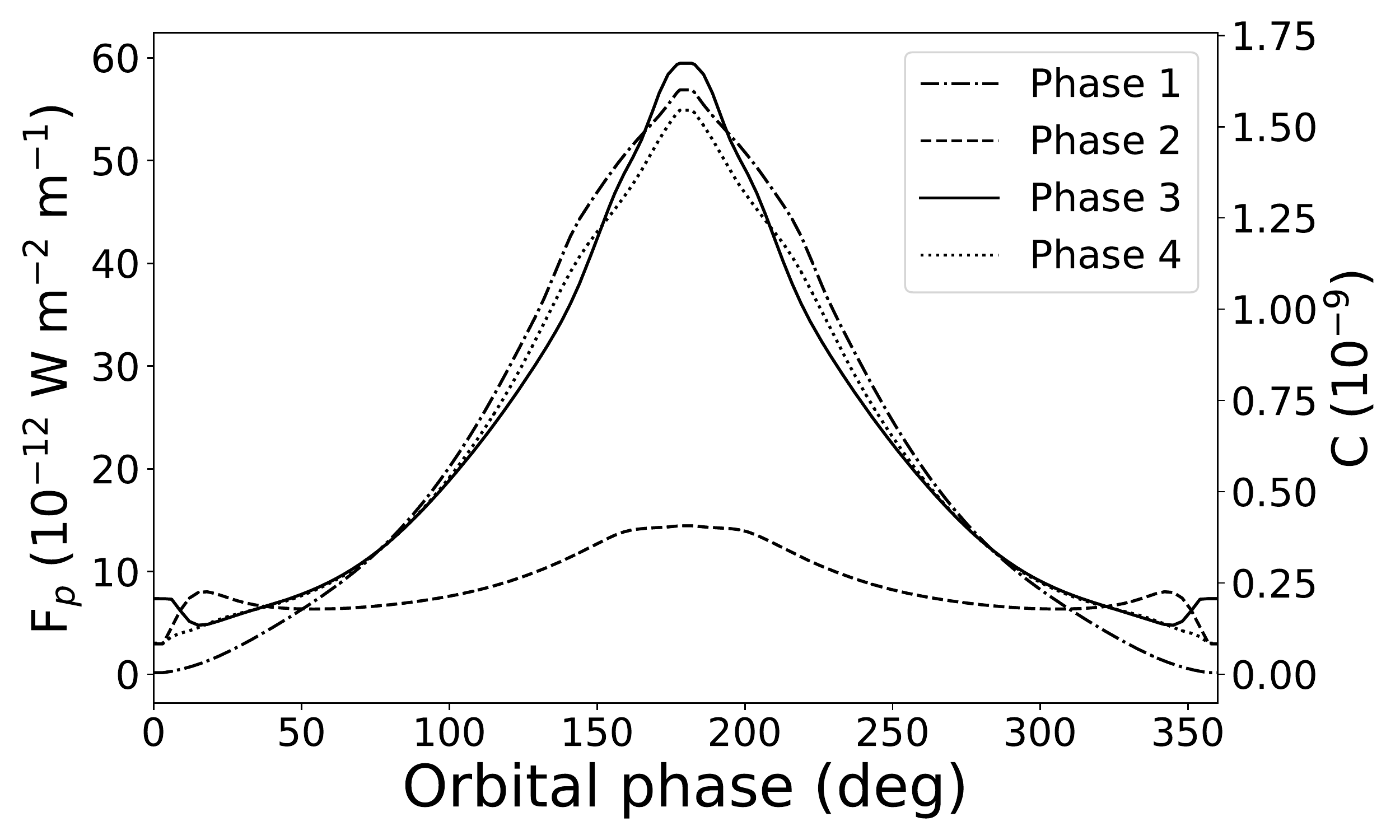}
    \includegraphics[width=0.32\textwidth]{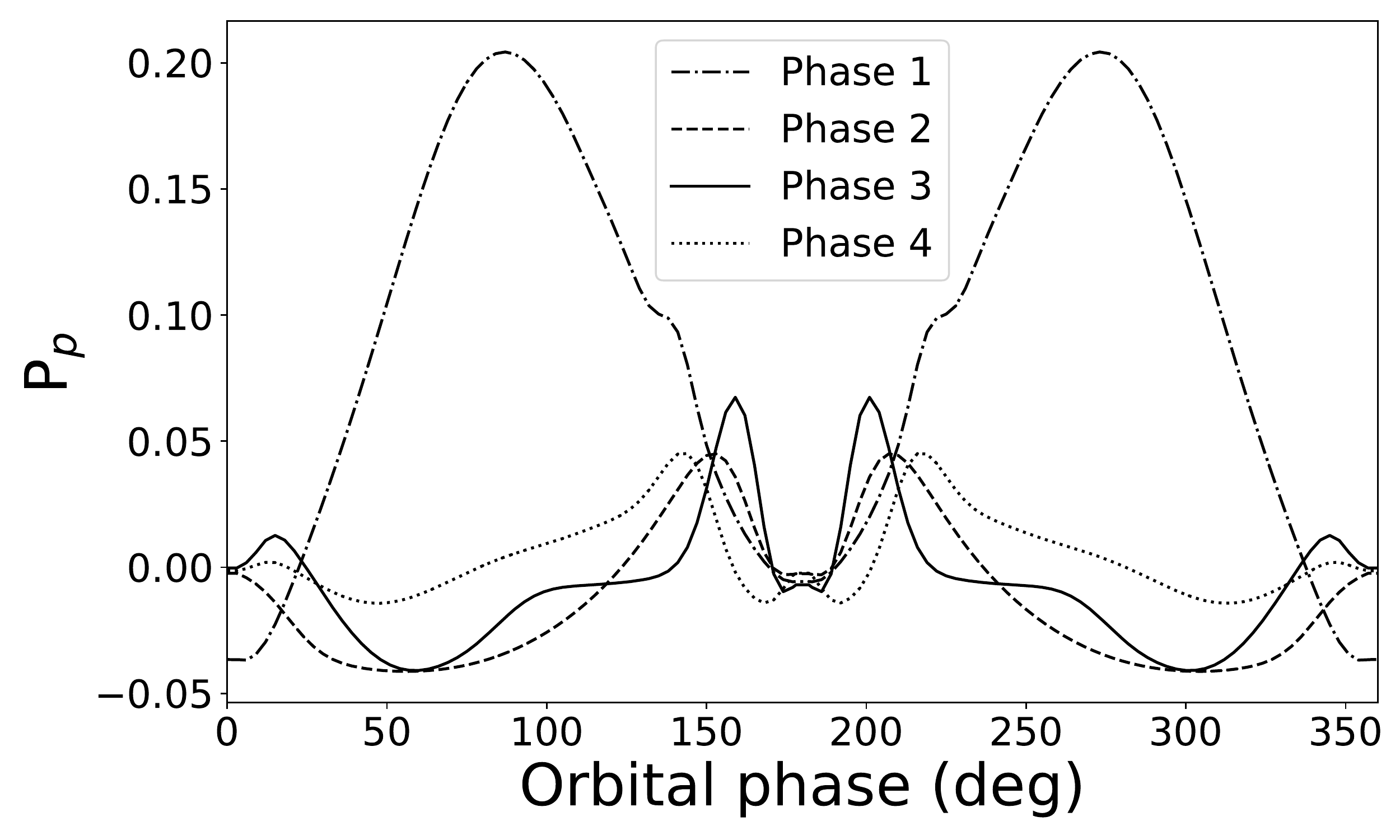}
    \includegraphics[width=0.32\textwidth]{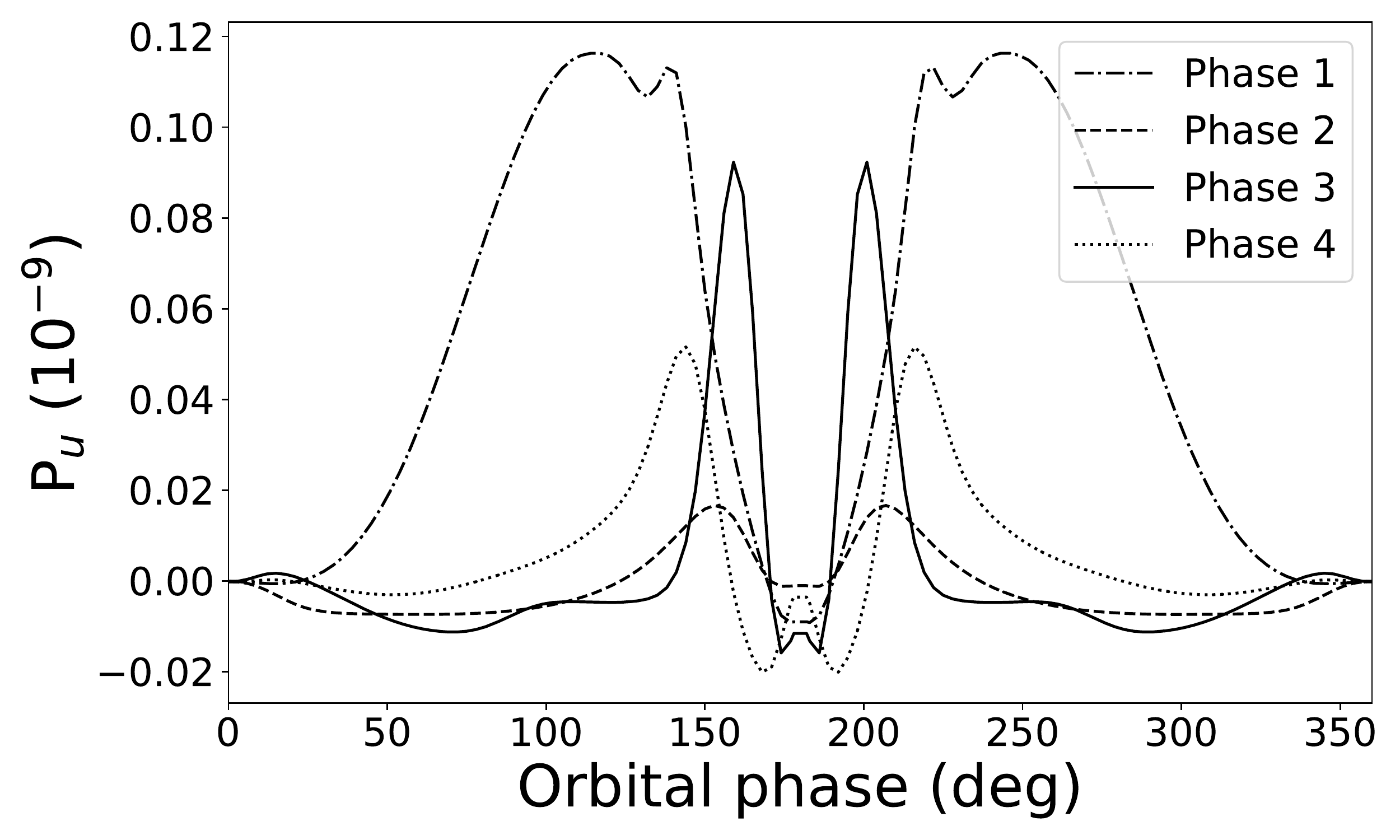}\\
    
    \includegraphics[width=0.32\textwidth]{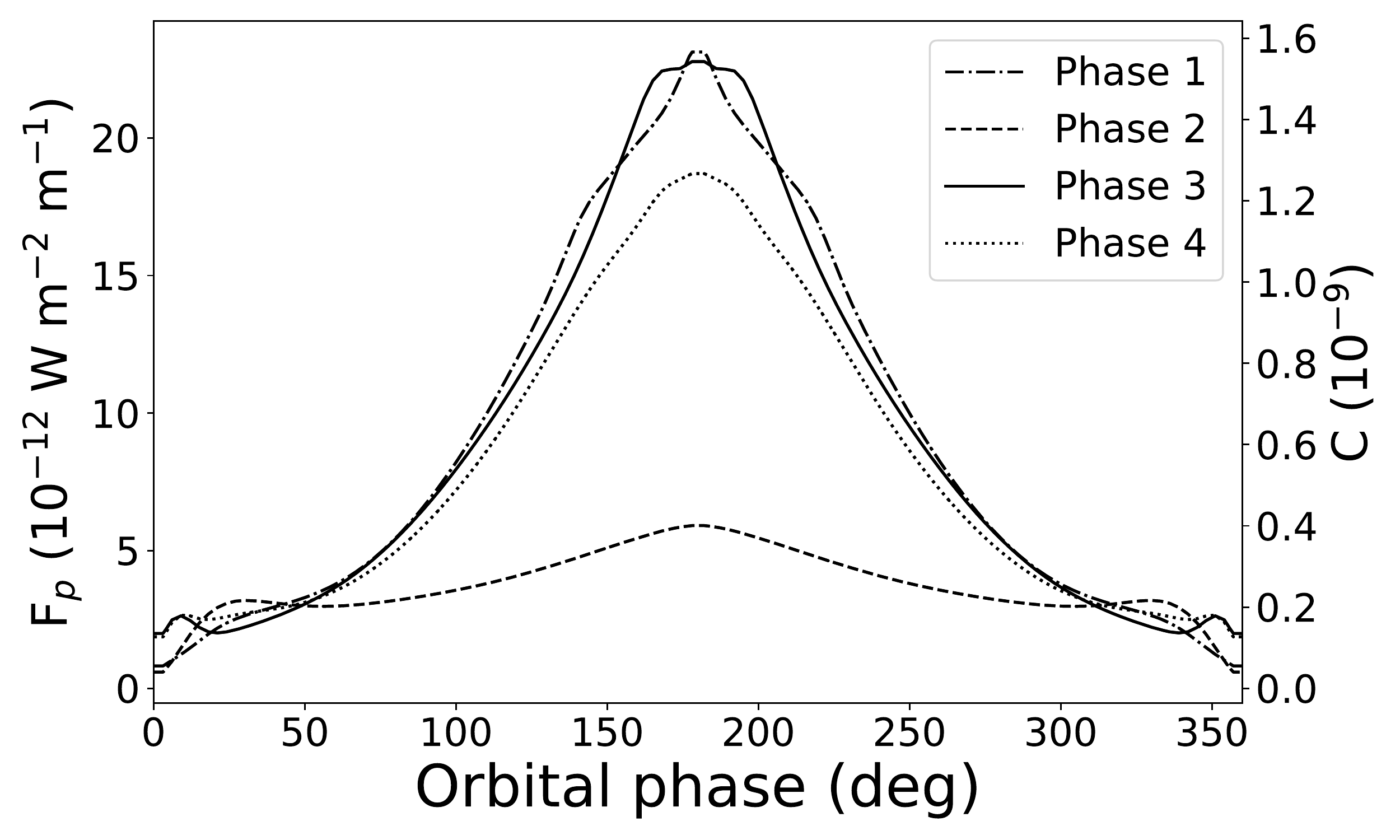}
    \includegraphics[width=0.32\textwidth]{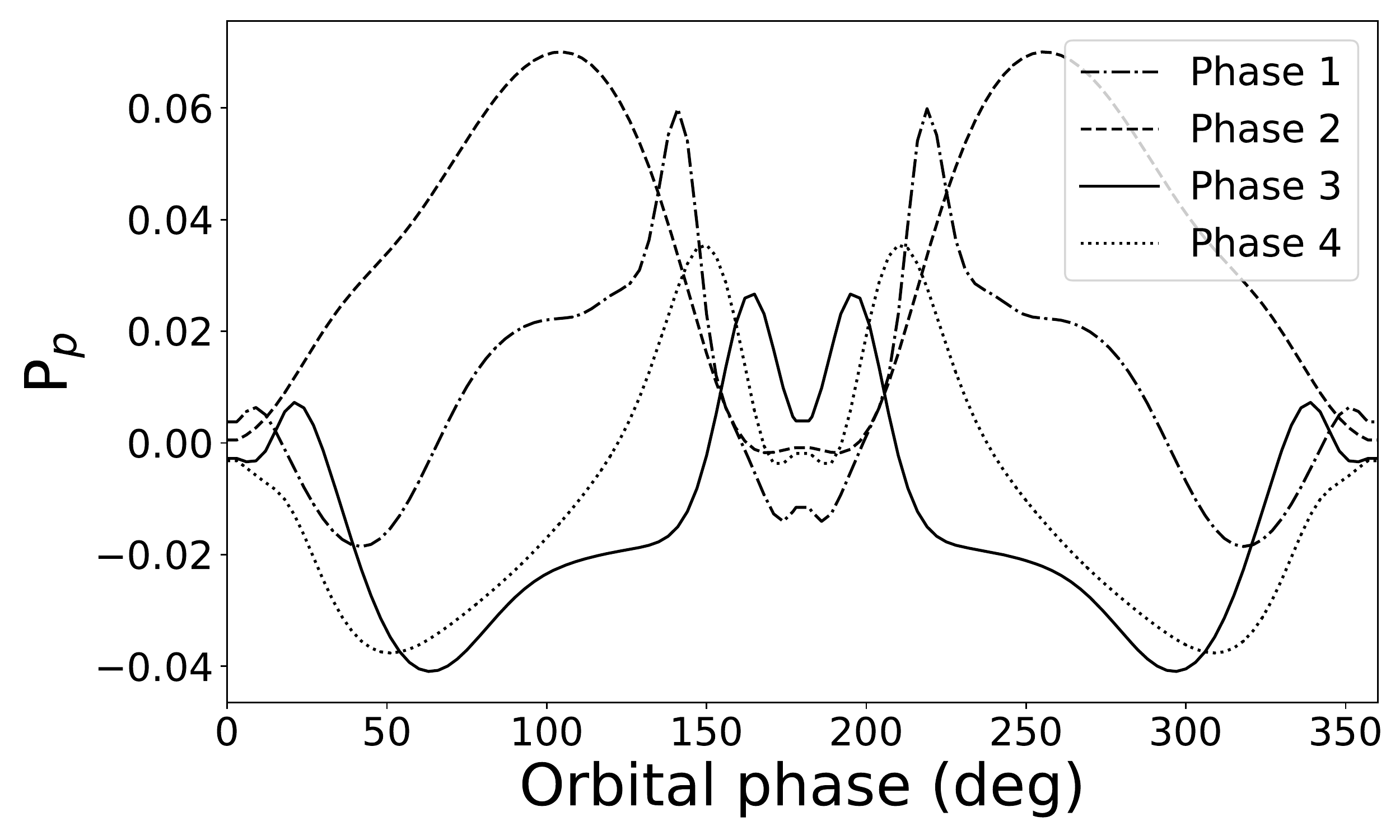}
    \includegraphics[width=0.32\textwidth]{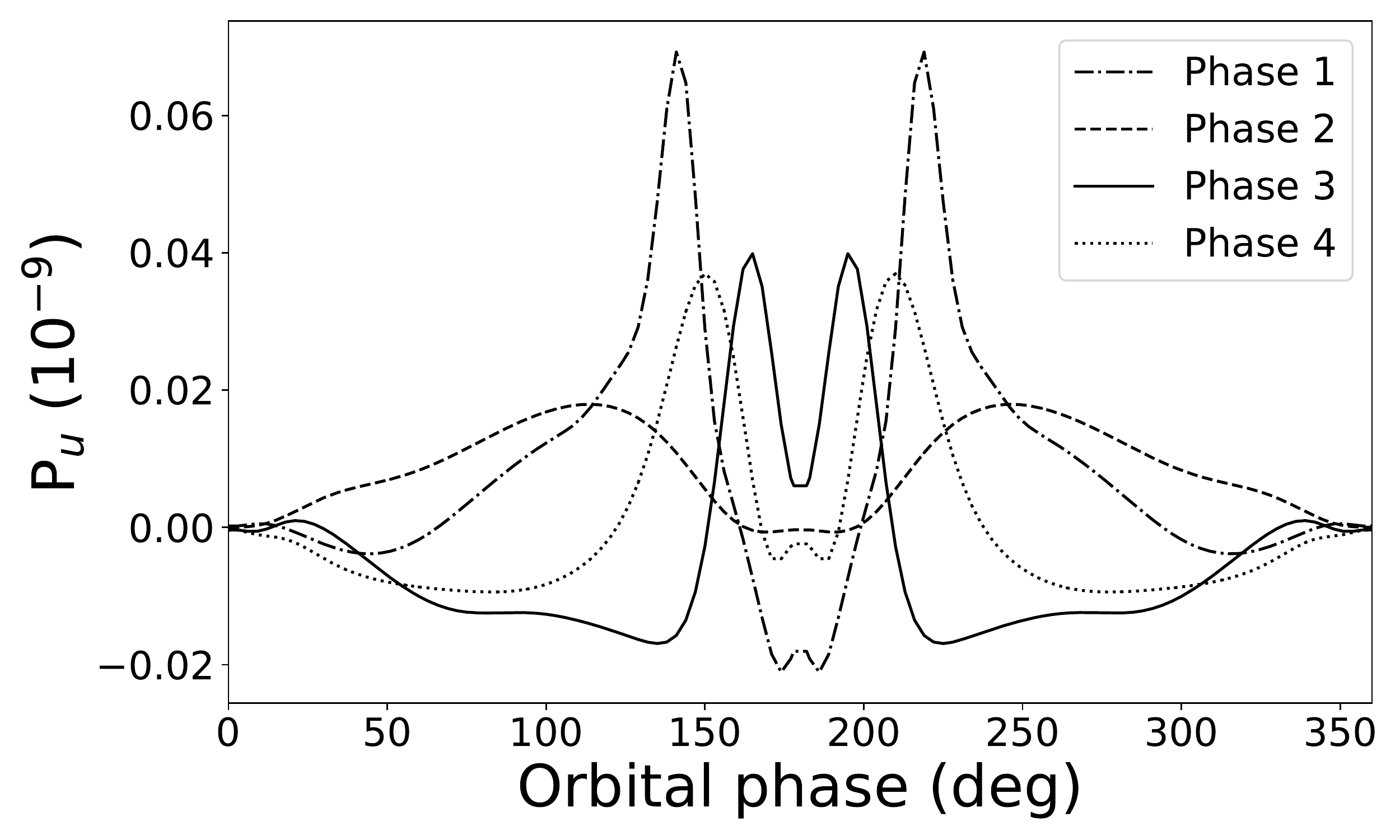}\\
    
    \includegraphics[width=0.32\textwidth]{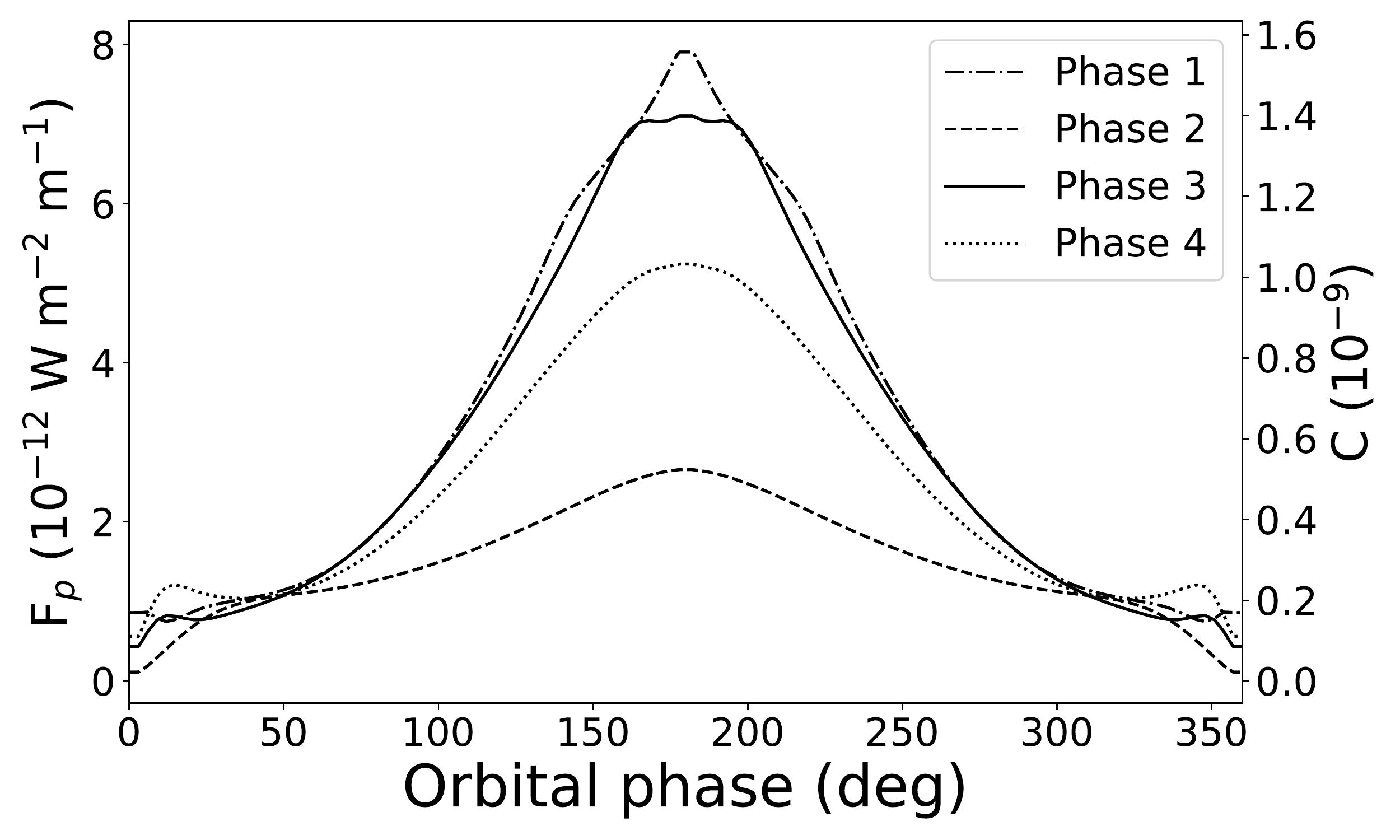}
    \includegraphics[width=0.32\textwidth]{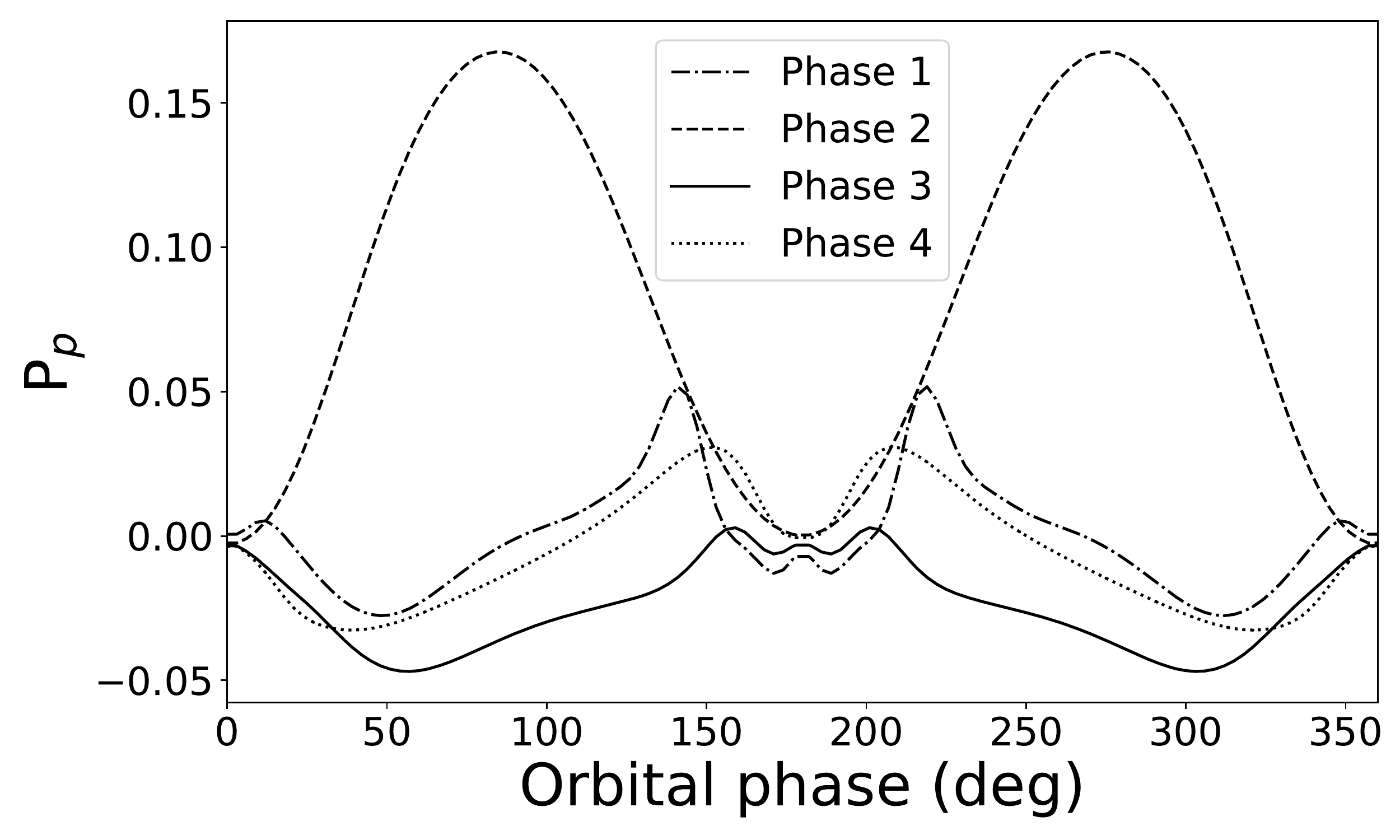}
    \includegraphics[width=0.32\textwidth]{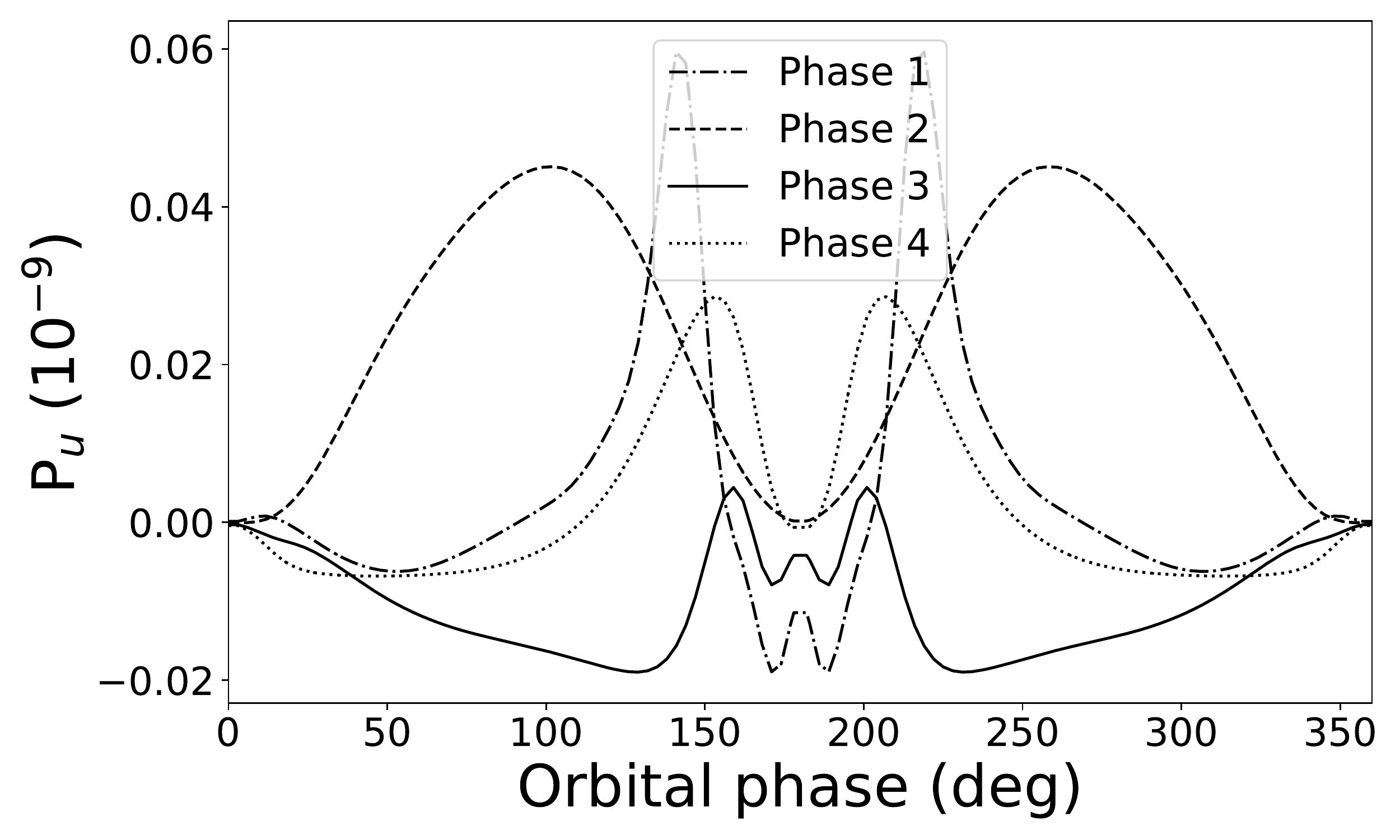}\\
    
    \includegraphics[width=0.32\textwidth]{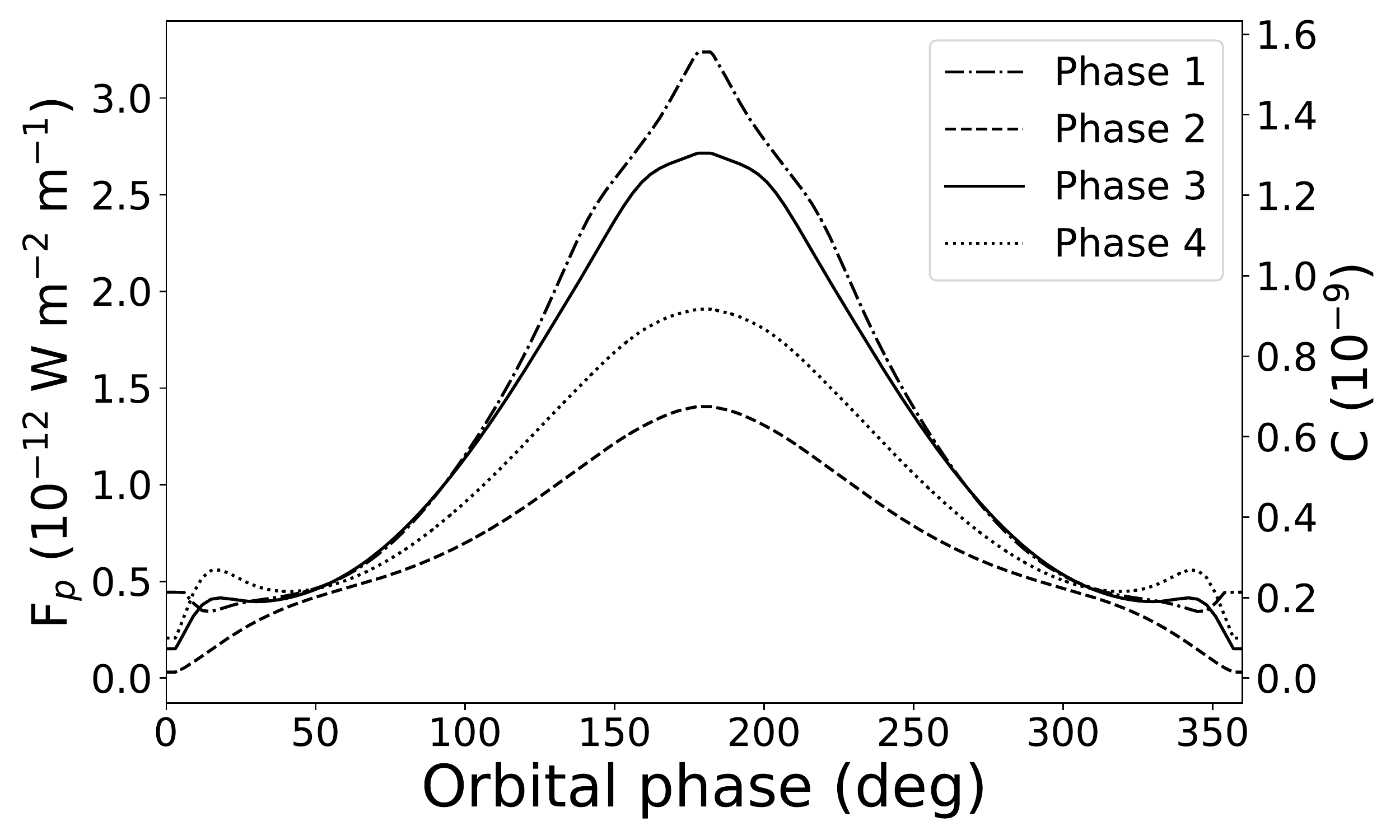}
    \includegraphics[width=0.32\textwidth]{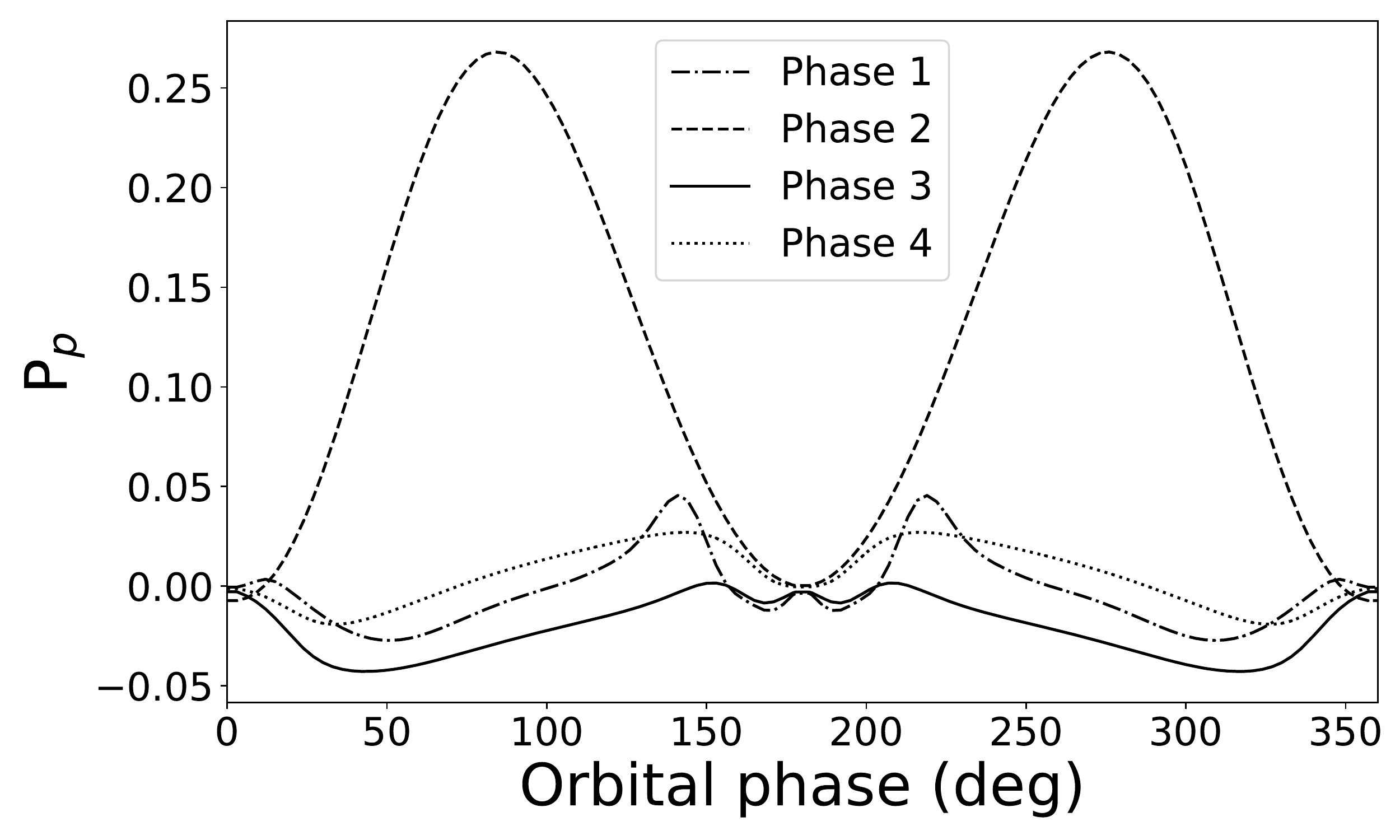}
    \includegraphics[width=0.32\textwidth]{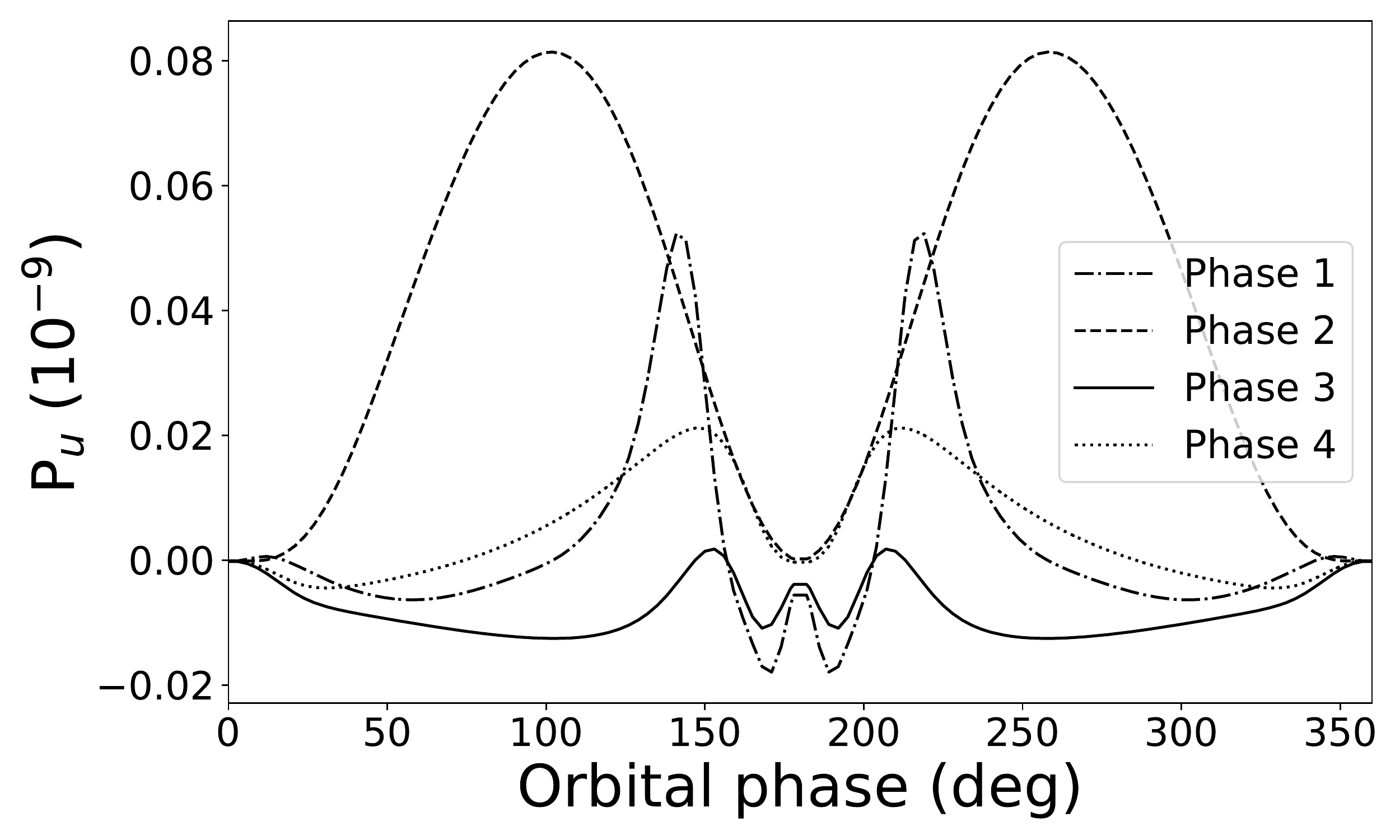}\\
\caption{Similar to Fig.~\ref{fig:orbitfluxes1} except for the model
         planets in the four evolutionary phases and 
         $\Omega= 0^{\circ}$ and $i_{\rm m}= 80^{\circ}$ 
         (thus for an edge--on orbit, $i=90^\circ$):
         Phase~1 ('Current Earth'), 
         Phase~2 ('Thin clouds Venus'),
         Phase~3 ('Thick clouds Venus'),
         Phase~4 ('Current Venus').
         The wavelengths $\lambda$ are like before: 
         $0.5~\mu$m (row 1); $1.0~\mu$m (row 2);
         $1.5~\mu$m (row 3); and $2.0~\mu$m (row 4).}
\label{fig:orbitfluxes3}
\end{figure*}
%-----------------------------------------------------------------------
%-----------------------------------------------------------------------
\begin{figure*}[!]
    \centering
    \includegraphics[width=0.32\textwidth]{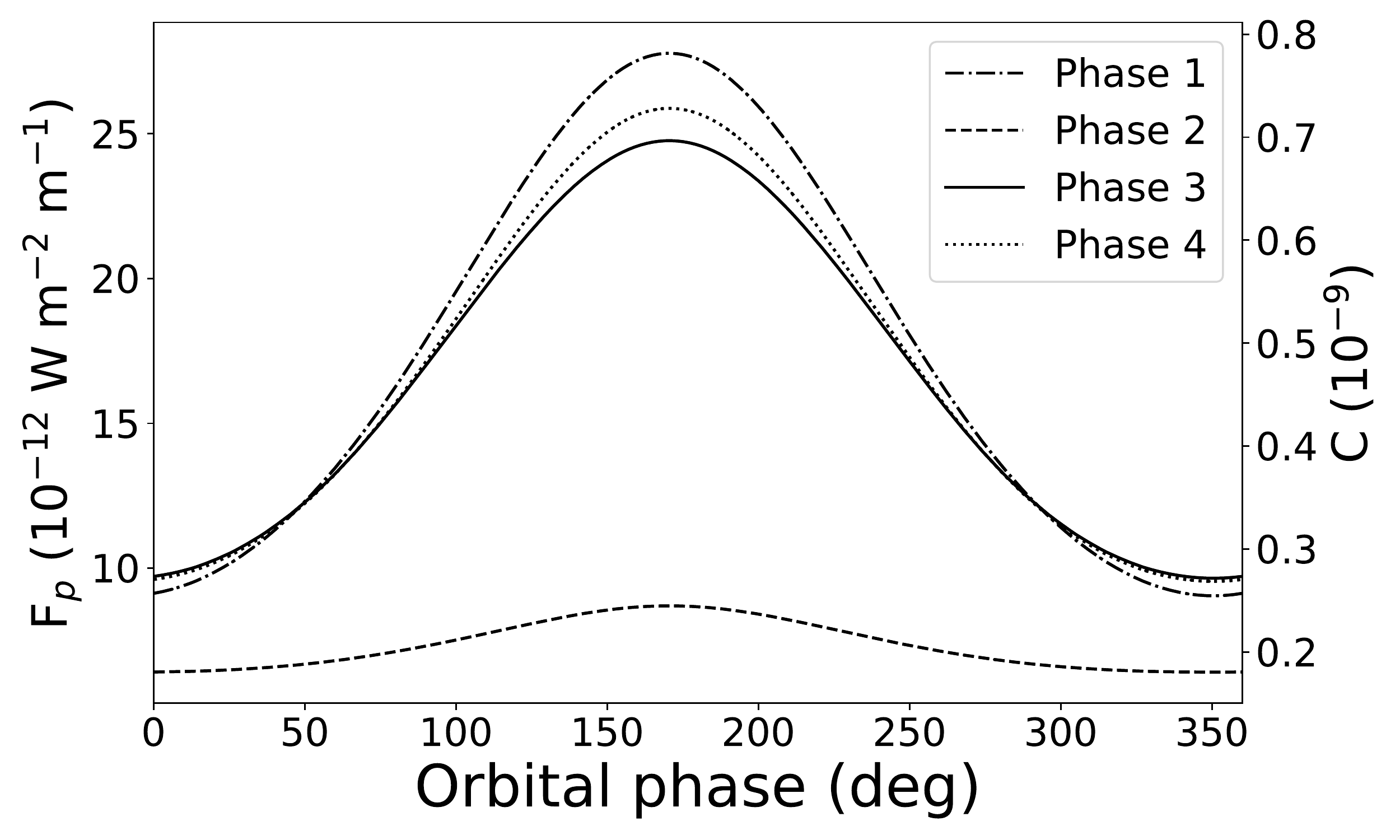}
    \includegraphics[width=0.32\textwidth]{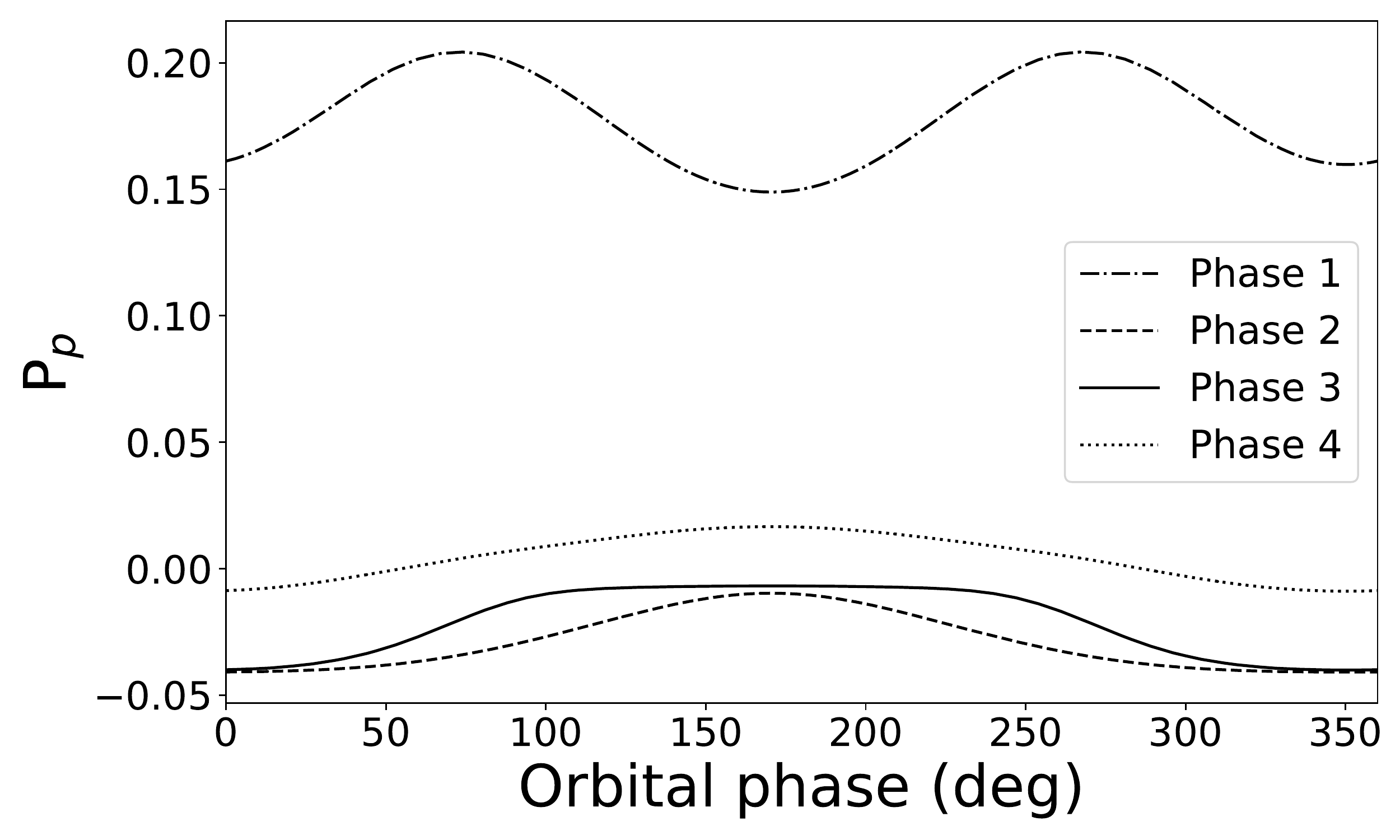}
    \includegraphics[width=0.32\textwidth]{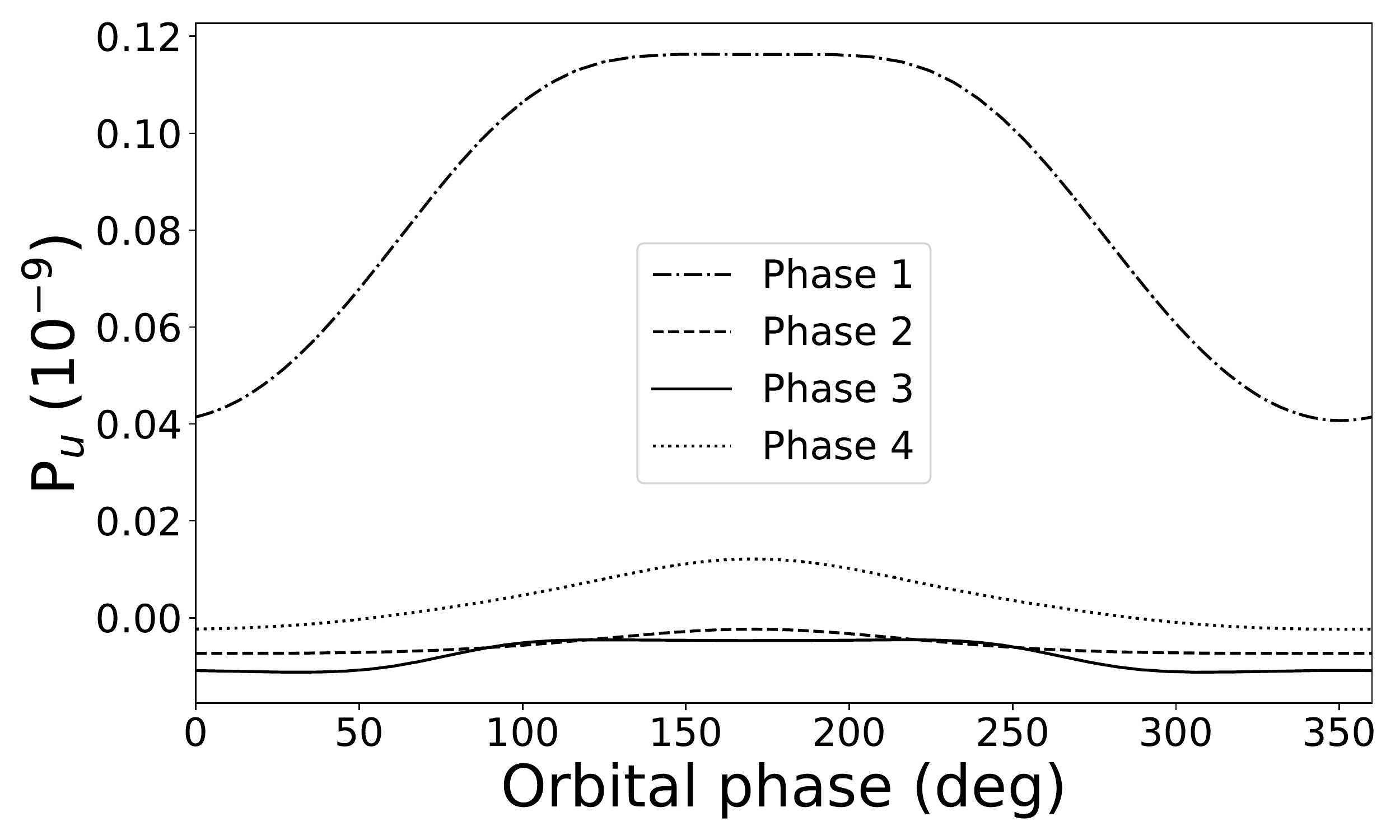}\\

    \includegraphics[width=0.32\textwidth]{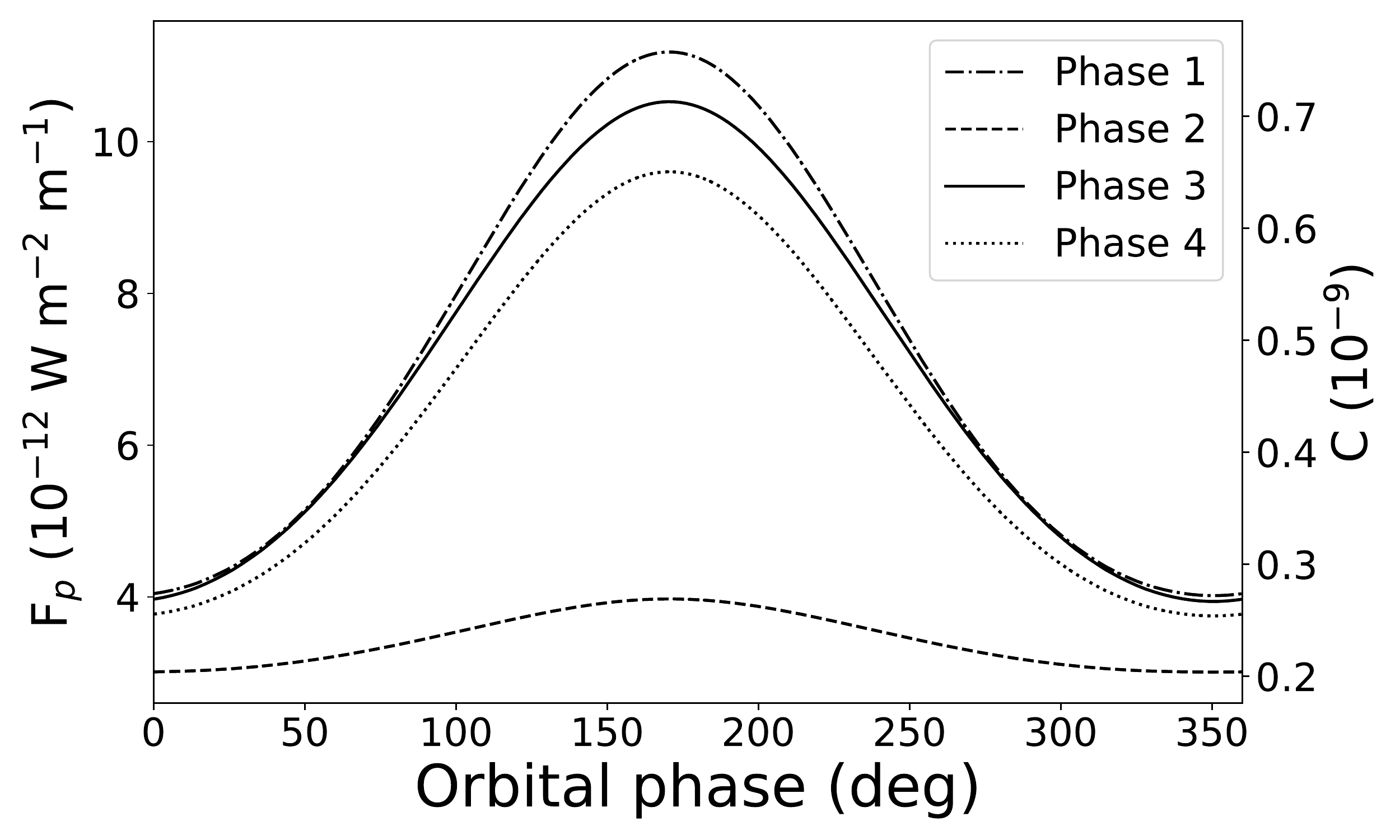}
    \includegraphics[width=0.32\textwidth]{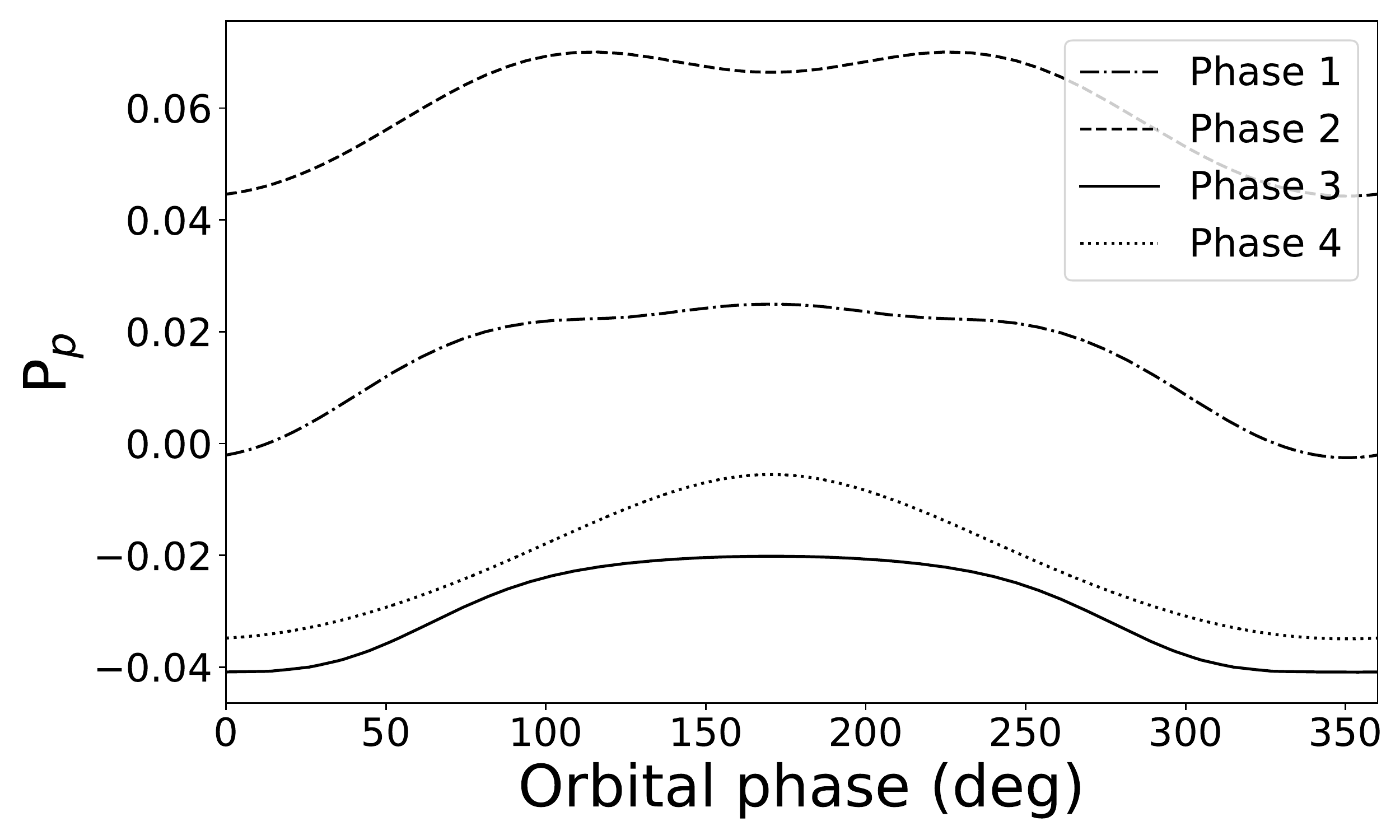}
    \includegraphics[width=0.32\textwidth]{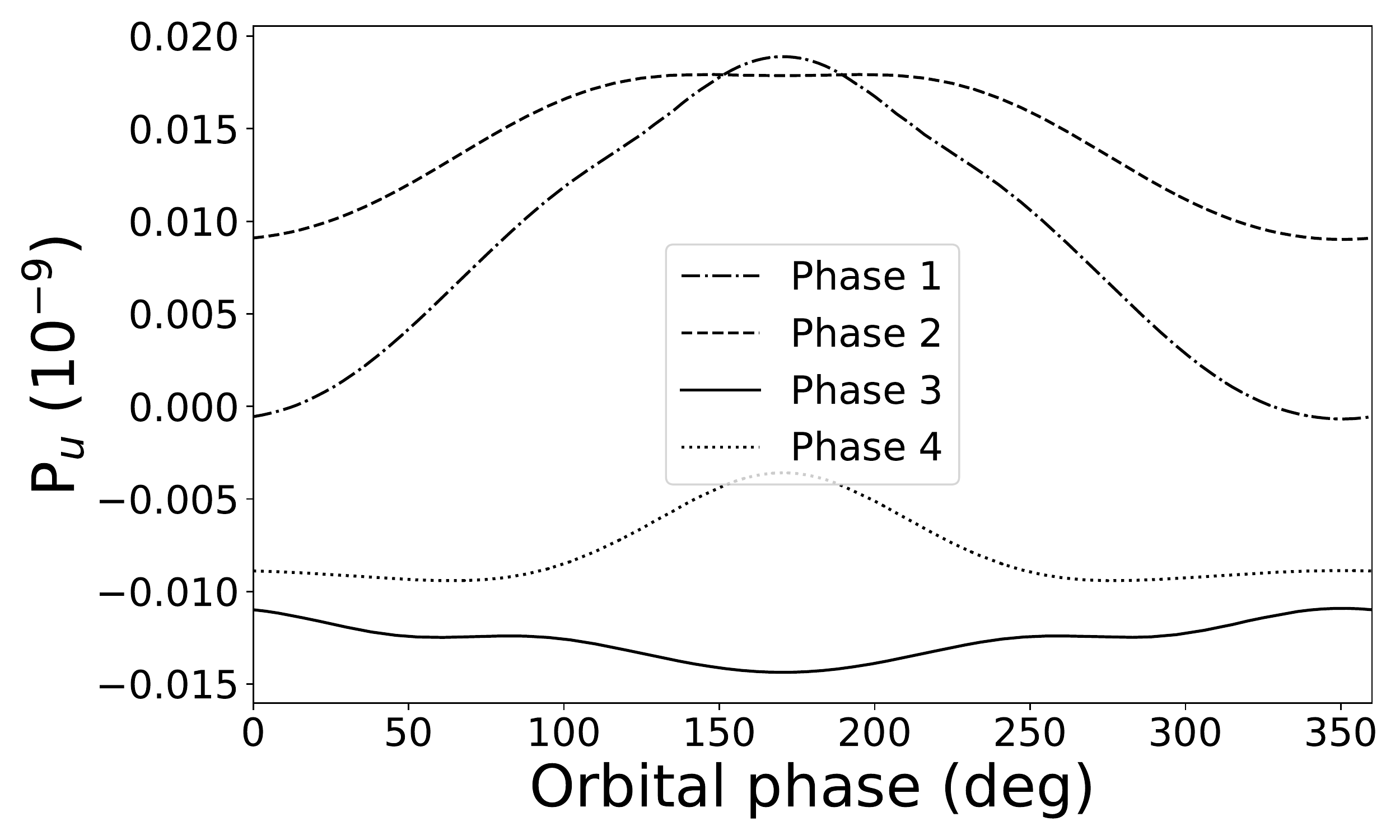}\\
    
    \includegraphics[width=0.32\textwidth]{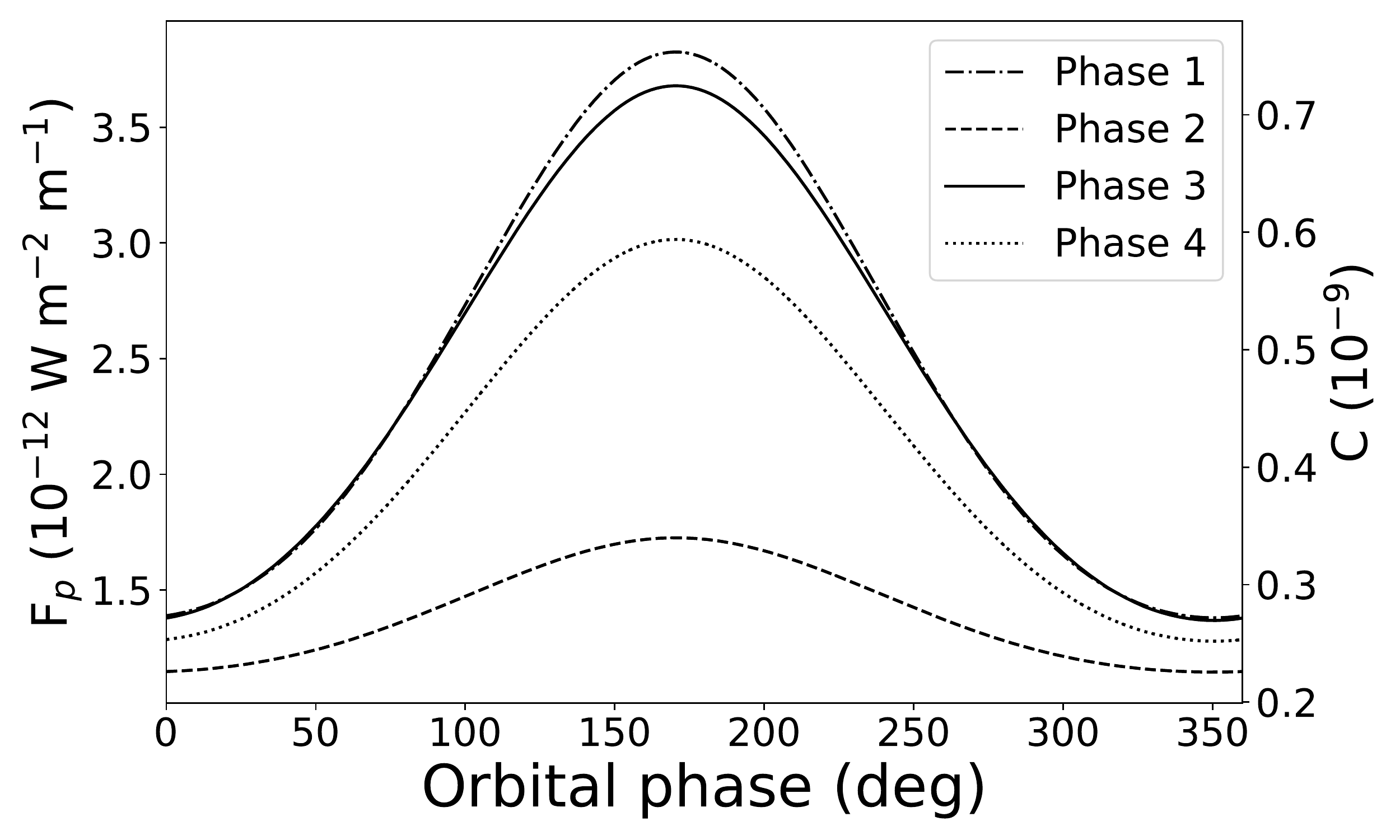}
    \includegraphics[width=0.32\textwidth]{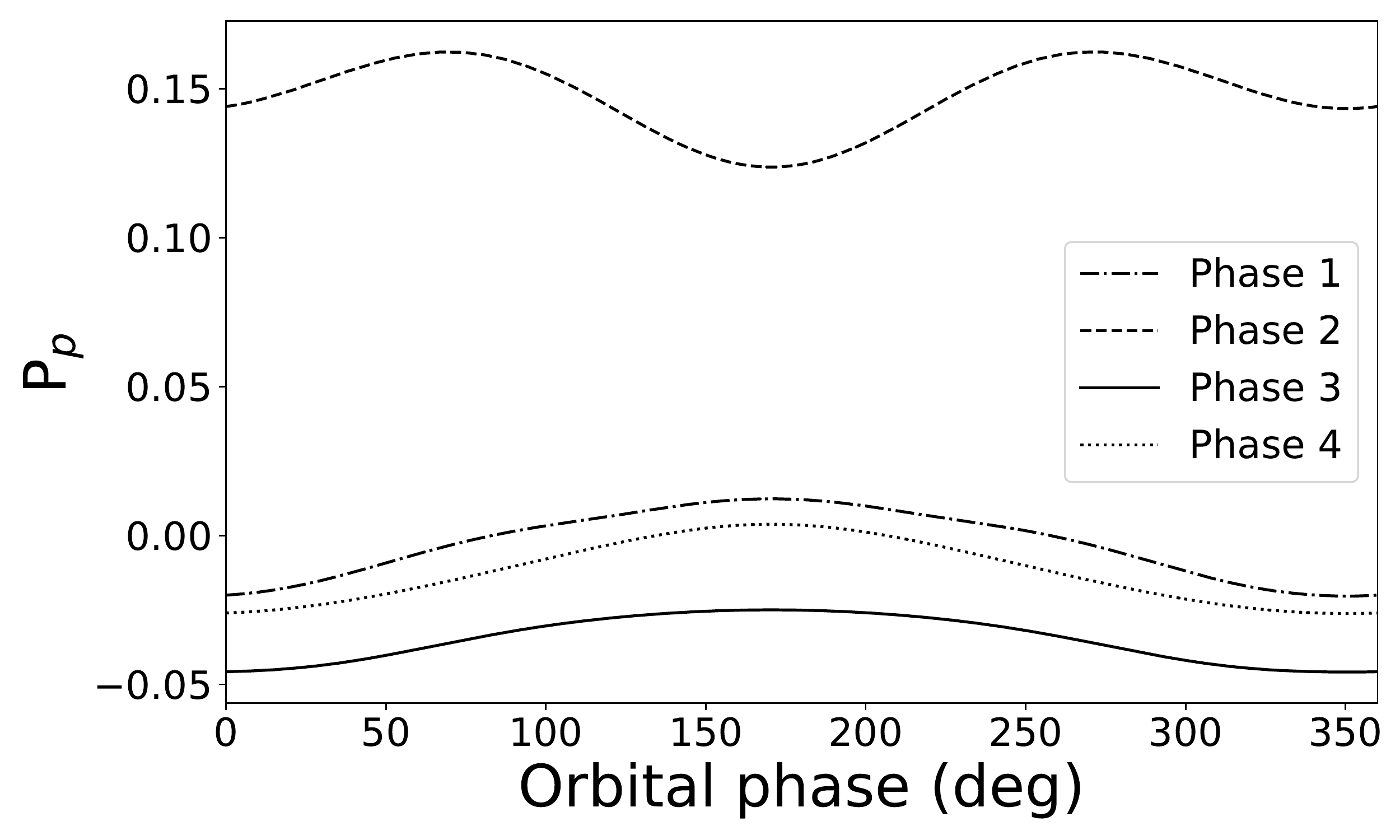}
    \includegraphics[width=0.32\textwidth]{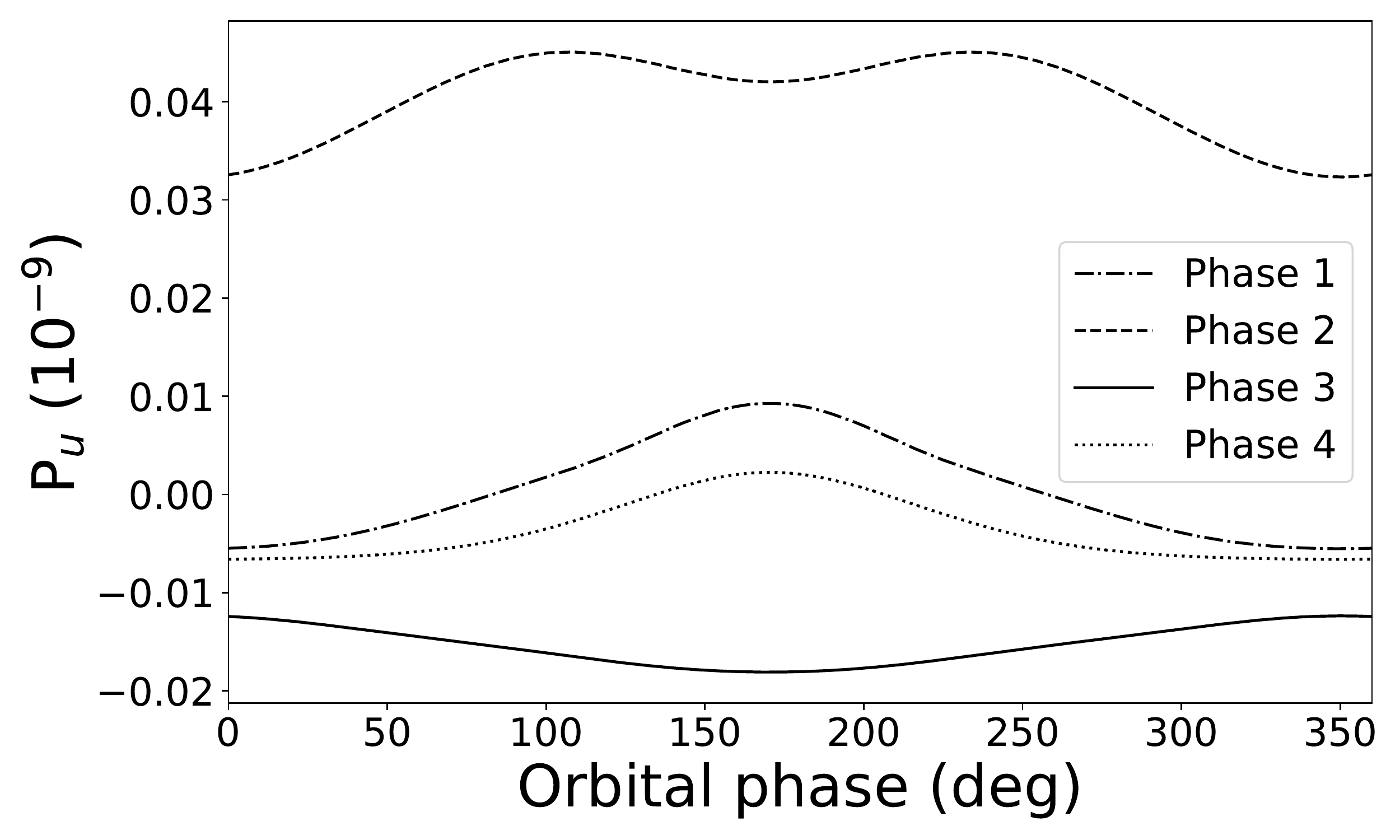}\\
    
    \includegraphics[width=0.32\textwidth]{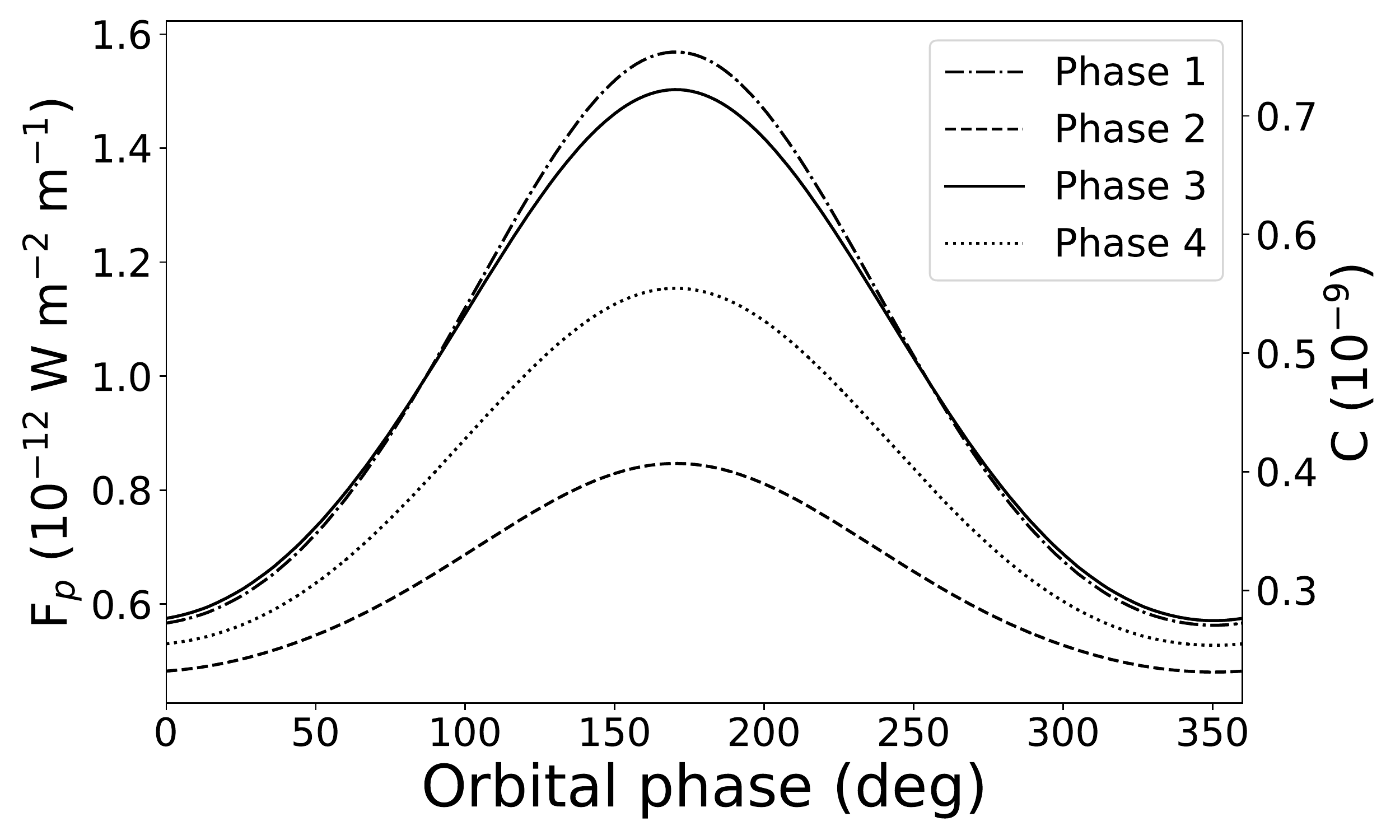}
    \includegraphics[width=0.32\textwidth]{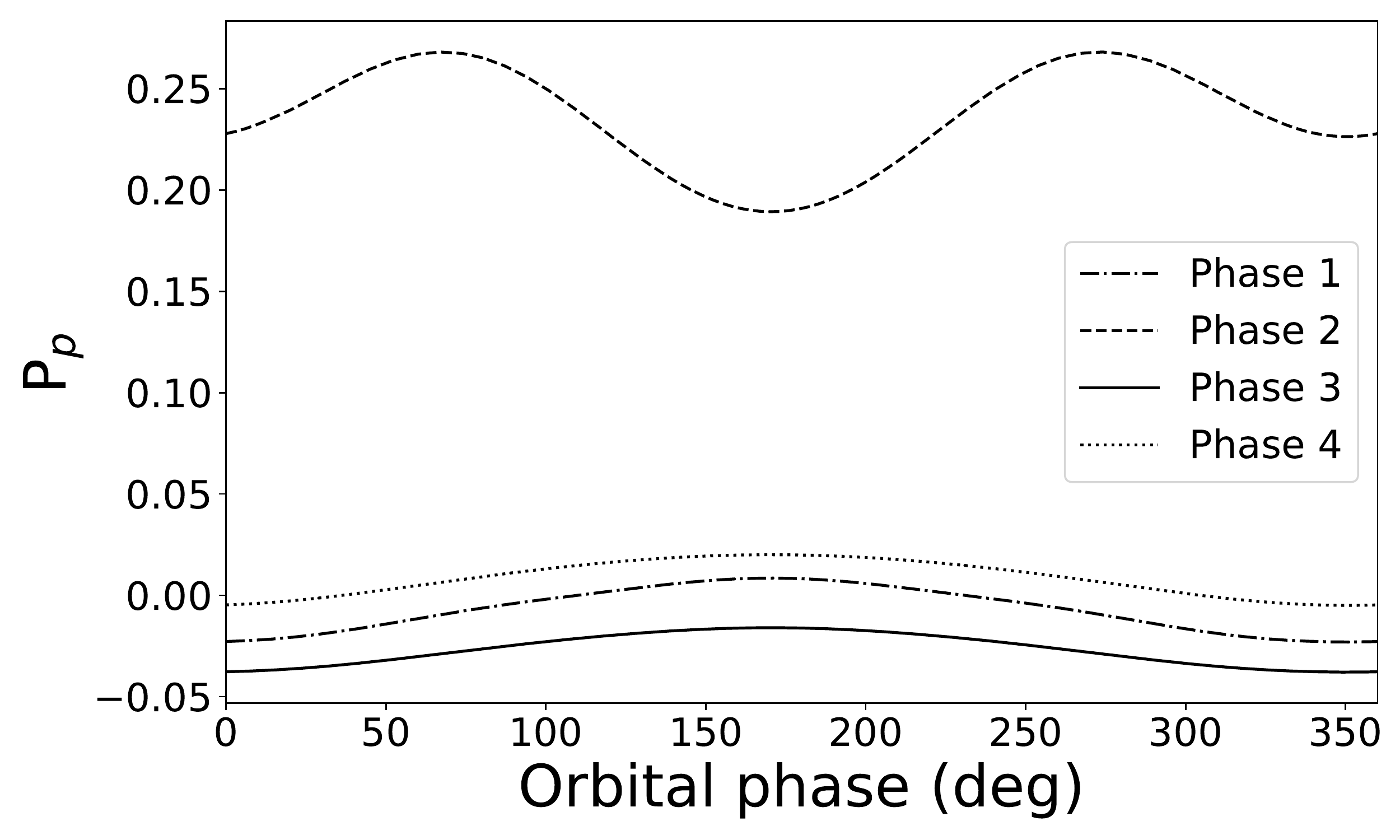}
    \includegraphics[width=0.32\textwidth]{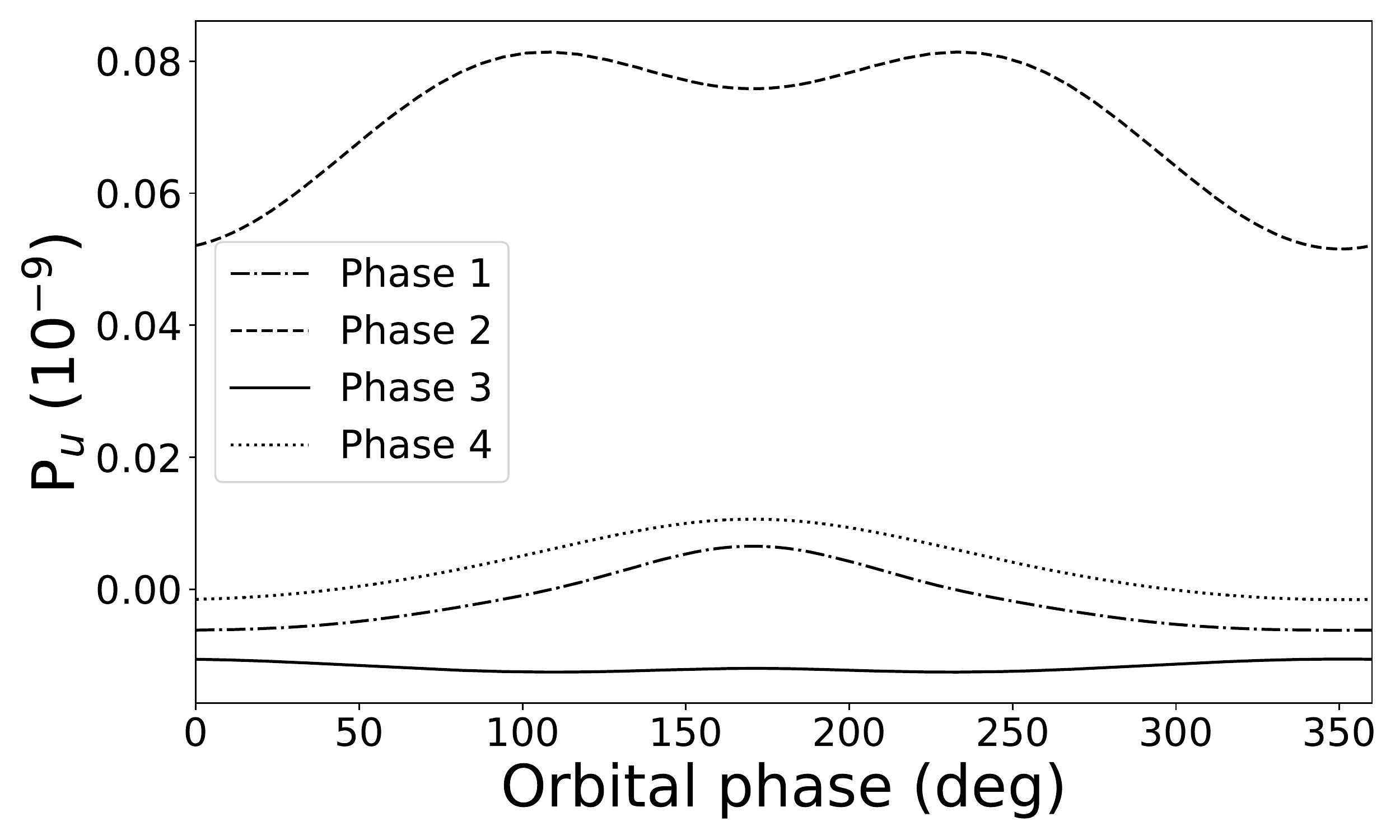}\\
\caption{Similar to Fig.~\ref{fig:orbitfluxes3} except for the 
         most probable, stable orbit around Alpha Centauri A, 
         i.e.\ for $\Omega=205^\circ$ and $i_{\rm m}$= 35$^{\circ}$
         \citep[][]{2016AJ....151..111Q}.
         The wavelengths are like before: 
         $0.5~\mu$m (row 1); $1.0~\mu$m (row 2);
         $1.5~\mu$m (row 3); and $2.0~\mu$m (row 4).}
\label{fig:orbitfluxes2}
\end{figure*}
%-----------------------------------------------------------------------

Figure~\ref{fig:orbitfluxes3} is similar to Fig.~\ref{fig:orbitfluxes1},
except for the four model planets in the different evolutionary phases
and all for $i_{\rm m}=80^\circ$ and $\Omega=0^\circ$. 
Because here the planetary orbits are seen edge-on ($i=90^\circ$), 
the full range of phase angles is covered, which makes it possible 
to explore the full extent of variation of flux and polarization signals. 
Because of this large phase angle range, $F_{\rm p}$ 
varies strongly with the orbital phase. 
The wavelength dependence of the total flux
can be traced back to Fig.~\ref{fig:phasecontour}, where
in particular the Phase 2 planet ('Thin cloud Venus')
is dark at all wavelengths, but relatively bright at the longest
wavelengths and small phase angles. As was the case in Fig.~\ref{fig:orbitfluxes1}, 
the variation of $C$ is the same as that of $F_{\rm p}$,
except for the off-set due to the stellar flux. 
The largest values of $C$ (about 1.6$\times 10^{-9}$),
are found for $\lambda=~0.5~\mu$m 
and around the orbital phase of 180$^\circ$ (at 180$^\circ$,
the planets would actually be behind the star).

Furthermore in Fig.~\ref{fig:orbitfluxes3}, $P_{\rm p}$ depends 
strongly on $\lambda$ and the planet's evolutionary phase.  
At 0.5~$\mu$m, the Phase~1 planet ('Current Earth') shows the 
largest values of $P_{\rm p}$ due to the Rayleigh scattering gas 
above the low altitude clouds.
At the longer wavelengths, where the Rayleigh scattering 
is less prominent, the curves for the Phase~1 planet 
clearly show the positive
polarization of the rainbow around the orbital phases of 
140$^\circ$ and 220$^\circ$ (see Fig.~\ref{fig:AlphaVsOrbitalPhase}).
For the Phase~2 planet ('Thin clouds Venus') and 0.5~$\mu$m, 
the small cloud particles cause positive 
polarization around 150$^\circ$ and 210$^\circ$, which connect
the rainbow and the Rayleigh scattering maximum in 
Fig.~\ref{fig:phasecontour}.
At longer wavelengths, the broad positive polarization 
signature of Rayleigh scattering by the cloud particles 
dominates the curves, while the curves for the Phase~3 and~4
planets show mostly negative polarization apart from the
orbital phases around 180$^\circ$.
When adding the starlight, the angular features of the polarization
$P_{\rm u}$ are suppressed along those parts of the orbits where 
$C$ is smallest, thus away from the orbital phase angle of 180$^\circ$. 
In particular within 180$^\circ$ $\pm 40^\circ$, $P_{\rm u}$ still shows 
distinguishing features, although they are very
small in absolute sense (smaller than 10$^{-10}$).

Figure~\ref{fig:orbitfluxes2} is similar to Fig.~\ref{fig:orbitfluxes3}
except here the model planets are in the most probable stable orbit 
around Alpha Centauri~A as predicted by 
\cite{2016AJ....151..111Q}, namely with 
$\Omega=~205^{\circ}$ and $i_{\rm m}=~35^{\circ}$. 
As can be seen in Fig.~\ref{fig:AlphaVsOrbitalPhase}, for this geometry 
$\alpha$ varies between about 60$^{\circ}$ and 120$^{\circ}$. 
In Fig.~\ref{fig:orbitfluxes2}, $F_{\rm p}$ 
shows a similar variation as the curves 
in Fig.~\ref{fig:orbitfluxes3}, although less prominent, 
as the planets do not reach a 'full' phase (where $\alpha=0^\circ$)
nor the full night phase ($\alpha=180^\circ$) along their orbit. 
The flux curves in Fig.~\ref{fig:orbitfluxes2} also miss
small angular features that appear in the single scattering phase 
functions of the cloud particles 
(see Fig.~\ref{fig:singlescattering}), such as the glory, 
again because the planets do not go through the related
phase angles.

In this particular orbital geometry,  
$P_{\rm p}$ shows less pronounced angular features than 
for the same model planets in edge-on orbits 
(Fig.~\ref{fig:orbitfluxes3}) because of the more limited
phase angle range.
For example, the 'Current Earth' (Phase~1) shows
no rainbow despite the H$_2$O clouds, because the 
phase angle of about 40$^\circ$ is not reached.
In the visible ($\lambda=0.5~\mu$m), $P_{\rm p}$ reaches the 
largest values for the 'Current Earth' (Phase~1). 
At longer wavelengths, $P_{\rm p}$ of the 'Thin clouds Venus' 
(Phase~2) strongly dominates because of the Rayleigh scattering 
by the small cloud particles.
The 'Thick clouds Venus' (Phase 3)
shows predominant negative polarization at all wavelengths 
and across the whole orbital phase angle range except at 
$\lambda=0.5~\mu$m around an orbital phase angle of 20$^{\circ}$. 

The polarization of the spatially unresolved planets, $P_{\rm u}$,
clearly shows the suppression of the polarization features due 
to the added starlight towards
the smaller and larger orbital phase angles, where the planets
are darker. While the Phase~2 planet ('Thin clouds Venus') is 
relatively dark ($C$ is very small), its Rayleigh scattering
polarization signal is so strong that its unresolved polarization
signal is larger than that of the other planets, except 
the Phase~1 planet ('Current Earth') at 1.0~$\mu$m and 
orbital phase angles close to 180$^\circ$.

%-----------------------------------------------------------------
\subsection{Evolutionary phases across $\alpha$ and $\lambda$}
\label{sect_phasewavel}

%-----------------------------------------------------------------------
\begin{figure*}[!]
\centering
\includegraphics[width=0.6\textwidth]{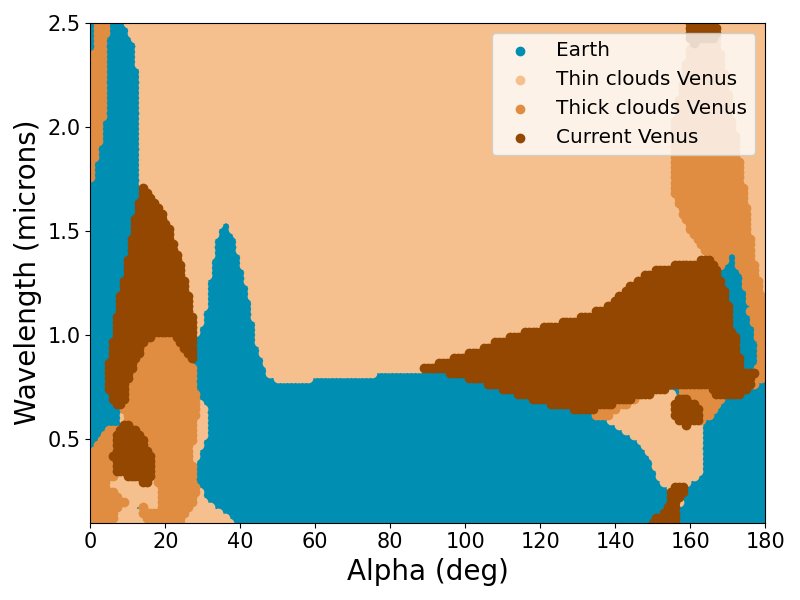}
\includegraphics[width=0.6\textwidth]{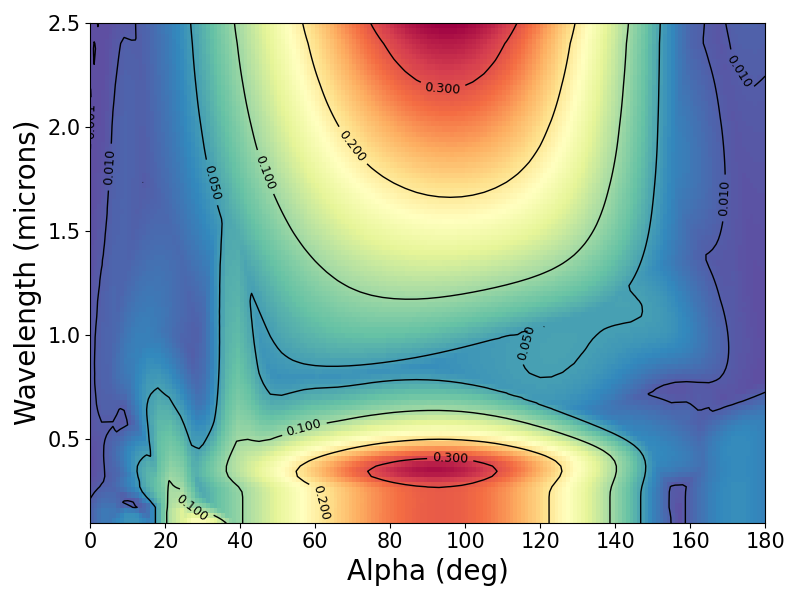}
\caption{Top: The planet models that yield the largest absolute degree of 
         polarization $|P_{\rm p}|$ over all phase angles 
         $\alpha$ and wavelengths $\lambda$:  
         Phase~1 - 'Current Earth' (blue); Phase~2 - 'Thin clouds Venus' 
         (light orange); Phase~3 - 'Thick clouds Venus' (dark orange);
         and Phase~4 - 'Current Venus' (brown). 
         Bottom: The maximum values of $|P_{\rm p}|$ of the four model
         planets as functions of $\alpha$ and $\lambda$.}
\label{fig:maxP}
\end{figure*}
%-----------------------------------------------------------------------

In Fig.~\ref{fig:maxP}, we show which evolutionary phase has the 
highest values of $|P_{\rm p}|$ across all phase 
angles $\alpha$ and wavelengths $\lambda$. 
We find that $|P_{\rm p}|$ of the Phase~1 planet ('Current Earth') 
dominates between 30$^\circ$ and 150$^\circ$, and mostly for 
$\lambda < 1.0~\mu$m. 
In particular, around $\alpha=40^\circ$ and up to 
$\lambda=2.0$~$\mu$m, the polarization signal of the rainbow 
produced by the large water cloud particles is about 0.1
(see the bottom plot of Fig.~\ref{fig:maxP}).
For $\lambda > 1.0~\mu$m, the Phase~2 planet ('Thin clouds Venus')
shows the strongest polarization due to the Rayleigh 
scattering by the small H$_2$O cloud particles, 
as can clearly be seen in the bottom plot. 
The small patches where the strongest polarization signal is 
from the 'Thick clouds Venus' (Phase~3), 
for example near $\alpha=20^\circ$ and $\lambda < 1.0~\mu$m, 
or from the 'Current Venus' (Phase~4) are due to the  
single scattering polarization features of the H$_2$SO$_4$
cloud particles, as can be seen in Fig.~\ref{fig:singlescattering}.

The accessible phase angle range for direct observations of
such exoplanets
obviously depends on the actual orientation of the 
planetary orbits and cannot be optimized by the observer. 
Precisely because of that, 
Fig.~\ref{fig:maxP} clearly indicates that measurements should be
performed across a broad wavelength range, including wavelengths 
below 1~$\mu$m to allow distinguishing between Earth-like and 
Venus-like planets in various evolutionary phases.

%--------------------------------------------------------------------
\section{Summary and conclusions}
\label{sect:discussion&summary}

We presented the total flux and linear polarization
of starlight that is reflected by model planets of 
various atmospheric types to investigate whether 
different phases in the evolution of planets like the Earth and 
Venus can be distinguished from each other. 
We have used four planet models to represent possible
evolutionary phases. Phase~1 ('Current Earth') has an 
Earth-like atmosphere and liquid water clouds; 
Phase~2 ('Thin clouds Venus') has a Venus-like CO$_2$ 
atmosphere and thin water clouds; Phase~3 ('Thick clouds Venus') 
has a Venus-like CO$_2$ atmosphere and thick sulphuric acid 
solution clouds; and Phase~4 ('Current Venus') 
has a CO$_2$ atmosphere and thin sulphuric acid solution clouds. 
We have computed the total flux and polarization signals 
specifically for model planets orbiting our neighbouring 
solar-type star Alpha Centauri~A using predicted stable orbits 
\citep[][]{2016AJ....151..111Q} in its habitable zone.

We have computed the reflected starlight for wavelengths 
$\lambda$ ranging from 0.3~$\mu$m to 2.5~$\mu$m and for 
planetary phase angles $\alpha$ from 0$^\circ$ to 180$^\circ$. 
We not only present the fluxes and polarization of 
spatially resolved model planets (thus without background 
starlight) but also of spatially unresolved planets, 
thus the combined signal of the planet and the star. For 
the latter cases, we also computed the planet-star 
contrast $C$ as a function of $\alpha$ and $\lambda$ 
to determine what would technically be required to 
detect the planetary signals upon the background starlight.

The range of planetary phase angles $\alpha$ at which a planet 
can be observed (spatially resolved or unresolved) depends on the
inclination of the planetary orbit with respect to the observer.
We have specifically studied the reflected light signals
of planets orbiting solar-type star Alpha Centauri~A.
This star is part of a double star system with
solar-type star Alpha Centauri B. The distance 
between the two stars varies from 35.6 to 11.2~AU
(M-dwarf Alpha Centauri~C or Proxima Centauri orbits 
the pair at a distance of about 13,000~AU).
Dynamical computations \citep[][]{2016AJ....151..111Q}
predict stable planetary orbits around 
Alpha Centauri~A in a narrow range of mutual inclination 
angles $i_{\rm m}$ between the orbital planes of the two stars 
and that of the planet. In particular, the most stable 
orbit has $i_{\rm m}= 35^{\circ}$, which provides an 
$\alpha$ range from 60$^{\circ}$ to 120$^{\circ}$. 
We find that with this orbital geometry the degree 
of polarization of the planet would be largest for the 
'Current Earth' (Phase~1) across the visible 
($\lambda < 1.0~\mu$m) due to Rayleigh scattering by the 
gas above the clouds. 
At near infrared wavelengths (1.0~$\mu$m $< \lambda < 2.5~\mu$m), 
the polarization of the 'Thin clouds Venus' (Phase~2) is highest,
because this planet has small cloud droplets that scatter
like Rayleigh scatterers at the longer wavelengths.

The well-known advantage of measuring the degree of polarization for the 
characterization of (exo)planets is that
the angular features in the signal of the planet as a whole
are similar to the angular features in the light that has
been singly scattered by the gas molecules and cloud particles, 
which are very sensitive to the microphysical properties 
(such as the size distribution, composition, shape)
of the scattering molecules and cloud particles and to the 
atmosphere's macrophysical properties 
(such as cloud altitude and thickness). 
The reflected total flux is much less sensitive to the 
atmospheric properties than the degree and direction of
polarization \citep[see e.g.][for several examples]{hansen1974light}. 
Indeed, the variations of the planetary flux $F_{\rm p}$
along a planet's orbit appear to be mostly due to the change of the 
fraction of the planetary disk that is illuminated and 
visible to the observer. They provide limited information
on the planet's atmospheric characteristics, especially if one takes
into account that with real observations, the planet 
radius will be unknown unless the planet happens to transit 
its star. The variation of the planetary flux $F_{\rm p}$ with 
phase angle $\alpha$ and wavelength $\lambda$ is similar that 
of planet-star contrast $C$,
the ratio of $F_{\rm p}$ to the stellar flux $F_{\rm s}$.
For a Venus-like planet orbiting Alpha Centauri A,
$C$ is on the order of 10$^{-9}$.

Our numerical simulations show that variations
in $P_{\rm p}$, the degree of polarization of the spatially 
resolved planets (thus without any starlight),
with $\alpha$ combined with variations with $\lambda$ 
could be used to distinguish the planetary evolutionary 
phases explored in this paper:
\begin{itemize}
    \item Phase~1 planets ('Current Earth') show strong
    positive (perpendicular to the reference plane 
    through the star, planet, and observer) 
    polarization around $\alpha=40^{\circ}$ due 
    to scattering by large water cloud droplets (the rainbow) 
    and also higher polarization for $\lambda < 0.5~\mu$m
    and $\alpha \approx 90^\circ$ due to 
    Rayleigh scattering by the gas above the clouds.
    \item Phase~2 planets ('Thin clouds Venus') polarize 
    light negatively (parallel to the reference plane) 
    across most phase angles and at visible wavelengths. 
    At near-infrared wavelengths, they have strong 
    positive polarization around $\alpha=90^\circ$
    due to Rayleigh scattering by the small cloud droplets, and a 
    'bridge' of higher polarization from the Rayleigh maximum
    to the rainbow angle ($\alpha \approx 40^\circ$)
    with decreasing $\lambda$.
    \item Phase~3 planets ('Thick clouds Venus')
    have predominantly negative polarization 
    from the visible to the near-infrared, with 
    small regions of positive polarization 
    for $\lambda < 1.0~\mu$m and for  
    $20^\circ \leq \alpha \leq 30^\circ$, that are characteristic 
    for the 75\% H$_2$SO$_4$ cloud particles with 
    $r_{\rm eff}=2.0~\mu$m.
    \item Phase~4 planets ('Current Venus') yield similar 
    polarization patterns as the Phase~3 planets, except with 
    more prominent negative polarization for 
    $10^\circ \leq \alpha \leq 30^\circ$
    and for 0.5~$\mu$m $\leq \lambda \leq 2~\mu$m.
    Rayleigh scattering by the small cloud particles 
    produces a maximum of positive 
    polarization around $\alpha=90^\circ$ and for
    $\lambda > 2~\mu$m.
\end{itemize}

Our simulations of the planetary polarisation $P_{\rm p}$ 
do not include any background starlight, and can thus reach several 
percent to even 20$\%$ for the 'Current Earth' (Phase~1) planet 
in an edge-on orbit (Fig.~\ref{fig:orbitfluxes3}).
Whether or not such polarization variations could be measured 
depends strongly on the techniques used to suppress the light of 
the parent star. 
If the background of the planet signal contains a fraction $x$ of 
the flux of the star, the degree of polarization of the light 
gathered by the detector pixel that contains the planet will 
equal $(C/(C + x)) P_{\rm p} \approx (C/x) P_{\rm p}$ 
with $C$ the contrast between the total fluxes of the planet 
and the star (Eq. \ref{eq_pux}).
For a Venus-like planet orbiting Alpha Centauri~A, $C$ is 
on the order of 10$^{-9}$, thus $x$ should be as small as 10$^{-4}$
to get a polarisation signal on the order of 10$^{-6}$,
assuming $P_{\rm p}$ is about 0.1. Not only excellent direct
starlight suppression techniques, but also a very high spatial 
resolution would help to decrease $x$.

Our simulations further show that temporal variations 
in the total flux $F_{\rm u}$ of Alpha Centauri~A with a
spatially unresolved terrestrial-type planet orbiting in its 
habitable zone would be less than 10$^{-12}$~W/m$^3$. The degree of polarization $P_{\rm u}$ of
the combined star and planet signals, would show variations
smaller than 0.05~ppb.
To identify this planetary flux on top of the stellar flux, 
a very high-sensitivity instrument would be required and an 
even higher sensitivity would be required to subsequently
characterize the planetary atmosphere. 
Recall that the orbital period of such a planet, and with that 
the period of the signal variation
and presumably the stability requirements of an instrument, 
would be about 0.76~years.
\cite{bailey2018polarized} computed the polarization signal
of spatially unresolved, hot, cloudy, Jupiter-like planet 
HD~189733b to be on the order of $\sim$20~ppm.
Because of their large size, Jupiter-like planets, and in 
particular those in close-in orbits that receive large
stellar fluxes, would clearly be less challenging  
observing targets than terrestrial-type planets. 

The HARPS instrument on ESO's 3.6 m telescope includes 
polarimetric observations with a polarimetric sensitivity 
of $10^{-5}$ \citep{snik2010harps}. 
Planetpol on the 4.2~m William Herschel Telescope (WHT) 
on La Palma achieved 
a sensitivity to fractional linear polarization (the ratio 
of linearly polarized flux to the total flux) 
of $10^{-6}$. While PlanetPol did not succeed in detecting exoplanets, 
it did provide upper limits on the albedo's of a number of 
exoplanets \citep{lucas2009planetpol}. 
The Extreme Polarimeter (ExPo), that was also mounted on 
the WHT, was designed to target young stars embedded in 
protoplanetary disks and evolved stars surrounded by dusty envelopes, 
with a polarimetric sensitivity better than $10^{-4}$ \citep{rodenhuis2012extreme}. 
The HIPPI-2 instrument uses repeated observations of bright stars in 
the SDSS g' band for achieving better than 3.5~ppm accuracy on the 3.9-m 
Anglo-Australian Telescope and better than 11~ppm on the 60-cm Western 
Sydney University’s telescope \citep{bailey2020hippi}. 
The POLLUX-instrument on the LUVOIR space telescope concept aims 
at high-resolution (R$\sim$120,000) spectropolarimetric 
observations across ultra-violet and visible wavelengths (100-400~nm) 
to characterize atmospheres of terrestrial-type exoplanets
\citep{luvoir2019luvoir,rossi2021spectropolarimetry}. The 
EPICS instrument planned to be mounted on the ELT telescope, 
is designed to achieve a contrast of 10$^{-10}$ depending the angular 
separation of the objects \citep{kasper2010epics}. 

In our simulations, we have neglected absorption by atmospheric gases. 
Including such absorption would yield lower total fluxes
in specific spectral regions, depending on the type and amount of
absorbing gas, its vertical distribution, and on the altitude
and microphysical properties of the clouds and hazes.
Including absorption by atmospheric 
gases could increase or decrease the degree of polarization,
depending on the amount and vertical distribution of the absorbing
gas and on the microphysical properties of the scattering
particles at various altitudes \citep[see e.g.][for examples of polarization spectra of Earth-like planets]{2022A&A...664A.172T,stam2008spectropolarimetric}.
While measuring total and polarized fluxes of reflected 
starlight across gaseous absorption bands is of obvious interest 
for the characterization of planets and their atmospheres, 
the small numbers of photons inside gaseous absorption 
bands would make such observations extremely challenging. 

We have also neglected any intrinsic polarization of  
Alpha Centauri~A. 
Measurements of the degree of linear polarization of FKG stars 
indicate that active stars like Alpha Centauri A have a typical mean 
polarization of 28.5~$\pm$~2.2~ppm \citep{cotton2017intrinsic}. 
This could add to the challenges 
in distinguishing the degree of polarization of the planet from that 
of the star if the planet is spatially unresolved, although the 
phase angle variation of the planetary
signal and the direction of polarization of the planet signal 
(i.e.\ usually either perpendicular or parallel to the plane
through the star, the planet, and the observer) could be 
helpful
provided of course that the instrument that is used for
the observations has the capability to measure the extremely small
variations in the signal as the planet orbits its star 
(on the order of 10$^{-9}$).

State-of-the-art instruments with sensitivity to polarization 
signals down to $10^{-6}$ (i.e.\ 1000~ppb) are still a few orders 
of magnitude away from detecting variations in polarization 
signals from spatially unresolved exo-Earths or exo-Venuses 
around nearby solar-type stars such as Alpha Centauri~A.
To be able to distinguish between the different planetary 
evolutionary phases explored in this paper, e.g.\ between 
water clouds or sulphuric acid clouds, variations on 
the order of 10$^{-9}$ and hence significant 
improvements in sensitivity would be needed if the planets 
are spatially unresolved. The variation in the degree of polarization of spatially 
resolved planets along their orbital phase should be detectable 
by instruments capable of achieving star-planet contrasts of 
10$^{-9}$ and that would allow to distinguish between water 
clouds or sulphuric acid clouds. 
Current high-contrast imaging instruments manage to directly image 
self-luminous objects such as young exoplanets and brown dwarfs 
in NIR total fluxes at contrasts of 10$^{-2}$-10$^{-6}$ 
\citep[]{bowler2016imaging,nielsen2019gemini,langlois2021sphere,van2021high}. 
Further, instruments such as EPICS on ELT and concepts for instruments on
future space observatories such as HabEx \citep{gaudi2020habitable} and 
LUVOIR \citep{luvoir2019luvoir} hold the promise 
for attaining contrasts of $\sim$10$^{-10}$. Reaching such extreme
contrasts would make it possible to directly  
detect terrestrial-type planets and to use polarimetry to 
differentiate between exo-Earths and exo-Venuses.

%--------------------------------------------------------------------
\begin{acknowledgements}
      This work was supported by the
      \emph{Netherlands Organisation for Scientific Research
      (NWO)} through the User Support Programme Space Research, 
      project number ALW GO 15-37. We thank the referee for 
      the valuable feedback that improved this paper.
\end{acknowledgements}

%--------------------------------------------------------------------

%--------------------------------------------------------------------

\begin{thebibliography}{51}
\expandafter\ifx\csname natexlab\endcsname\relax\def\natexlab#1{#1}\fi

\bibitem[{{Bailey}(2007)}]{2007AsBio...7..320B}
{Bailey}, J. 2007, Astrobiology, 7, 320

\bibitem[{Bailey {et~al.}(2020)Bailey, Cotton, Kedziora-Chudczer, De~Horta, \&
  Maybour}]{bailey2020hippi}
Bailey, J., Cotton, D.~V., Kedziora-Chudczer, L., De~Horta, A., \& Maybour, D.
  2020, Publications of the Astronomical Society of Australia, 37

\bibitem[{Bailey {et~al.}(2018)Bailey, Kedziora-Chudczer, \&
  Bott}]{bailey2018polarized}
Bailey, J., Kedziora-Chudczer, L., \& Bott, K. 2018, Monthly Notices of the
  Royal Astronomical Society, 480, 1613

\bibitem[{Bowler(2016)}]{bowler2016imaging}
Bowler, B.~P. 2016, Publications of the Astronomical Society of the Pacific,
  128, 102001

\bibitem[{Bullock \& Grinspoon(2001)}]{bullock2001recent}
Bullock, M.~A. \& Grinspoon, D.~H. 2001, Icarus, 150, 19

\bibitem[{Cotton {et~al.}(2017)Cotton, Marshall, Bailey, Kedziora-Chudczer,
  Bott, Marsden, \& Carter}]{cotton2017intrinsic}
Cotton, D.~V., Marshall, J.~P., Bailey, J., {et~al.} 2017, Monthly Notices of
  the Royal Astronomical Society, 467, 873

\bibitem[{de~Haan {et~al.}(1987)de~Haan, Bosma, \& Hovenier}]{de1987adding}
de~Haan, J.~F., Bosma, P., \& Hovenier, J. 1987, Astronomy and Astrophysics,
  183, 371

\bibitem[{De~Rooij \& Van~der Stap(1984)}]{de1984expansion}
De~Rooij, W. \& Van~der Stap, C. 1984, Astronomy and Astrophysics, 131, 237

\bibitem[{Donahue {et~al.}(1982)Donahue, Hoffman, Hodges, \&
  Watson}]{donahue1982venus}
Donahue, T., Hoffman, J., Hodges, R., \& Watson, A. 1982, Science, 216, 630

\bibitem[{Garc{\'\i}a-Mu{\~n}oz {et~al.}(2014)Garc{\'\i}a-Mu{\~n}oz,
  P{\'e}rez-Hoyos, \& S{\'a}nchez-Lavega}]{munoz2014glory}
Garc{\'\i}a-Mu{\~n}oz, A., P{\'e}rez-Hoyos, S., \& S{\'a}nchez-Lavega, A. 2014,
  Astronomy \& Astrophysics, 566, L1

\bibitem[{Gaudi {et~al.}(2020)Gaudi, Seager, Mennesson, Kiessling, Warfield,
  Cahoy, Clarke, Domagal-Goldman, Feinberg, Guyon,
  {et~al.}}]{gaudi2020habitable}
Gaudi, B.~S., Seager, S., Mennesson, B., {et~al.} 2020, arXiv preprint
  arXiv:2001.06683

\bibitem[{Hale \& Querry(1973)}]{hale1973optical}
Hale, G.~M. \& Querry, M.~R. 1973, Applied optics, 12, 555

\bibitem[{Hansen \& Hovenier(1974)}]{hansen1974interpretation}
Hansen, J.~E. \& Hovenier, J. 1974, Journal of the Atmospheric Sciences, 31,
  1137

\bibitem[{Hansen \& Travis(1974)}]{hansen1974light}
Hansen, J.~E. \& Travis, L.~D. 1974, Space science reviews, 16, 527

\bibitem[Jordan et al.(2021)]{2021ApJ...922...44J} Jordan, S., Rimmer, P.~B., Shorttle, O., et al.\ 2021, \apj, 922, 44. doi:10.3847/1538-4357/ac1d46

\bibitem[{Kane {et~al.}(2020)Kane, Vervoort, Horner, \&
  Pozuelos}]{kane2020could}
Kane, S.~R., Vervoort, P., Horner, J., \& Pozuelos, F.~J. 2020, The Planetary
  Science Journal, 1, 42

\bibitem[{Karalidi {et~al.}(2011)Karalidi, Stam, \&
  Hovenier}]{karalidi2011flux}
Karalidi, T., Stam, D., \& Hovenier, J. 2011, Astronomy \& Astrophysics, 530,
  A69

\bibitem[{Karalidi {et~al.}(2012)Karalidi, Stam, \&
  Hovenier}]{karalidi2012looking}
Karalidi, T., Stam, D., \& Hovenier, J. 2012, Astronomy \& Astrophysics, 548,
  A90

\bibitem[{Kasper {et~al.}(2010)Kasper, Beuzit, Verinaud, Gratton, Kerber,
  Yaitskova, Boccaletti, Thatte, Schmid, Keller, {et~al.}}]{kasper2010epics}
Kasper, M., Beuzit, J.-L., Verinaud, C., {et~al.} 2010, in Ground-based and
  Airborne Instrumentation for Astronomy III, Vol. 7735, SPIE, 948--956

\bibitem[{Kasting(1988)}]{kasting1988runaway}
Kasting, J.~F. 1988, Icarus, 74, 472

\bibitem[{Keller {et~al.}(2010)Keller, Schmid, Venema, Hanenburg, Jager,
  Kasper, Martinez, Rigal, Rodenhuis, Roelfsema, {et~al.}}]{keller2010epol}
Keller, C.~U., Schmid, H.~M., Venema, L.~B., {et~al.} 2010, in Ground-based and
  airborne instrumentation for astronomy III, Vol. 7735, International Society
  for Optics and Photonics, 77356G

\bibitem[{Kemp {et~al.}(1987)Kemp, Henson, Steiner, \&
  Powell}]{kemp1987optical}
Kemp, J.~C., Henson, G., Steiner, C., \& Powell, E. 1987, Nature, 326, 270

\bibitem[{{Kliore} {et~al.}(1985){Kliore}, {Moroz}, \&
  {Keating}}]{kliore1985venus}
{Kliore}, A.~J., {Moroz}, V.~I., \& {Keating}, G.~M. 1985, Advances in Space
  Research, 5

\bibitem[{Knollenberg \& Hunten(1980)}]{knollenberg1980microphysics}
Knollenberg, R.~G. \& Hunten, D.~M. 1980, Journal of Geophysical Research:
  Space Physics, 85, 8039

\bibitem[{Langlois {et~al.}(2021)Langlois, Gratton, Lagrange, Delorme,
  Boccaletti, Bonnefoy, Maire, Mesa, Chauvin, Desidera,
  {et~al.}}]{langlois2021sphere}
Langlois, M., Gratton, R., Lagrange, A.-M., {et~al.} 2021, Astronomy \&
  Astrophysics, 651, A71

\bibitem[{Laven(2008)}]{laven2008effects}
Laven, P. 2008, Applied optics, 47, H133

\bibitem[{Lincowski {et~al.}(2018)Lincowski, Meadows, Crisp, Robinson, Luger,
  Lustig-Yaeger, \& Arney}]{lincowski2018evolved}
Lincowski, A.~P., Meadows, V.~S., Crisp, D., {et~al.} 2018, The Astrophysical
  Journal, 867, 76

\bibitem[{Lucas {et~al.}(2009)Lucas, Hough, Bailey, Tamura, Hirst, \&
  Harrison}]{lucas2009planetpol}
Lucas, P.~W., Hough, J.~H., Bailey, J.~A., {et~al.} 2009, Monthly Notices of
  the Royal Astronomical Society, 393, 229

\bibitem[{Lustig-Yaeger {et~al.}(2019{\natexlab{a}})Lustig-Yaeger, Meadows, \&
  Lincowski}]{lustig2019detectability}
Lustig-Yaeger, J., Meadows, V.~S., \& Lincowski, A.~P. 2019{\natexlab{a}}, The
  Astronomical Journal, 158, 27

\bibitem[{Lustig-Yaeger {et~al.}(2019{\natexlab{b}})Lustig-Yaeger, Meadows, \&
  Lincowski}]{lustig2019mirage}
Lustig-Yaeger, J., Meadows, V.~S., \& Lincowski, A.~P. 2019{\natexlab{b}}, The
  Astrophysical Journal Letters, 887, L11

\bibitem[{Markiewicz {et~al.}(2014)Markiewicz, Petrova, Shalygina, Almeida,
  Titov, Limaye, Ignatiev, Roatsch, \& Matz}]{markiewicz2014glory}
Markiewicz, W.~J., Petrova, E., Shalygina, O., {et~al.} 2014, Icarus, 234, 200

\bibitem[{Markiewicz {et~al.}(2018)Markiewicz, Petrova, \&
  Shalygina}]{markiewicz2018aerosol}
Markiewicz, W.~J., Petrova, E.~V., \& Shalygina, O.~S. 2018, Icarus, 299, 272

\bibitem[{Nielsen {et~al.}(2019)Nielsen, De~Rosa, Macintosh, Wang, Ruffio,
  Chiang, Marley, Saumon, Savransky, Ammons, {et~al.}}]{nielsen2019gemini}
Nielsen, E.~L., De~Rosa, R.~J., Macintosh, B., {et~al.} 2019, The Astronomical
  Journal, 158, 13

\bibitem[{Palmer \& Williams(1975)}]{palmer1975optical}
Palmer, K.~F. \& Williams, D. 1975, Applied Optics, 14, 208

\bibitem[{{Quarles} \& {Lissauer}(2016)}]{2016AJ....151..111Q}
{Quarles}, B. \& {Lissauer}, J.~J. 2016, \aj, 151, 111

\bibitem[{Ragent {et~al.}(1985)Ragent, Esposito, Tomasko, Marov, Shari, \&
  Lebedev}]{ragent1985particulate}
Ragent, B., Esposito, L., Tomasko, M., {et~al.} 1985, Advances in Space
  Research, 5, 85

\bibitem[{{Rodenhuis} {et~al.}(2012){Rodenhuis}, {Canovas}, {Jeffers}, {de Juan
  Ovelar}, {Min}, {Homs}, \& {Keller}}]{rodenhuis2012extreme}
{Rodenhuis}, M., {Canovas}, H., {Jeffers}, S.~V., {et~al.} 2012, in Society of
  Photo-Optical Instrumentation Engineers (SPIE) Conference Series, Vol. 8446,
  Ground-based and Airborne Instrumentation for Astronomy IV, ed. I.~S.
  {McLean}, S.~K. {Ramsay}, \& H.~{Takami}, 84469I

\bibitem[{Rossi {et~al.}(2021)Rossi, Berzosa-Molina, Desert, Fossati,
  Mu{\~n}oz, Haswell, Kabath, Kislyakova, Stam, \&
  Vidotto}]{rossi2021spectropolarimetry}
Rossi, L., Berzosa-Molina, J., Desert, J.-M., {et~al.} 2021, Experimental
  Astronomy, 1

\bibitem[{Rossi {et~al.}(2018)Rossi, Berzosa-Molina, \&
  Stam}]{rossi2018pymiedap}
Rossi, L., Berzosa-Molina, J., \& Stam, D.~M. 2018, Astronomy \& Astrophysics,
  616, A147

\bibitem[{Rossi {et~al.}(2015)Rossi, Marcq, Montmessin, Fedorova, Stam,
  Bertaux, \& Korablev}]{rossi2015preliminary}
Rossi, L., Marcq, E., Montmessin, F., {et~al.} 2015, Planetary and Space
  Science, 113, 159

\bibitem[{Rossi \& Stam(2018)}]{rossi2018circular}
Rossi, L. \& Stam, D.~M. 2018, Astronomy \& Astrophysics, 616, A117

\bibitem[{{Russell} {et~al.}(1992){Russell}, {Brown}, {Chandos}, {Fincher},
  {Kubel}, {Lacis}, \& {Travis}}]{1992SSRv...60..531R}
{Russell}, E.~E., {Brown}, F.~G., {Chandos}, R.~A., {et~al.} 1992, \ssr, 60,
  531

\bibitem[Snik et al.(2011)]{2011ASPC..437..237S} Snik, F., Kochukhov, O., Piskunov, N., et al.\ 2011, Solar Polarization 6, 437, 237. doi:10.48550/arXiv.1010.0397

\bibitem[{Stam(2008)}]{stam2008spectropolarimetric}
Stam, D. 2008, Astronomy \& Astrophysics, 482, 989

\bibitem[{Stam \& Hovenier(2005)}]{stam2005errors}
Stam, D. \& Hovenier, J. 2005, Astronomy \& Astrophysics, 444, 275

\bibitem[{The Luvoir Team (2019)}]{luvoir2019luvoir}
Team, L. {et~al.} 2019, arXiv preprint arXiv:1912.06219

\bibitem[{Titov {et~al.}(2018)Titov, Ignatiev, McGouldrick, Wilquet, \&
  Wilson}]{titov2018clouds}
Titov, D.~V., Ignatiev, N.~I., McGouldrick, K., Wilquet, V., \& Wilson, C.~F.
  2018, Space Science Reviews, 214, 1

\bibitem[{{Trees} \& {Stam}(2022)}]{2022A&A...664A.172T}
{Trees}, V.~J.~H. \& {Stam}, D.~M. 2022, \aap, 664, A172

\bibitem[{{Tselioudis}(2001)}]{isccp_particle}
{Tselioudis}, G. 2001, {ISCCP Definition of Cloud Types}

\bibitem[{Turbet {et~al.}(2021)Turbet, Bolmont, Chaverot, Ehrenreich, Leconte,
  \& Marcq}]{turbet2021day}
Turbet, M., Bolmont, E., Chaverot, G., {et~al.} 2021, Nature, 598, 276

\bibitem[{van Holstein(2021)}]{van2021high}
van Holstein, R. 2021, PhD thesis, Leiden University

\bibitem[{Way \& Del~Genio(2020)}]{way2020venusian}
Way, M.~J. \& Del~Genio, A.~D. 2020, Journal of Geophysical Research: Planets,
  125, e2019JE006276

\end{thebibliography}
\end{document}